\documentclass[aps,twocolumn,amsmath,amssymb,showpacs,prb,floatfix]{revtex4}
\usepackage{ulem}
\usepackage{dsfont}
\usepackage{amsmath}
\usepackage{color}
\usepackage{graphicx}
\usepackage{epsfig}
\usepackage{SIunits}
\usepackage{mathtools}

\def\joinrelde{\mathrel{\mkern-11mu}}
\def\joinrel{\mathrel{\mkern-9mu}}

\def\joinrelw{\mathrel{\mkern-6mu}}

\def\relbd{\mathrel{{\bf\smash{{\phantom- \above1pt \phantom-
}}}}}
\def\ltdash{\raise-1.8pt\hbox{$\scriptscriptstyle |$}}
\newlength{\upit}\upit=0.1truein
\newcommand{\raiser}[1]{\raisebox{\upit}[0cm][0cm]{#1}}
\newcommand{\ltappr}{{{\lower4pt\hbox{$<$} } \atop \widetilde{ \ \ \
}}}
\newcommand{\gtappr}{{{\lower4pt\hbox{$>$} } \atop \widetilde{ \ \ \ }}}
\newlength{\bxwidth}\bxwidth=1.5 truein
\newcommand\frm[1]{\epsfig{file=#1,width=\bxwidth}}
\newcommand{\zmatrix}[4]{\left(\begin{matrix}#1 & #2\cr #3&#4\end{matrix}\right)}

\newcommand{\tr}{{\hbox{Tr}}}

\newcommand{\dg}{^{\dagger }}

\newcommand{\up}{\uparrow}
\newcommand{\dw}{\downarrow}
\newcommand{\rarrow}{\rightarrow}
\newcommand{\pmat}[1]{\begin{pmatrix} #1 \end{pmatrix}}
\newcommand{\Hast}{{\Psi}}
\newcommand{\hast}{{\Psi}}

\newlength{\figwidth}
\newlength{\shift}
\newlength{\fight}
\newcommand{\fg}[3]
{
\begin{figure}[ht]

\vspace*{-0cm}
\[
\includegraphics[width=\fight]{#1}
\]
\vskip -0.2cm
\caption{\label{#2}
\small #3
}
\end{figure}}
\newcommand{\fgb}[3]
{
\begin{figure}[b]
\vskip 0.0cm
\begin{equation}\label{}
\includegraphics[width=\figwidth]{#1}\nonumber
\end{equation}
\vskip -0.2cm
\caption{\label{#2}
\small #3
}
\end{figure}}

\newcommand \bea {\begin{eqnarray} }
\newcommand \eea {\end{eqnarray}}
\newcommand{\bk}{{\bf{k}}}
\newcommand{\bx}{{\bf{x}}}
\newcommand{\bQ}{{\bf{Q}}}
\newcommand{\bR}{{\bf{R}}}
\newcommand{\urs}{URu$_{2}$Si$_{2}$\ }
\newcommand{\ursp}{URu$_{2}$Si$_{2}$}
\newcommand{\Det}[1]{{\rm det}\left[ {#1} \right]}
\newcommand{\psic}{{c}}

\newcommand{\contract}[1]{\stackrel{\mathclap{\displaystyle \ltdash
\joinrel\relbd
\joinrelw\relbd
\joinrelw\relbd
\joinrelw\relbd\joinrel
 \ltdash
\joinrelde\relbd
\joinrel\relbd
\joinrel\relbd
\joinrel\relbd
\joinrel\relbd\joinrel
 \ltdash
}}{#1}}
\newcommand{\contracty}[1]{\stackrel{\mathclap{\displaystyle
\ltdash
\joinrel\relbd
\joinrelw\relbd
\joinrelw\relbd
\joinrelw\relbd\joinrel
\ltdash
}}{#1}}

\begin{document}
\title{Hastatic Order in \urs:  Hybridization with a Twist}
\author{Premala Chandra,$^1$ Piers Coleman,$^1$ and Rebecca Flint$^2$}
\affiliation{$^1$ Center for Materials Theory, Department of 
Physics and Astronomy, Rutgers University, 
Piscataway, NJ 08854}
\affiliation{$^2$ 
Department of Physics and Astronomy, Iowa State University, Ames, Iowa 50011, USA}
\date{\today}
%
\begin{abstract}
The broken symmetry that develops below 17.5K in the heavy fermion compound
\urs has long eluded identification. Here we argue 
that the recent observation of Ising quasiparticles in \urs
results from a spinor hybridization order parameter 
that breaks {\sl double} time-reversal symmetry by mixing states
of integer and half-integer spin. Such ``hastatic order'' ({\sl hasta}:[Latin]{spear}
) hybridizes Kramers conduction electrons with Ising, non-Kramers  $5f^{2}$ states of the uranium
atoms to produce Ising quasiparticles.  The development of a spinorial hybridization at 17.5K accounts for both the large entropy of condensation and the
magnetic anomaly observed in torque magnetometry.  This paper develops
the theory of hastatic order in detail, providing the mathematical 
development of its key concepts.
Hastatic order predicts a tiny transverse moment
in the conduction sea, a collosal Ising anisotropy in the
nonlinear susceptibility anomaly and a resonant
energy-dependent nematicity in the tunneling density of states.
\end{abstract}
\maketitle

\tableofcontents
\section{Introduction}

%
%


The heavy fermion superconductor \urs  exhibits a large
specific heat anomaly at $T_{0}=$17.5K signalling the development of long-range
order with an associated entropy of condensation, 
$ \int_{0}^{T_{0}}\frac{C_{V}}{T}dT\approx \frac{1}{2}R\ln 2$
per mole formula unit\cite{Palstra85,Schlabitz86}. This sizable ordering entropy in conjunction
with the sharpness of the specific heat anomaly suggests underlying
itinerant ordering; however the anisotropic bulk spin 
susceptibility (cf. Fig. 1) 
of \urs displays Curie-Weiss behavior down to $T \sim 70 K$ indicative 
of local moment behavior.
Initially, the hidden order was attributed to a spin density wave with
a tiny c-axis moment\cite{broholm91,walter93}, 
which was later shown to be extrinsic\cite{takagi07}. 
However, spin ordering in the form of 
antiferromagnetism is observed at pressures exceeding 0.8GPa\cite{Amitsuka99,Amitsuka07,Butch2010}.
Despite thirty years of intense experimental effort, no laboratory 
probe has yet coupled directly to the order parameter in \urs at 
ambient pressure, though there have been a wide variety 
of theoretical proposals for this ``hidden order'' (HO) 
problem\cite{Amitsuka94,Haule09,Santini94,Varma,pepin,Morr10,Dubi10,Fujimoto11,Ikeda11,Mydosh11,hastatic,Flint14,Chandra14}. We point 
the interested reader to a recent review on \urs for more details.\cite{Mydosh11} 

Expanding on our recent proposal\cite{hastatic,Flint14,Chandra14} of ``hastatic order'' 
in \urs, here we argue that the failure to observe the nature of its ``hidden 
order'' is not due to its intrinsic complexity but instead results from a 
fundamentally new kind of broken time-reversal symmetry 
associated with an order parameter of spinorial, half-integer spin character. 
Key evidence supporting this conjecture is the observation of quasiparticles 
with an  Ising anisotropy characteristic of integer spin f-moments\cite{Ohkuni99,Altarawneh11,Hc2,Altarawneh2}. 
Hastatic order accounts for this unusual feature as a consequence of a 
spinor order parameter that coherently hybridizes  the integer 
spin, Ising f-moments with half-integer spin conduction electrons; the 
observed quasiparticle and the magnetic anisotropies thus have the same origin.

\fight=\columnwidth
\fg{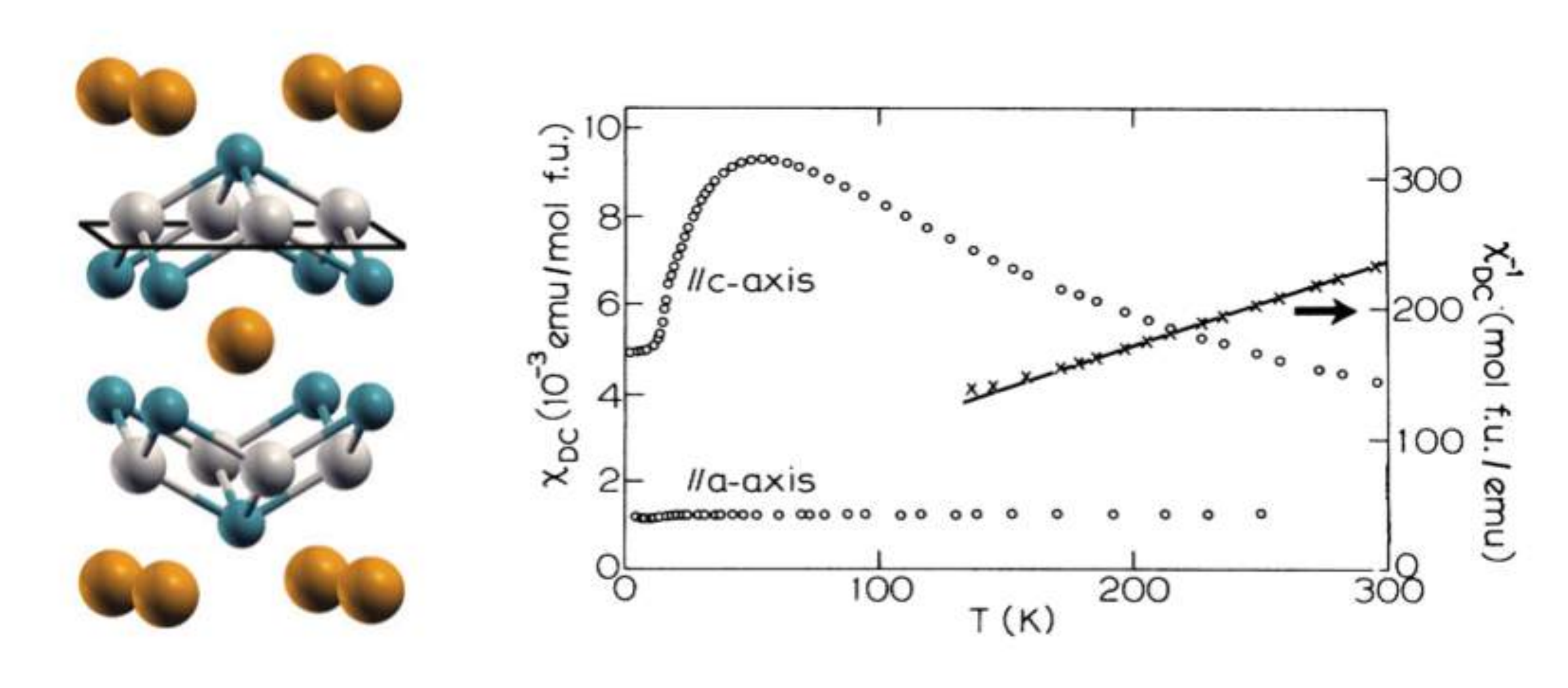}{fig1}{(a) Tetragonal structure of \urs (b)
Temperature dependence of magnetic susceptibility in \urs after \cite{palstra86}.}

\subsection{Experimental Motivation for Hastatic Order}

Above the hidden order
transition, \urs is an incoherent 
heavy fermion metal, with an large, anisotropic, linear 
resistivity\cite{palstra86},
 and a linear specific heat with $\gamma =
\frac{C_{V}}{T}\sim 200 $mJmol$^{-1}$K$^{-2}$ . The development of
hidden order results in a significant reduction in the specific heat to 
$\gamma_{0}= \frac{C_{V}}{T}\sim 60 $
mJmol$^{-1}$K$^{-2}$, corresponding to the 
the loss of about two-thirds of the heavy Fermi surface\cite{Mydosh11}.  
At $T_c = 1.5K$, the remaining heavy quasiparticles go superconducting.
Under a modest pressure of 0.8GPa, the hidden order ground-state 
of \urs undergoes a first order transition into an Ising
antiferromagnet with an staggered ordered moment of order 0.4
$\mu_{B}$  aligned along the c-axis\cite{Amitsuka07}. 
de Haas-van Alphen (dHvA) shows that the quasiparticles in the hidden order
phase form small, highly coherent heavy electron pockets with an
effective mass up to $8.5 m_{e}$\cite{Ohkuni99}.
Remarkably, 
these small heavy electron pockets survive across the first order
transition into the high-pressure antiferromagnetic phase, 
leading many groups to conclude that the (commensurate) ordering wavevectors ${\bf Q}= (0,0,1)$ 
of the antiferromagnetic and the HO phases
are the same\cite{Jo07,Villaume08,Hassinger10,Haule09,Haule10}. 

Perhaps the most dramatic feature of these heavy electron pockets is
the essentially perfect 
Ising magnetic anisotropy 
in the magnetic g-factors of the itinerant
heavy f-electrons in the HO state of URS\cite{Altarawneh11}. This Ising quasiparticle
anisotropy has been determined by measuring the Fermi surface magnetization
in an angle-dependent magnetic field in the HO state;
this magnetization is a periodic function of the ratio 
of the Zeeman and the cyclotron energies, where the former
is defined 
through an angle-dependent g-factor $g (\theta )$ 
\begin{equation}
\Delta E(\theta) = g(\theta) \mu_B |B|.
\end{equation} 
Interference of Zeeman-split 
orbits in tilted fields leads to spin zeroes in the quantum oscillation 
measurements (cf. Fig. 2) satisfying the condition
\begin{equation} 
g(\theta_n) \frac{m^*}{m_e} = 2n + 1
\end{equation}
where $n$ is a positive integer and 
$\theta_n$ is the (indexed) angle with respect to the c-axis.  
Sixteen such spin zeroes were identified in the HO state of URS\cite{Ohkuni99,Altarawneh11}, 
and the experimentalists found that 
\begin{equation}
\frac{g_\perp}{g_c} < \frac{1}{30}
\end{equation}
where $g_\perp = g(\theta_n \sim \frac{\pi}{2})$ and $g_c = g(\theta_n = 0)$, 
indicating that $\Delta E(\theta_n)$ depends {\sl solely} 
on the c-axis component of the applied magnetic field ($B_c$), namely that
\begin{equation}
g(\theta_n) = g^* \cos \theta_n
\end{equation}
where $g^* = 2.6$ in contrast to the isotropic $g=2$ for free electrons. 

In these high-field measurements, the quasiparticle anisotropy manifests 
itself through the
appearance of a rapid modulation in the amplitude of the dHvA oscillations generated by the heavy $\alpha $ pockets of \urs as
the magnetic field is tilted from the c-axis into the basal plane.  
The {\sl same}
magnetic anisotropy is also observed in the angular dependence of the
upper-critical field of the superconducting state which develops at
low temperatures (cf. Fig. 2)\cite{Hc2,Altarawneh2}.  
Whereas the dHvA measurements could in principle belong to a
select region of the Fermi surface, the upper-critical field, $H_{c2}(\theta)$ is
sensitive to the entire  heavy fermion pair condensate, proving
crucially that the Ising quasiparticle anisotropy pervades the 
entire Fermi surface of hidden order state. 
{We note that while $H_{c2}(\theta)$ matches the anisotropy of the g-factor for angles near the c-axis, where $H_{c2}$ is Pauli limited, when the field is in-plane, $H_{c2}(\theta)$ is larger than expected, likely due to orbital contributions.}  
We note that since the Pauli susceptibility $\chi^P$ 
scales with the {\sl square} of 
the g-factor, these resolution-limited measurements of $\frac{g_c}{g_\perp}$ 
suggest that 
\begin{equation}
\chi^P (\theta) = \chi^{P*}\cos^2 \theta \qquad
\frac{\chi^P_c}{\chi^P_\perp} > 900.
\end{equation}
Such a large anisotropy should be observable in electron spin 
resonance measurements that probe the Pauli susceptibility directly
in contrast to bulk susceptibility measurements where  
Van Vleck contributions are also present.

\fight=\columnwidth
\fg{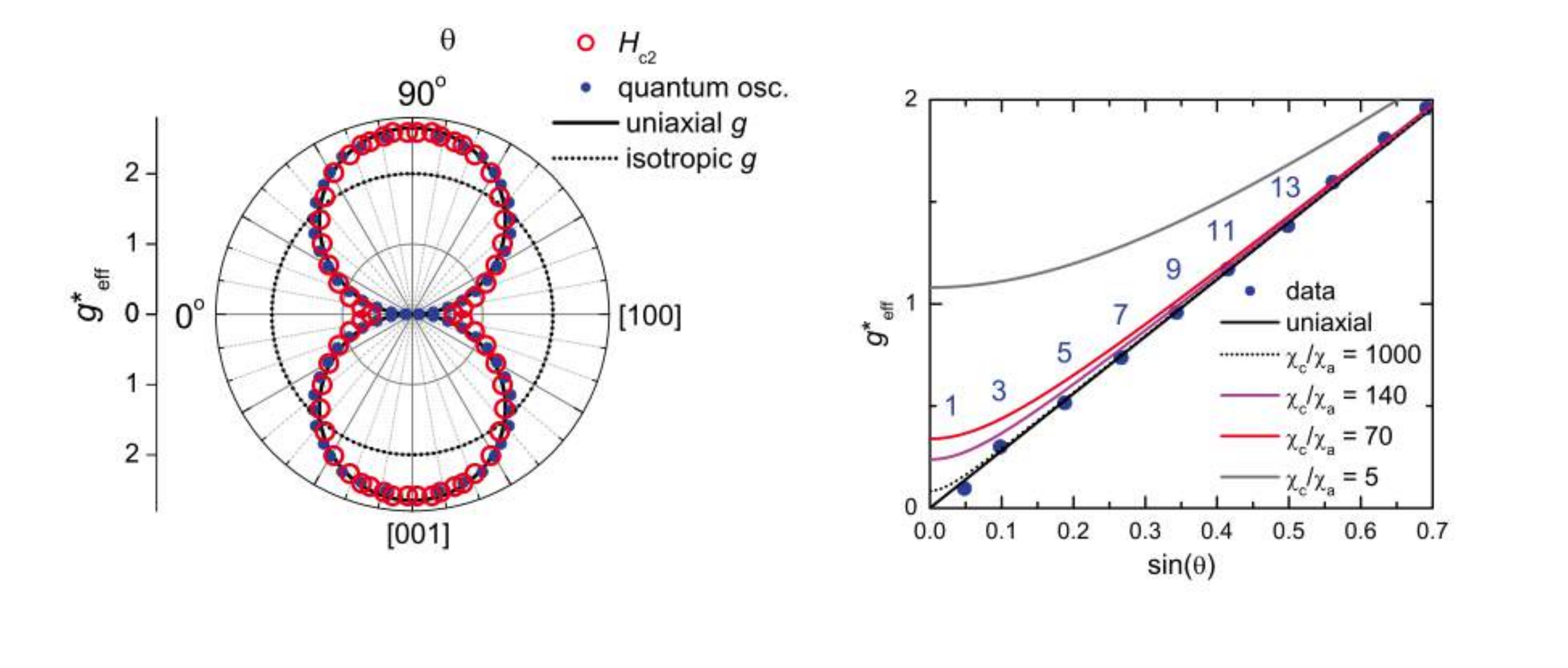}{fig2}{(a) Anisotropy of the measured 
g-factor\cite{Altarawneh11}
plotted (a) in polar co-ordinates derived from spin zeros in quantum oscillation
measurements and the anisotropy of the upper critical field 
(b) versus sine of the angle out of the
basal plane, showing that the data\cite{Altarawneh11} requires a Pauli
susceptibility anisotropy in excess of a thousand.}



\subsection{Ground-State Configuration of the Uranium Ion}

Bulk susceptibility measurements (see Fig. 1) of \urs  reveal that the
magnetic U ions have an Ising anisotropy, with a magnetic moment
that is consistent either with a U$^{3+}$ ($5f^{3}$)  configuration or
a non-Kramers $U^{4+}$ ($5f^{2}$) ion\cite{Palstra85,Ramirez92}. A natural explanation for 
the quasiparticle Ising anisotropy is that the Ising character of 
the uranium (U) ions has been transferred to the quasiparticles
via hybridization, and this is a key element of the hastatic proposal.
The giant anisotropy in $\frac{g_\perp}{g_c}$, places a strong constraint 
on the energy-splitting $\Delta$ between the two Ising states
\begin{equation}
\frac{2 \Delta}{(g_c \mu_B B_{\rm dHvA})} < \frac{1}{30}
\end{equation}
requiring that in a transverse field, $B_{\rm dHvA} = 11 \rm{T}$ the U ion is doubly degenerate to
within $\Delta \sim 1 K$.   {Further support for a very small $\Delta$ comes from the dilute limit,
U$_x$Th$_{1−x}$Ru$_2$Si$_2$ ($x \le .07$), where the Curie-like single-ion behavior
crosses over to a critical logarithmic temperature-dependence\cite{Amitsuka94} 
below $10K$,
$−\log T/T_K$, where $T_K \approx 10K$. This physics
has been attributed to two channel Kondo criticality, again requiring a splitting $\Delta \ll 10K$.}
Such a degeneracy is of course natural for
in a Kramers U ($5f^{3}$) ion,  containing an odd-number of f-electrons.  
However it can also occur in a $5f^{2}$ ``non-Kramers doublet'' with a 
two-fold orbital degeneracy protected by tetragonal symmetry\cite{Amitsuka94,Ohkawa99}.

Motivated and constrained by the bulk spin susceptibility and the 
quantum spin zeroes, we therefore require the U ion to be in an Ising 
doublet in the tetragonal environment of URS (cf. Fig. 1); such a magnetic 
doublet
of \urs has the form    
\begin{equation}
|\Gamma_\pm \rangle  = \sum_n a_n |\pm (J_z - 4n)\rangle,
\end{equation}
where the addition and subtraction of angular momentum in units of
$4\hbar $ is a consequence of the four-fold symmetry of the tetragonal
crystal. 
However, the presence of a perfect Ising anisotropy requires an 
{\sl Ising selection rule}
\begin{equation}\label{sum}
\langle \Gamma_\pm |J_\pm
|\Gamma_\mp\rangle  = 0
\end{equation}
that, in the absence of fine-tuning of the coefficients $a_n$, 
leads to the condition that $-(J_{z}+4n')\neq (J_{z}+4n)\pm 1$, or
$J_{z}\neq 2(n-n')\pm \frac{1}{2}$, requiring $J_z \in \mathbb{Z}$
{\sl must be an integer}.
For any
half-integer $J_z$, corresponding to a Kramers doublet, the
selection rule is absent and the ion  will
develop a generic basal-plane moment. 
As an aside we note that the fine-tuned case will produce an Ising Kramers
doublet, but corresponds to the complete absence of tetragonal 
mixing, highly unlikely
in a tetragonal environment. 

Let us apply this argument specifically to the case of \urs.  
We first suppose that the U ion is in a $5f^3$ ($J = \frac{9}{2}$)
configuration, predominantly in a $|\pm \frac{7}{2}\rangle$ state.  
The presence of tetragonal symmetry results in a crystal-field ground-state
\begin{equation}
|\pm\rangle = a|\pm\frac{7}{2}\rangle + b|\mp\frac{1}{2}\rangle + c|\mp\frac{9}{2}\rangle
\end{equation}
so that $|\langle - | J_+ |+\rangle|^2 = 5b^2 + 6ac$ and perfect Ising
anisotropy is only achieved with fine-tuning of the tetragonal mixing
coefficients such that the condition
$5b^2 + 6ac = 0$
is satisfied.  By contrast, for U ion in a $5f^2$ ($J = 4$)
configuration, its ground-state may be a non-Kramers $\Gamma_5$ doublet
\begin{equation}
|\pm\rangle = a |\pm 3\rangle + b|\mp 1\rangle,
\end{equation}
where Ising anisotropy exists for arbitrary mixing 
between the $|\pm 3\rangle$
and $|\mp 1\rangle$ states since this tetragonally-stabilized 
$\Gamma_5$ state is dipolar in the c-direction
and quadrupolar in the a-b plane.  Because the phase 
space associated with the non-Kramers doublet
is significantly larger than that for its finely-tuned Kramers 
counterpart, we take the $\Gamma_5$
doublet to be the more natural ground-state configuration of the U ion in URS. 
However we have proposed a direct benchtop test
to distinguish between these two single-ion U ground-state 
configurations in \urs using the basal-plane nonlinear susceptibility to check
this key assumption in the hastatic proposal\cite{flint12}.  

The combination of the observed Ising anisotropy and tetragonal symmetry
are crucial towards pointing us to the non-Kramers $\Gamma_5$ doublet. By contrast in a hexagonal system, like 
CeAl$_3$\cite{Goremychkin00}, the six-fold symmetry mixes terms that differ by $6\hbar$ units of angular momentum, so
a pure doublet $|\pm M\rangle$ mixes with $|\pm M'\rangle$ states only if $M' = M - 6 n \;( n \in \mathbb{Z})$.  For $J < 7/2$, 
the maximum $M - M'$ is $\frac{5}{2} - (-\frac{5}{2}) = 5$ meaning there is no valid choice of $M$ and $M'$;
thus there are two Ising doublets for the Ce ($J=5/2$) case: 
$|\pm 5/2\rangle$ and $|\pm 3/2\rangle$.  These Kramers doublets can
undergo a {\sl single channel} Ising Kondo effect
\cite{Goremychkin00,sikkema96}
that will differ substantially from the two-channel Kondo
physics associated with a non-Kramers doublet\cite{CoxZawad}.

\subsection{Hybridization, Hastatic Order and Double-Time Reversal Symmetry}

\fight=\columnwidth
\fg{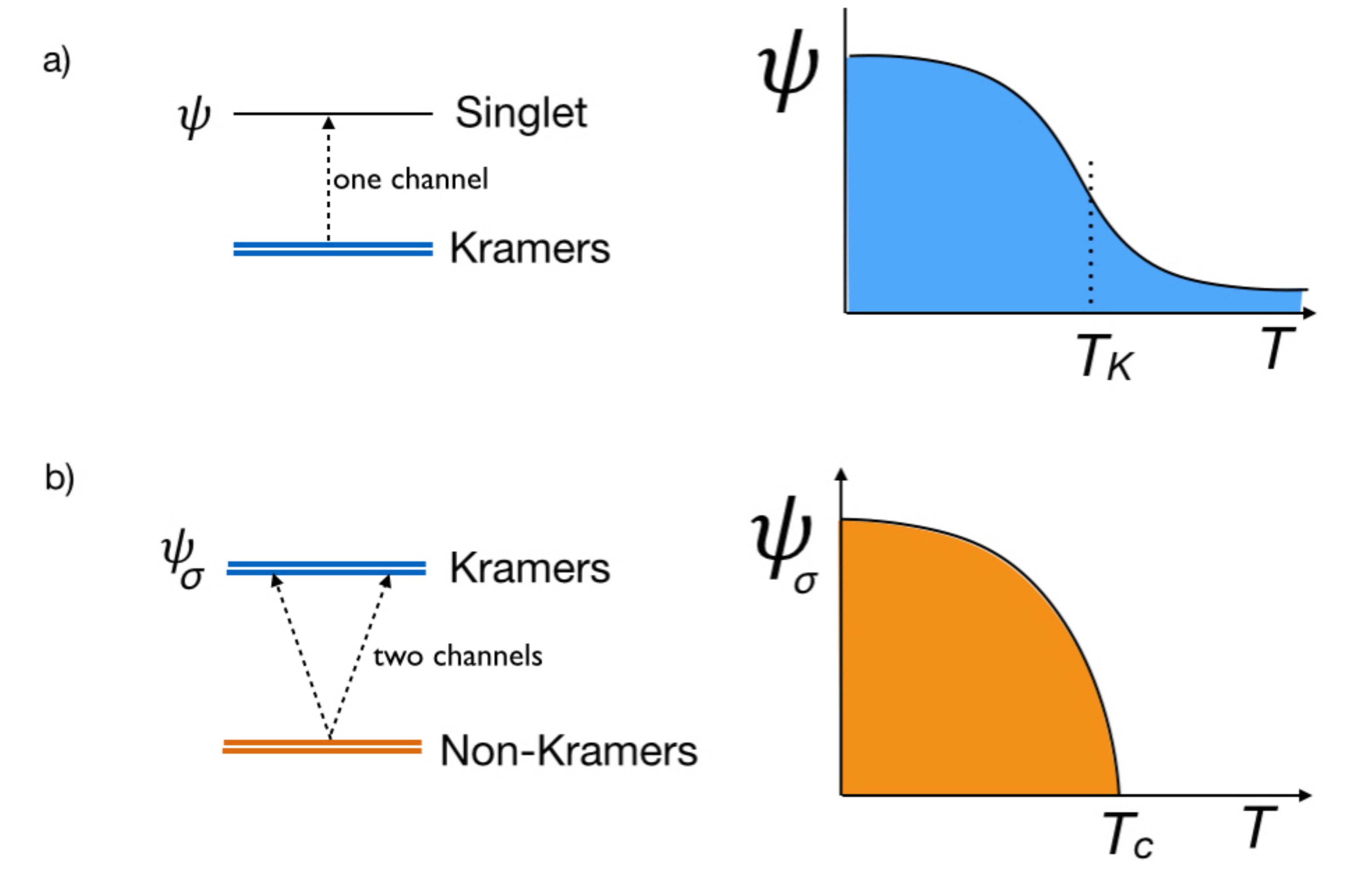}{fig3}{Schematic of (a) conventional (scalar) vs  (b)
spinorial hybridization where the hybridization is a) a crossover
and b) breaks spin-rotational and time-reversal symmetries and thus
develops discontinuously as a phase transition.}

The formation of heavy f-bands  in heavy fermion systems involves the
formation of f-resonances within the conduction sea.  Hybridization
between these many-body resonances and the conduction electrons
produces charged heavy f-electrons that inherit the magnetic
properties of the local moments. When this process involves a
Kramers doublet (the usual case), the hybridization can develop without any broken
symmetry and thus is associated with a crossover.   At first sight, the most straightforward explanation of hidden order
is to attribute it to the formation of a ``heavy density wave'' 
within a pre-formed  heavy electron fluid. Since 
there is no observed magnetic moment or charge density 
observed in the hidden order phase, such a
density wave must necessarily involve a higher order multipole
of the charge or spin degrees of freedom and various theories in this category
have indeed been advanced. In each of these scenarios, 
the heavy electrons develop coherence via cross-over at
 higher temperatures, and the essential hidden order is then a
multipolar charge or spin density wave. 
However such multipolar order can not account for the emergence of
heavy Ising quasiparticles.  The essential point here, is that
conventional quasiparticles have
half-integer spin, lacking the lack the essential Ising protection 
required by experiment. 

Moreover,
in \urs both optical \cite{Timusk11} and 
tunnelling \cite{Schmidt10,Aynajian10,Park11}
probes suggest that hybridization develops abruptly at the HO
transition, leading to proposals\cite{Morr10,Dubi10}
that the hybridization is an order parameter. The
associated global broken symmetry and phase transition
is naturally described within the hastatic proposal. As we 
now describe, hybridization with a
non-Kramers doublet requires the development of an order parameter
that breaks double time-reversal symmetry, a requirement that leads us
to conclude that the order parameter has a spinorial quality. (See
Fig. \ref{fig3})





The observation of heavy quasiparticles with Ising anisotropy in the
tetragonal environment of URS implies an underlying hybridization of 
half-integer spin electrons with integer-spin doublets that 
has important symmetry implications for the nature of the hidden order.  
More specifically such hybridization requires a quasiparticle mixing terms 
in the low-energy fixed-point Hamiltonian of the form
\begin{equation}\label{mixing}
{\cal H} = |k \sigma\rangle V_{\sigma \alpha} (k) \langle \alpha | + H.C.
\end{equation}
where $|k \sigma \rangle$ and $|\alpha\rangle$ refer to the half-integer 
spin conduction and the integer-spin doublet states respectively and 
H.C. is the Hermitian conjugate;
here $k$ and $\sigma$ are the momentum and the spin components respectively.
Because time-reversal, $\hat{\Theta}$, is an anti-unitary quantum operator, it has
no associated eigenvalue.
However {\sl double} time-reversal $\hat{\Theta}^2$,
equivalent to a $2\pi$ rotation, is a unitary operator whose quantum number is the
Kramers index, $K = (-1)^{2J}$ where $J$ refers to the total angular momentum of
the quantum state; $K$  
defines the phase factor
acquired by its wavefunction after two successive time-reversals:
\begin{equation}
\hat\Theta^{2}\vert \psi \rangle = \vert \psi^{2\pi}\rangle 
= K \vert \psi \rangle.
\end{equation}
For an integer-spin state $\vert \alpha\rangle$, $K=1$ since 
$\vert \alpha^{2\pi}\rangle = +\vert \alpha\rangle$, indicating that
it is unchanged by a $2\pi$ rotation.
By contrast, for conduction electrons with half-integer spin states, $\vert k \sigma \rangle $, 
$\vert k \sigma^{2\pi}\rangle = - \vert k \sigma \rangle $, so that 
$K=-1$.
Therefore, by mixing half-integer and integer spin states,  the quasiparticle hybridization ${\cal H}$ does 
not conserve Kramers index.  
Indeed application of two successive time-reversals to ${\cal H}$ yields
\begin{equation}
\hat{\Theta}^2 (V \vert k \sigma \rangle \langle \alpha \vert) =  V^{{2\pi}}\vert k \sigma ^{2\pi}\rangle \langle \alpha^{2\pi}\vert
= - V^{2\pi}\vert k \sigma \rangle \langle \alpha \vert.
\end{equation}
Since the microscopic Hamiltonian must be time-reversal invariant, 
it follows that 
\begin{equation}
V=-V^{2\pi}
\end{equation}  
so the hybridization transforms
as a half-integer spin state and is therefore a spinor.
It then follows that this spinorial hybridization breaks both single-
and double-time reversal symmetries distinct from conventional
magnetism where Kramers index is conserved; we call this new 
state of matter ``hastatic (Latin: {\sl spear}) order''.

Before we proceed to discuss the theory of spinorial hybridization and its consequences,
let us pause briefly to elaborate on the distinction between spinors and vectors, expanding
the previous discussion with emphasis on time-reversal symmetry properties.  Quantum mechanically, 
the non-relativistic evolution of a wavefunction $\psi (x,t)$ is described by the Schr\"odinger equation 
$H \psi = i\hbar \frac{\partial \psi}{\partial t}$ so that
\begin{equation}
\Theta \psi (x,t) = \psi^* (x, -t)
\end{equation}
where $\Theta$ is the time-reversal operator; this is an anti-unitary operation.  For vector spins,
$\vec{S} \rightarrow - \vec{S}$.  The spin $\frac{1}{2}$ wavefunction is a {\sl spinor} 
\begin{equation}
\Psi  = \left(\begin{matrix}
\psi_\up\cr
\psi_\dw\end{matrix}
\right)
\end{equation}
so that
\begin{equation}
\Theta \Psi(x,t) = 
\left(\begin{matrix}
-\psi^*_\dw(x,-t)\cr
\psi^*_\up(x,-t)\end{matrix}
\right)
\end{equation}
and
\begin{equation}
\Theta^2 \Psi(x,t) = 
\left(\begin{matrix}
-\psi^*_\up(x,t)\cr
-\psi^*_\dw(x,t)\end{matrix}
\right) = - \Psi(x,t)
\end{equation}
Bosons, with integer spins, are vectors with $\Theta^2 = +1$, whereas fermions with half-integer spins
have $K = - 1$.  
Because of its spinorial character, we can think of hastatic order as ``hybridization with a twist'' since
this is a simple way of visualization its behavior 
under $2 \pi$ (inverted) and $4 \pi$ (restored) rotations.

Hybridization
in heavy fermion compounds 
is usually driven by valence fluctuations mixing
a
ground-state Kramers doublet and an excited singlet (cf. Fig. 1a). 
In this case, the hybridization
amplitude is a scalar that develops 
via a crossover, leading to mobile heavy fermions.
However valence fluctuations from a $5f^{2}$ ground-state create
excited states with an odd number of electrons and hence a Kramers
degeneracy (cf. Fig. 1b). 
Then the quasiparticle hybridization 
has two components, $\hast_{\sigma }$,
that determine the mixing of the excited Kramers doublet into the
ground-state.  These two amplitudes form a spinor defining the 
``hastatic'' 
order parameter
\begin{equation}\label{}
\Hast= \left(\begin{matrix} \hast_{\up }\cr \hast_{\dw}\end{matrix}
\right).
\end{equation}
Loosely speaking, the hastatic spinor is the square root of a multipole
\begin{equation}\label{}
\Psi \sim \sqrt{\hbox{multipole}}.
\end{equation}
Moreover, the presence of distinct up/down hybridization
components indicates that $\Psi$ 
carries a half-integer spin quantum number;
its development must 
now break double time-reversal and spin rotational
invariance via a phase transition.

Under pressure, \urs  undergoes a first-order phase transition from
the hidden order (HO) state to an antiferromagnet (AFM) 
\cite{Amitsuka07}.  These two states are remarkably close in energy
and share
many key features\cite{Jo07,Villaume08,Hassinger10} including
common Fermi surface pockets; this motivated
the recent proposal that despite the first order transition
separating the two phases, they are linked by ``adiabatic 
continuity,''\cite{Jo07} corresponding to a notional rotation
of the HO in internal parameter space \cite{Haule09,Haule10}.
In the magnetic phase, this spinor points along the c-axis
\begin{equation}\label{}
\Hast_A\sim \left(\begin{matrix} 1 
\cr 0 
\end{matrix} \right)  
\hbox{,  }
\Hast_B\sim 
 \left(\begin{matrix} 0
\cr 1 
\end{matrix} \right)  
\end{equation}
corresponding to time-reversed configurations on alternating layers A
and B, leading to a staggered Ising moment.
For the HO state, 
the spinor points in the basal plane
\begin{equation}\label{}
\Hast_A \sim  \frac{1}{\sqrt{2}}\left(\begin{matrix} 
e^{-i\phi/2} \cr  e^{i \phi/2 } \end{matrix} \right)
\hbox{,  }
\Hast_B \sim  \frac{1}{\sqrt{2}}\left(\begin{matrix} 
- e^{-i\phi/2} \cr  e^{i \phi/2 } \end{matrix} \right),
\end{equation}
where again, $\Psi_{B}= \Theta \Psi_{A}$. This state
is protected from developing
large moments by the pure Ising character of the $5f^{2}$
ground-state.

\subsection{Two-Channel Valence Fluctuation Model}

We next present a model that relates hastatic order to the particular valence fluctuations 
in \ursp,
based on a two-channel Anderson lattice model.  The uranium ground state is a $5f^2$ Ising $\Gamma_5$ doublet\cite{Amitsuka94}, which then fluctuates to an excited $5f^3$ or $5f^1$ state via valence fluctuations.  The lowest lying excited state is most likely the $5f^3$ ($J=9/2$) state, but for simplicity here we take it to be the symmetry equivalent $5f^1$ state, and assume that fluctuations to the $5f^3$ are suppressed - in this sense, we take an infinite-U two-channel Anderson model.

As we can write the $f^2$ $\Gamma_5$ doublet, $\vert \pm  \rangle = a  \vert \pm 3 \rangle +b \vert \mp
1\rangle$ in terms of combinations of two $J = 5/2$ f-electrons in the three tetragonal orbitals, $\Gamma_7^{\pm}$ and $\Gamma_6$, 
\begin{eqnarray}\label{nk2}
\vert +\rangle&=& ({p} f\dg_{\Gamma_{7}^{-}\dw}f\dg_{\Gamma_{7}^{+}\dw }
+ {q} f\dg_{\Gamma_{6}\up}f\dg_{\Gamma_{7}^{+}\up}
+ {s} f\dg_{\Gamma_{6}\up}f\dg_{\Gamma_{7}^{-}\up}
)\vert \Omega \rangle
\cr
\vert -\rangle&=& ({p} f\dg_{\Gamma_{7}^{-}\up
}f\dg_{\Gamma_{7}^{+}\up }
+ {q} f\dg_{\Gamma_{6}\dw}f\dg_{\Gamma_{7}^{+}\dw}
+ {s}  f\dg_{\Gamma_{6}\dw}f\dg_{\Gamma_{7}^{-}\dw}
)\vert \Omega \rangle,\cr &&
\end{eqnarray}
if we assume a $5f^1$ $\Gamma_7^+$ excited state, we can now read off the valence fluctuation matrix elements directly.   Valence fluctuations occur in two orthogonal conduction electron channels, $\Gamma_7^-$ and $\Gamma_6$, and we find 
\begin{eqnarray}\label{l} H_{VF}(j) 
&=& V_6 \psic_{\Gamma_6 \pm}\dg(j) |\Gamma_7^+
\pm\rangle \langle \Gamma_5 \pm| \cr &+& V_7 \psic_{\Gamma_7 \mp}\dg(j)
|\Gamma_7^+ \mp \rangle\langle \Gamma_5 \pm| + \mathrm{H.c.}.
\end{eqnarray}
where $\pm$ denotes the ``up'' and ``down''
states of the coupled Kramers and non-Kramers doublets. The field
$\psic\dg_{\Gamma \sigma}(j) = \sum_\bk
\left[\Phi\dg_\Gamma(\bk)\right]_{\sigma \tau} c\dg_{\bk\tau}
\mathrm{e}^{-i\bk\cdot \bR_j}$ creates a conduction electron at
uranium site
$j$ with spin $\sigma$, in a Wannier orbital with symmetry $\Gamma\in
\{6,7\}$, while $V_6$ and $V_7$ are the corresponding hybridization
strengths. 
The full model is then written 
\begin{equation}\label{}
H = \sum_{ \bk \sigma }\epsilon_{\bk }c\dg_{\bk \sigma }c_{\bk \sigma}
+ \sum_{j}\left[H_{VF} (j)+ H_{a} (j) \right]
\end{equation}
while  $H_a(j) = \Delta E \sum_\pm|\Gamma_7 \pm,j\rangle \langle \Gamma_7 \pm,j| $
is the atomic Hamiltonian.

\fg{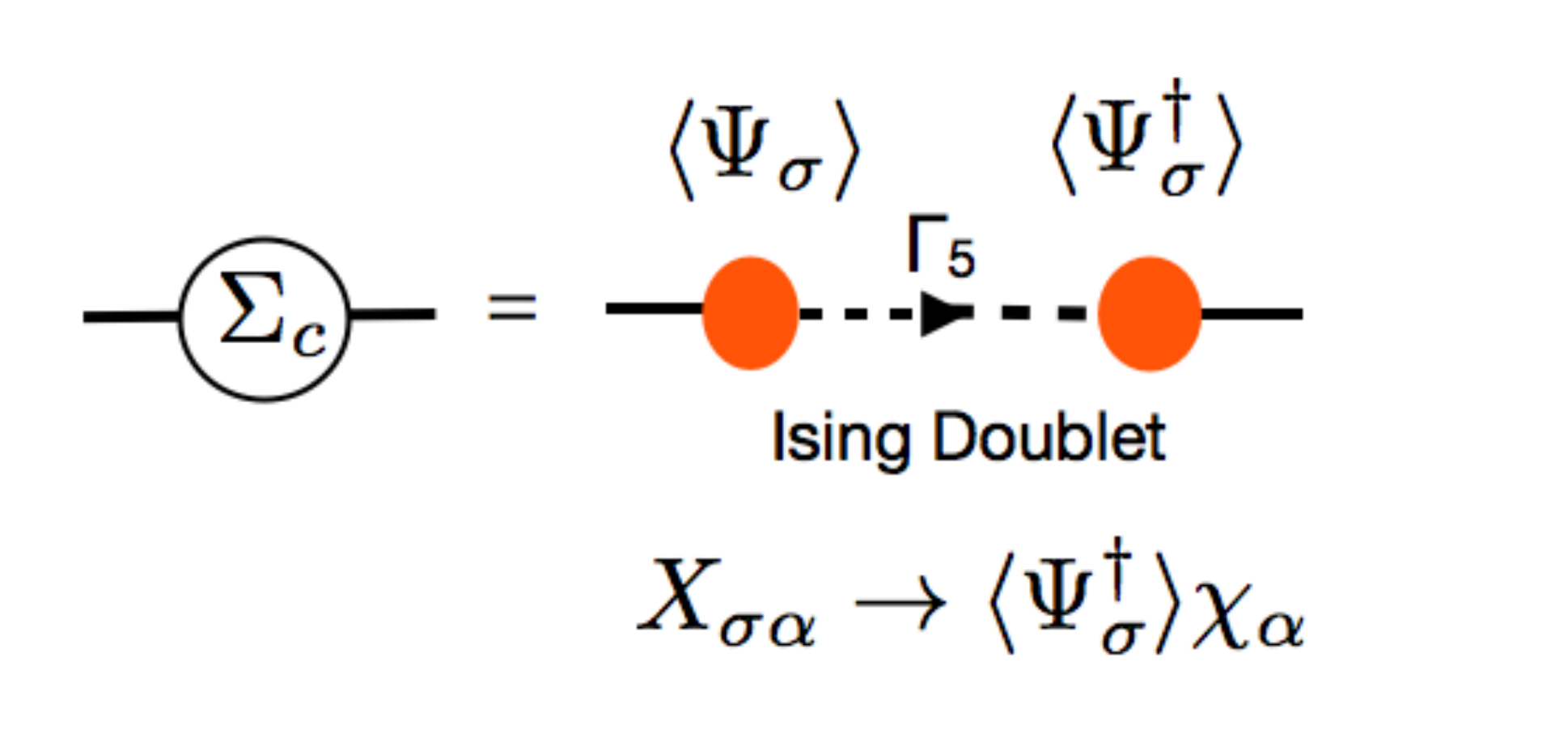}{fig4}{Showing conduction electron self-energy
$\Sigma_{c}$. 
Hybridization with spinorial order
parameter $\langle  \Psi_{\sigma }\rangle$ permits the development of
a $\Gamma_{5}$ Ising resonance inside the conduction sea, 
represented by the above Feynman diagram. 
}

To encompass the Hubbard operators in a field-theory description, we
factorize them as follows
\begin{equation}\label{}
X_{\sigma \alpha }= |\Gamma_7^+ \sigma\rangle\langle \Gamma_5 \alpha| = \hat \hast\dg_\sigma \chi _\alpha.
\end{equation}
Here $|\Gamma_5 \alpha\rangle = \chi\dg_\alpha
|\Omega\rangle$ is the non-Kramers doublet, represented by the
pseudo-fermions
$\chi \dg_{\alpha }$, while
$\hat \hast_\sigma\dg$ are {slave} bosons{\cite{Coleman83}} representing
the excited $f^{1}$ doublet
$|\Gamma_7^+ \sigma\rangle
=\hat\hast\dg_\sigma|\Omega\rangle$.
Hastatic order is realized as then condensation of this bosonic spinor
\begin{equation}\label{}
\hast_{\sigma }\dg\chi_{\alpha} \rightarrow \langle\hat  \hast_{\sigma }\rangle 
\chi_{\alpha },
\end{equation}
generating a
hybridization between the conduction electrons and
the Ising $5f^{2}$ state while also breaking double time reversal (see
Fig. \ref{fig4}).  These slave bosons play a dual role in capturing the hybridization while also acting as Schwinger bosons describing a $5f^1$ magnetic moment with a reduced amplitude, $2-n_f$.  
The tensor 
product $Q_{\alpha \beta } \equiv \Psi_{\alpha }\Psi \dg_{\beta }$
describes the development of composite order between 
the non-Kramers doublet and the spin density of 
conduction electrons. Composite order has been 
considered by several earlier authors
in the context of 
two channel Kondo lattices\cite{Cox96,CATK,Hoshino11} in which the
valence fluctuations have been integrated out.
However, by factorizing the composite order in terms 
of the spinor  $\Psi_{\alpha }$, we are able to directly understand
the development of coherent Ising quasiparticles and the broken double
time-reversal. 


There are two general aspects of this condensation that deserve
special comment. First, the two-channel Anderson impurity model is
known to possess a non-Fermi liquid ground-state 
with an entanglement entropy of $\frac{1}{2} k_{B}\ln  2$\cite{Bolech02}. 
The development of hastatic order 
in the lattice
liberates
this zero-point entropy, accounting naturally for the large entropy
of condensation.
As a slave boson, $\Psi $ carries both
the charge $e$ of the electrons and the local gauge charge $Q_{j}=
\Psi\dg _{j}\Psi_{j}+\chi \dg_{j}\chi_{j}$ of constrained valence fluctuations,
its condensation gives a mass to their relative
phase via the Higgs mechanism\cite{marston03}.  
But as a Schwinger
boson,  the condensation of $\Psi $
this process breaks the SU (2) spin symmetry. 
In this way
the hastatic boson can be regarded as a magnetic analog of the Higgs boson.

\subsection{Structure of the Paper}

To recap, here we are arguing that the observation of an anisotropic conduction fluid in \urs indicates
the coherent admixture of spin $\frac{1}{2}$ electrons with integer spin doublets, leading us to
propose that the order parameter in \urs is spinorial hybridization that breaks both single- and
double-time reversal.  In conventional heavy fermion materials, hybridization is driven by valence fluctuations
between a Kramers doublet and an excited singlet in a single channel.  The hybridization carries no quantum numbers
and develops as a crossover resulting in heavy mobile electrons.  However if the ground-state is a non-Kramers
doublet, the Kondo effect occurs via an excited state with an odd number of elections that is a Kramers doublet.
The quasiparticle hybridization then carries a global spin quantum number and has two distinct amplitudes that form
a spinor defining the hastatic order parameter
\begin{equation}
\Psi  = \left(\begin{matrix}
\psi_\up\cr
\psi_\dw\end{matrix}
\right).
\end{equation}
The onset of hybridization must break spin rotational invariance in addition
to double time-reversal invariance via a phase transition; we note that
optical, spectroscopic and tunneling probes in \urs indicate the hybridization
occurs abruptly at the hidden order transition in contrast to the crossover
behavior observed in other heavy fermion systems.

We now describe the structure of this paper. The microscopic
basis of hastatic order is presented using a two-channel
Anderson lattice model in section II, along with the mean field solution.  
In section III, we develop the Landau-Ginzburg theory of hastatic order, including the appearance of pressure induced antiferromagnetism and the nonlinear susceptibility, while in section IV, we derive and discuss a number of experimental consequences of hastatic order, showing the consistency of hastatic order with a number of experiments, including the large entropy of condensation and tetragonal symmetry breaking observed in torque magnetometry, and then making a number of key predictions, including a tiny staggered basal plane moment in the conduction electrons.  We end with discussion and future implications in section V.



%





%
%
%


\section{Landau theory: Pressure-induced antiferromagnetism}

\figwidth=\columnwidth
\fgb{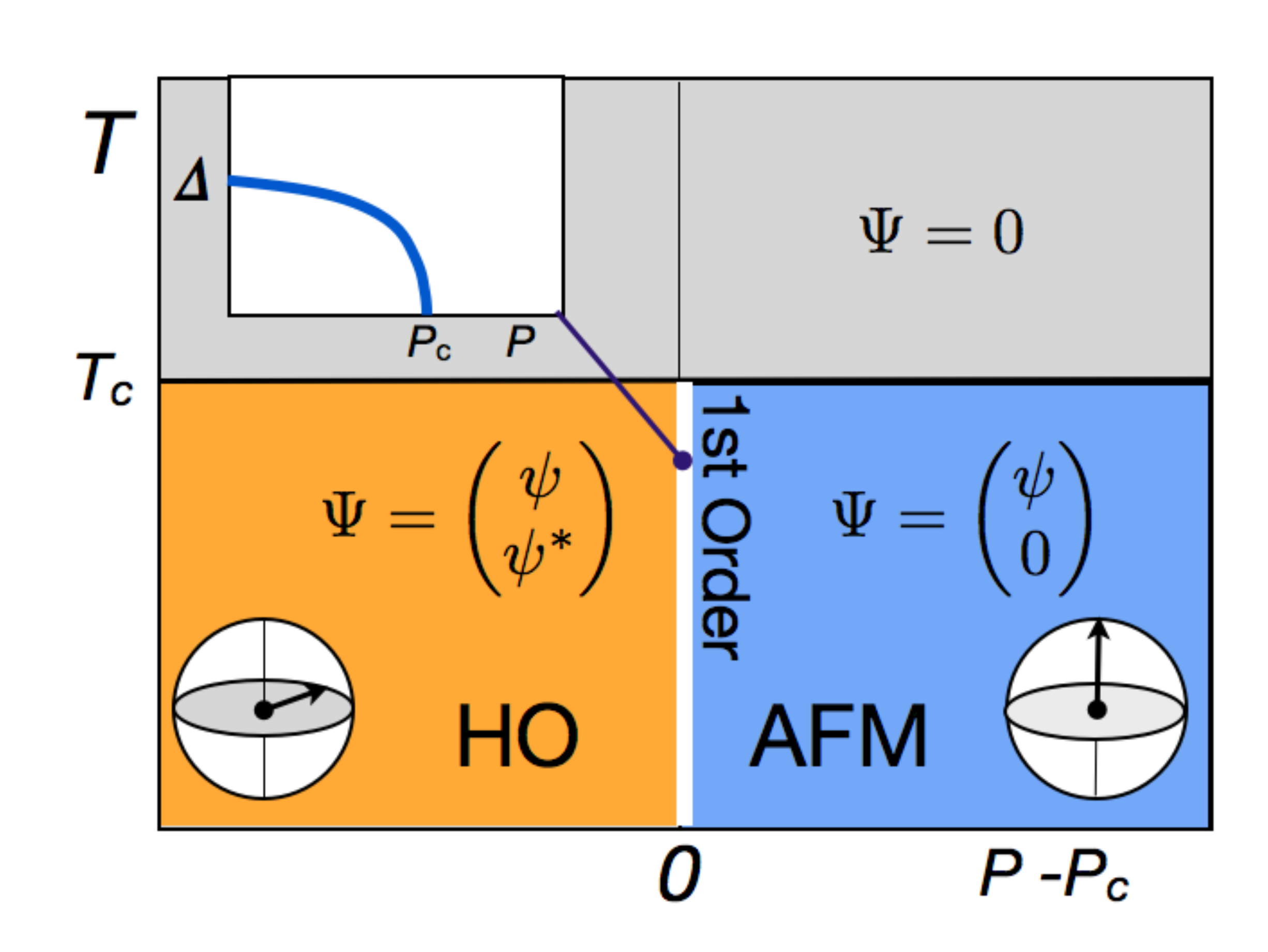}{landy}{Global phase diagram predicted by
Landau theory.} 

\subsection{Thermodynamics}\label{}

Hastatic order captures the key features of the observed pressure-induced 
first-order phase transition in \urs between the hidden order and
an Ising antiferromagnetic (AFM) phases.
The most general Landau functional for the free energy density of a
hastatic state with a spinorial order parameter $\Psi $ as a function
of pressure and temperature is 
\begin{equation}\label{}
f[\Psi] = 
\alpha (T_{c}-T)\vert \Psi \vert^{2}+ \beta \vert \Psi \vert^{4} - 
\gamma (\Psi \dg \sigma_{z}\Psi )^{2}
\end{equation}
where $\gamma = \delta(P-P_{c})$ is a pressure-tuned anisotropy term and 
\begin{equation}\label{}
\Psi  = r \left(\begin{matrix}
\cos (\theta/2 )e^{i\phi /2}\cr
\sin(\theta/2)e^{-i\phi /2}\end{matrix}
\right), 
\end{equation}
where $\theta $ is the disclination of $\Psi\dg  \vec{\sigma }\Psi $
from the c-axis, and $\vert \Psi \vert^2 = r^2$.
Experimentally the $T_{AFM}(P)$ line is almost vertical, indicating
by the Clausius-Clapeyron relation that the there will be negligible
change in entropy between the HO and the AFM states.  Indeed
these two phases share a number of key features, including
common Fermi surface pockets; this has prompted the proposal that
they are linked by ``adiabatic continuity'', associated by
a notational rotation in the space of internal parameters.
This is easily accommodated with a spinor order parameter; for
the AFM phase ($P > P_c$), there is a large staggered Ising 
f-moment with
\begin{equation}
\Psi_A  \propto \left(\begin{matrix}
1\cr
0\end{matrix}
\right), \qquad 
\Psi_B  \propto \left(\begin{matrix}
0\cr
1\end{matrix}
\right)
\end{equation}    
corresponding to time-reversed spin configurations on alternating
layers $A$ and $B$.  For the HO state ($P < P_c$), the spinor
points in the basal plane 
\begin{equation}
\Psi_A  \propto \frac{1}{\sqrt{2}}\left(\begin{matrix}
e^{-i\phi /2}\cr
e^{i\phi /2}\cr\end{matrix}
\right), \qquad 
\Psi_B  \propto \frac{1}{\sqrt{2}}\left(\begin{matrix}
-e^{-i\phi /2}\cr
e^{i\phi /2}\cr\end{matrix}
\right)
\end{equation}    
where $\Psi_B = \Theta \Psi_b$ and there is no Ising f-moment, consistent
with experiment, but Ising fluctuations do exist.
 
According to the above expression for $\Psi $, $(\Psi \dg \sigma_{z}\Psi) = r^2 \cos \theta$
so that the Landau functional can be re expressed as  
\begin{equation}\label{}
f= - \alpha  (T-T_{c})r^{2}+ (\beta  - \gamma \cos^2 \theta) r^{4}.
\end{equation}
If $P< P_{c}$, then $\gamma <0$ and the minimum of the free energy
occurs for $\theta  = \pi/2$, corresponding to the hidden order state
ordered state. By contrast, if $P>P_{c}$, then $\gamma > 0$ and the
minimum of the free energy occurs at $\theta  = \{0, \pi\}$, corresponding
to the antiferromagnet.  The ``spin flop'' in $\theta $ at $P=P_{c}$
corresponds to a first order phase transition between the hidden order
and antiferromagnet (See Fig. \ref{landy}).

\subsection{Soft modes and dynamics}\label{}\label{}

{The adiabatic continuity between the hastatic and antiferromagnetic
phases allows for a simple interpretation of the soft longitudinal
spin fluctuations that have been observed to develop in the HO state
\cite{Broholm91,Wiebe07}, and even to go soft, but not critical upon
approaching $T_{HO}$\cite{Niklowitz11}.  These longitudinal modes can
be thought of as incipient Goldstone excitations between the two
phases\cite{Haule10}.  Specifically, in the HO state, rotations of the
hastatic spinor out of the basal plane will lead to a gapped Ising
collective mode like the one observed.  Within our Landau theory, we
can study the evolution of this mode with pressure.}

In order to study the soft modes of the hastatic order, we need to generalize
the Landau theory to a time-dependent Landau Ginzburg theory for the
action, with action $S= \int L dt d^{3}x$, where the Lagrangian
\[
-L[\Psi ]  = f [\Psi] + \rho \left(|\nabla \Psi |^{2}  - c^{-2}|\dot{\Psi }|^{2}\right),
\]
and $\rho $ is the stiffness.
Expanding $\Psi$ around its equilibrium value $\Psi_{0}$, we take $\phi
=0$ for convenience and write
\begin{equation}\label{}
\Psi (x,t)=  \Psi_{0}e^{i\delta \theta (x)\sigma_{y}/2} = 
(1 + i/2\sum_{q}\delta \theta (q)e^{i \vec{ q}\cdot
\vec{x}-\omega t}\sigma_{y})\Psi_{0} .
\end{equation}
This gives rise to a change in 
$\Psi \dg \vec{\sigma}\Psi = \hat x |\Psi_{0}|^{2}+ \delta \theta (x)
\hat z |\Psi_{0}|^{2}$
corresponding to a fluctuation in the longitudinal magnetization.
This rotation in $\Psi $ does not affect the first two isotropic 
terms in $f[\Psi ]$. 
The variation in the action is then given by 
\begin{equation}\label{}
\delta S = \rho |\Psi_{0}|^{2} \sum_{q} \vert  \theta (q)\vert^{2}
\left(\vec{q}^{\ 2}- \frac{\omega^{2}}{c^{2}}+\frac{2\delta}{\rho}  (P_{c}-P)|\Psi_{0}|^{2} \right)
\end{equation}
The dispersion is therefore 
\begin{equation}\label{}
\omega^{2} = (c q)^{2}+ \Delta^{2}
\end{equation}
where 
\begin{equation}\label{}
\Delta^{2}= \frac{2 \delta (P_{c}-P)}{\rho }|\Psi_{0}|^{2},
\end{equation}
so that even though the phase transition at $P=P_{c}$ is first order, 
the gap for longitudinal spin fluctuations is
\[
\Delta \propto  |\Psi_{0} | \sqrt{P_{c}-P}.
\]
Since $dP_c/dT_c$ is finite, close to the transition, $\sqrt{P_c-P} \approx \sqrt{dP_c/dT_c(T-T_c)}$, and $\Delta \propto \sqrt{T-T_c}$.  Inelastic neutron scattering experiments can measure this gap a function of temperature at a fixed pressure where there is a finite temperature first order transition.  The iron-doped compound, URu$_{2-x}$Fe$_x$Si$_2$ can provide an attractive alternative to hydrostatic pressure, as iron doping acts as uniform chemical pressure and tunes the hidden order state into the antiferromagnet\cite{butch}.

{However, higher order terms, like $\lambda (\Psi\dg \sigma_z \Psi)^2|\Psi|^2$ are also allowed by symmetry, and will introduce a weak discontinuity of order $|\Psi_0|^4$,
\begin{equation}\label{}
\Delta^{2}= \frac{2 \delta (P_{c}-P)}{\rho }|\Psi_{0}|^{2} + \frac{2 \lambda}{\rho} |\Psi_0|^4 + O(|\Psi_0|^6).
\end{equation}
So there will likely always be a discontinuity in the longitudinal spin fluctuation mode at the first order phase transition, although it will be of a lower order ($|\Psi_0|^4$) than a generic first order transition ($|\Psi_0^2|$).}

\section{Microscopic Model: Two-Channel Anderson Lattice }

Hastatic order emerges as a spinorial hybridization between a
non-Kramers doublet ground state and a Kramers doublet excited state.
In our picture of \ursp, a lattice of $5f^2$ ($J = 4$) U$^{4+}$ ions
provide the non-Kramers doublet ($\Gamma_5$), which are then
surrounded by a sea of conduction electrons that facilitate valence
fluctuations between the $5f^2$ non-Kramers doublet and a $5f^1$ or
$5f^3$ excited Kramers doublet.  The two-channel Anderson lattice
model has three components:
\begin{equation}
H = H_c + \sum_j \left[ H_{VF}(j) + H_{a}(j) \right],
\end{equation}
a conduction electron term, $H_c$, a valence fluctuation term capturing how the conduction electrons hop on and off the U site, $H_{VF}$ and an atomic Hamiltonian capturing the different energy levels of the U ion, $H_a$.

\subsection{The Valence Fluctuation Hamiltonian}

\subsubsection{The $5f^1$ model}

The $5f^2$ $\Gamma_5$ non-Kramers doublet is given by
\begin{equation}\label{nk1}
\vert 5f^2: \Gamma_ 5 \pm  \rangle = a \vert \pm 3 \rangle + b \vert \mp 1\rangle,
\end{equation}
and all energies are measured relative
to the energy of this isolated doublet.
In principle, 
valence fluctuations may either occur to $5f^1$ ($J = 5/2$) or to $5f^3$ ($J = 9/2$), and in fact the lowest lying excited state is likely to be
the $5f^3$ state.  However, 
for technical simplicity, we 
take the lowest lying valence fluctuation excitation to be the $5f^{1}$ $\Gamma_{7}^{+}$ excited state, 
\begin{equation}\label{}
\vert 5f^{2}: \Gamma_5 \pm \rangle  \rightleftharpoons\vert
5f^{1}:\Gamma_{7}^{+} \pm \rangle  + e^{-}
\end{equation}
where 
\begin{equation}\label{}
\vert 5f^{1}:\Gamma_{7}^{+}\pm \rangle  = \eta  \vert \pm 5/2\rangle
+ \delta \vert  \mp 3/2\rangle 
\end{equation}
is the excited Kramers  doublet. The following section will show
how particle-hole
symmetry can be used to formulate the equivalent model with
fluctuations into a $5f^{3}$ Kramers doublet. 

To evaluate the matrix elements for valence fluctuations
we need to express the $5f^{2}$ state in
terms of one-particle states. 
The $\Gamma_5$ state can be rewritten as a product of the one particle
$J=5/2$ f-orbitals, $\vert 5/2,m\rangle \equiv
f\dg_{m}\vert \Omega \rangle $, using the Clebsch Gordan decomposition 
\begin{eqnarray}\label{l}
\vert \mp 1\rangle  &=&  \left(
\sqrt{\frac{5}{7}}f\dg_{\pm 1/2}f\dg_{\mp
3/2}+\sqrt{\frac{2}{7}}f\dg_{\pm 3/2}f\dg_{\mp 5/2}
 \right)\vert  \Omega \rangle,\cr
\vert \pm 3\rangle  &=& f\dg_{\pm 5/2}f\dg_{\mp 1/2}
\vert  \Omega \rangle.
\end{eqnarray}
Next we decompose the one-particle f-states in terms of the 
one-particle crystal field eigenstates,  $\vert \Gamma \pm \rangle  \equiv f\dg_{\Gamma, \pm
}\vert \Omega \rangle $; writing $f\dg_{m} = f\dg_{\Gamma \beta }\langle
\Gamma \beta \vert m\rangle $, or more explicitly, 
\begin{eqnarray}\label{l}
f\dg _{\pm 1/2} &=& f\dg_{\Gamma_{6}\pm }\cr
f\dg _{\mp 3/2} &=& 
\delta  f\dg_{\Gamma_{7}^{+}\pm }- \eta f\dg_{\Gamma_{7}^{-}\pm }\cr
f\dg _{\pm 5/2} &=& 
\eta  f\dg_{\Gamma_{7}^{+}\pm }+ \delta f\dg_{\Gamma_{7}^{-}\pm }
\end{eqnarray}
the non-Kramer's doublet can be written in the form 
\begin{eqnarray}\label{nk2}
\vert \Gamma_5 +\rangle&=& ({p} f\dg_{\Gamma_{7}^{-}\dw}f\dg_{\Gamma_{7}^{+}\dw }
+ {q} f\dg_{\Gamma_{6}\up}f\dg_{\Gamma_{7}^{+}\up}
+ {s} f\dg_{\Gamma_{6}\up}f\dg_{\Gamma_{7}^{-}\up}
)\vert \Omega \rangle
\cr
\vert \Gamma_5 -\rangle&=& ({p} f\dg_{\Gamma_{7}^{-}\up
}f\dg_{\Gamma_{7}^{+}\up }
+ {q} f\dg_{\Gamma_{6}\dw}f\dg_{\Gamma_{7}^{+}\dw}
+ {s}  f\dg_{\Gamma_{6}\dw}f\dg_{\Gamma_{7}^{-}\dw}
)\vert \Omega \rangle,\cr &&
\end{eqnarray}
where  ${p} = b \sqrt{\frac{2}{7}}
$, ${q}  = b\delta  \sqrt{\frac{5}{7}}- a \eta  $,
${s}  = -b\delta  \sqrt{\frac{5}{7}}- a \eta  $. 
Valence fluctuations from the ground state
({5f$^2$} $\Gamma_5$) to the excited state ({5f$^1$} $\Gamma_7^+$) are
described by a 
a one-particle Anderson model with an on-site hybridization term 
\begin{equation}\label{hybterm}
H_{VF}(j) = \sum_{\Gamma = \Gamma_{6}, \Gamma_{7}^{\pm }; \sigma }
\left[ 
v_{\Gamma}{c}
\dg_{\Gamma \sigma } (j) f_{\Gamma \sigma } (j)+ {\rm H. c}\right],
\end{equation}
where the $v_{\Gamma}$ are the hybridization matrix elements in the
three orthogonal crystal field channels and ${c}\dg_{\Gamma \sigma } (j)$ creates a conduction electron
in a Wannier state with symmetry $\Gamma \sigma$ on site $j$.

Now we need to project this Hamiltonian down into the reduced subspace of
the ground-state 
$\vert 5f^{2}:\Gamma_{5}\alpha \rangle $ 
and 
$\vert 5f^{1}:\Gamma_{7}^{-}\sigma \rangle $ 
excited state doublets, 
replacing 
\begin{eqnarray}\label{l}
f_{\Gamma\sigma } (j)\rightarrow 
\sum_{\sigma ',\alpha = \pm }
\vert \Gamma_{7}^{+}\sigma '
\rangle 
\Biggl(
\langle \Gamma_{7}^{+}\sigma' \vert  
f_{\Gamma \sigma } (j) 
\vert \Gamma_{5}\alpha \rangle 
\Biggr)\langle \Gamma_{5}\alpha \vert 
\end{eqnarray}
in (\ref{hybterm}).
Using (\ref{nk2}), the only non-vanishing matrix elements between the
$5f^{2}$ and $5f^{1}$ states are 
\begin{eqnarray}\label{l}
\langle \Gamma_{7}^{+}\mp
\vert 
f_{\Gamma_{7}^{-}\mp}\vert  \Gamma_{5}\pm \rangle &=& p,
\cr
\langle \Gamma_{7}^{+}\pm
\vert 
f_{\Gamma_{6}\pm}\vert  \pm  \rangle &=& q. 
\end{eqnarray}
Matrix elements for the $\Gamma_{7}^{+}$ channel identically vanish
$\langle \Gamma_{7}^{+}\sigma
\vert 
f_{\Gamma_{7}^{+}\pm }\vert  \sigma' \rangle =0$,
so 
the third term in (\ref{nk2}) does not contribute to the projected
Hamiltonian. 
The final projected model is then written 
\begin{equation}\label{}
H = \sum_{ \bk \sigma }\epsilon_{\bk }c\dg_{\bk \sigma }c_{\bk \sigma}
+ \sum_{j}\left[H_{VF} (j)+ H_{a} (j) \right]
\end{equation}
where $c\dg_{\bk \sigma }$ creates a conduction electron of 
momentum
$\bk $ spin $\sigma $, with energy $\epsilon_{\bk }$, and 
\begin{eqnarray}\label{l}
\label{VF} H_{VF}(j) &=& V_6 {c}_{\Gamma_6 \pm}\dg(j) |\Gamma_7^{+}
\pm\rangle \langle \Gamma_5 \pm|\cr
&+& V_7 {c}_{\Gamma^{-}_7 \mp}\dg(j)
|\Gamma_7^+ \mp \rangle\langle \Gamma_5 \pm| + \mathrm{H.c.}
\end{eqnarray}
describes the valence fluctuations between the $\Gamma_{5}$ doublet
and the excited $\Gamma_{7}^{+}$ Kramers doublet. Here $V_{6
}=v_{\Gamma_{6}}q$ and $V_{7}= v_{\Gamma_{7}^{-}}p $ while
\begin{equation}\label{}
H_a(j) = \Delta E \sum_\pm|\Gamma_7 \pm\rangle \langle \Gamma_7 \pm| 
\end{equation}
is the atomic Hamiltonian for the excited $5f^{1}:\Gamma_{7}^{+}$
Kramers doublet.  Notice that, as we have projected out most of the possible U states, we are working in terms of the Hubbard operators, $|\Gamma_5 \pm\rangle$ and $|\Gamma_7 \pm\rangle $ to describe the allowed U states.

To further develop the valence fluctuation term, we need to determine the form factors relating the conduction electron
Wannier states in terms of Bloch waves,
\begin{equation}
{c}\dg_{\Gamma \alpha} = \sum_\bk c\dg_{\bk\beta} \left[\Phi_{\Gamma
\bk}\right]_{\beta \alpha } \mathrm{e}^{-i \bk\cdot \bR_j}.
\end{equation}
For a single site interacting with a plane wave, these are given by 
$\left[\Phi_{\Gamma \bk}\right]_{\alpha \beta} = y^{\Gamma}_{\alpha
\beta} 
(\bk )$, where
\begin{eqnarray}\label{l}
y^{\Gamma}_{\alpha \beta}(\bk )
&=& \sum_{m=-5/2}^{5/2}
Y_{3m-\frac{\alpha }{2}} (\hat
\bk ) 
\langle 3m-\frac{\alpha }{2}, \frac{1}{2} \frac{\alpha }{2}\vert 5/2 m \rangle 
{\langle m\vert \Gamma,\ \beta\rangle }\cr
&=& \alpha \sum_{m=-5/2}^{5/2}
\sqrt{\frac{1}{2}- \frac{m\alpha }{7}}
\ Y_{3m-\frac{\alpha }{2}} (\hat\bk ) 
{\langle m\vert \Gamma\beta\rangle }.\cr&&
\end{eqnarray}
where $\alpha ,\beta  \in \pm $.
However, in \ursp,  the uranium 
atoms are located on a body centered tetragonal(bct)  lattice 
at
relative locations, $\bR_{NN} = ( \pm a/2, \pm a/2, \pm
c/2)$, and the correct form factor must respective the lattice
symmetries. Our model treats  the conduction band as
a single band of s-electrons located at the U sites. 
Moreover, 
we assume that the f-electrons 
hybridize via electron states
at the nearby 
silicon atoms located at  
${\bf{a}}_{NN} = ( \pm a/2, \pm a/2, \pm z)$ where $z =
.371 c$ is the height of the silicon atom above the U
atom\cite{Oppeneer10}. 
\begin{widetext}
The form-factor is then
\begin{eqnarray}\label{l}
\left[\Phi_{\Gamma \bk}\right]_{\alpha \beta} 
= 
\sum_{\{{\bf{R}}_{NN}, {\bf{a}}_{NN}\}
} 
e^{-i \bk
\cdot ({\bf R}_{NN}-{\bf a}_{NN})}
\times 
e^{-i \bk
\cdot{\bf a}_{NN}}
 y^{\Gamma}_{\alpha \beta} 
({\bf{a}}_{NN})
= 
\sum_{\{{\bf{R}}_{NN}, {\bf{a}}_{NN}\}
} e^{-i \bk
\cdot{\bf R}_{NN}} y^{\Gamma}_{\alpha \beta} 
({\bf{a}}_{NN}).
\end{eqnarray}
\end{widetext}
Here,  the term $e^{-i \bk
\cdot{\bf a}_{NN}}
 y^{\Gamma}_{\alpha \beta} 
({\bf{a}}_{NN})
$ is the amplitude to hybridize with the silicon site at site 
${\bf a}_{NN}$, while
the prefactor $e^{-i \bk
\cdot ({\bf R}_{NN}-{\bf a}_{NN})}$ is the additional 
phase factor for hopping from the silicon site 
${\bf a}_{NN}$
to the U atom, ${\bf R}_{NN}$ directly above it. 
This form of the hybridization can also be derived using
Slater-Koster\cite{slater} methods, under the assumption that the important part of the
hybridization potential is symmetric about the axis between the U and
Si atom. 
Notice that this function has the following properties:
$\left[\Phi_{\Gamma \bk+{\bf G}}\right]_{\alpha \beta} 
=\left[\Phi_{\Gamma \bk}\right]_{\alpha \beta} $ and 
$\left[\Phi_{\Gamma \bk+\bQ}\right]_{\alpha \beta} 
=-\left[\Phi_{\Gamma \bk}\right]_{\alpha \beta}$. 
We should note that this model of the hybridization is overly simplified, in that the U most likely hybridizes with the d-electrons sitting on the Ru site.  Such a d-f hybridization can be treated in a similar fashion, and is the subject of future work.

\subsubsection{The $5f^3$ case}

For simplicity we have discussed the two channel Anderson model involving fluctuations from a $5f^2$ $\Gamma_5$ ground state to $5f^1$ ($J=5/2$).
However, the more realistic case involves fluctuations to $5f^3$, whose low energy states have $J=9/2$, and are split into five Kramers doublets by the tetragonal crystal field,
\begin{eqnarray}\label{l}
|\Gamma_6^{\lambda} \pm \rangle & = & c^{\lambda}|\pm 9/2\rangle +
d^{\lambda}|\pm 1/2\rangle+ e^{\lambda}|\mp 7/2\rangle, 
\cr
|\Gamma_7^1 \pm\rangle & = & a |\pm 5/2\rangle + b|\mp 3/2\rangle \cr
|\Gamma_7^2 \pm\rangle & = & -b|\pm 5/2\rangle +a|\mp 3/2\rangle ,
\end{eqnarray}
where $\lambda\in ( 1,2,3)$ labels the three $\Gamma_{6}$ Kramer's
doublets. 
There are two generic situations: either a $\Gamma_7$ doublet is 
lowest in energy, and the valence fluctuations are then determined by the overlap,
\begin{equation}
|\Gamma_7 \pm\rangle = \alpha\psi\dg_{6 \mp}|\Gamma_5 \pm\rangle + \beta\psi\dg_{7 \mp}|\Gamma_5 \mp \rangle,
\end{equation}
or alternatively, a  $\Gamma_6$ doublet is 
lowest in energy, with the relevant overlap,
\begin{equation}
|\Gamma_6 \pm\rangle = \alpha \psi\dg_{7 \mp}|\Gamma_5 \pm\rangle + \beta\psi\dg_{6 \mp}|\Gamma_5 \mp \rangle,
\end{equation}
where the form-factors are as above.  In both cases fluctuations will involve conduction electrons in both $\Gamma_6$ and $\Gamma_7$ symmetries.  When the lowest excited state is a $\Gamma_7$, the valence fluctuation Hamiltonian is given by,
\begin{eqnarray}\label{l}
H_{VF3}(j) &=&  V_6 (\psi_{j\Gamma_6 \mp}\dg |\Gamma_5 \pm\rangle
\langle \Gamma_7 \pm| +{\rm H.c})\cr
&+&
 V_7 (\psi_{j\Gamma_7 \mp}\dg |\Gamma_5 \mp \rangle\langle \Gamma_7 \pm| + \mathrm{H.c.}).
\end{eqnarray}
The only difference between the $5f^3$ and the $5f^{1}$ model
is that the excited state requires adding a particle, so 
the valence fluctuation term here 
is the particle-hole conjugate of the $5f^1$ case.

\subsection{Slave particle treatment}

Hubbard operators 
cannot be directly treated with quantum field theory techniques since 
they do not satisfy Wick's theorem. We follow the standard approach 
and introduce a slave particle factorization of the Hubbard operators 
that permits a field theoretic treatment,
\begin{equation}\label{}
|\Gamma_7^+ \sigma\rangle\langle \Gamma_5 \alpha| = \hat \hast\dg_\sigma \chi _\alpha.
\end{equation}
where 
\begin{equation}\label{}
|\Gamma_5 \alpha\rangle = \chi\dg_\alpha
|\Omega\rangle
\end{equation}
is the non-Kramers doublet, represented by the
pseudo-fermion $\chi \dg_{\alpha }$, while
$\hat \hast_\sigma\dg$ is a {slave} boson{\cite{Coleman83}} representing
the excited $f^{1}$ doublet
\begin{equation}\label{}
|\Gamma_7^+ \sigma\rangle
=\hat\hast\dg_\sigma
|\Omega\rangle
\end{equation}
that carries a positive charge and a spin quantum
number.
Condensation of the spin-$\frac{1}{2}$ boson  then gives rise to the
hastatic order parameter
\begin{equation}\label{}
{\bf \hast} =
\pmat{\langle \hat \hast_{\up
}\rangle \cr
\langle \hat \hast _{\dw}\rangle}.
\end{equation}
This condensation process that we may now replace the 
Hubbard operator 
$\hat X_{\sigma \alpha }$ by a single 
bound-state fermion 
\begin{equation}\label{}
\hast_{\sigma }\dg\chi_{\alpha} \rightarrow \langle\hat  \hast_{\sigma }\rangle 
\chi_{\alpha }.
\end{equation}
We may interpret this replacement as a kind of multi-particle
contraction of the many body Hubbard operator into a single fermionic
bound-state.  Once this bound-state forms, a symmetry-breaking
hybridization develops between the conduction electrons and
the Ising $5f^{2}$ state. 

The dual Schwinger/slave character of   
the boson $\hat\hast\dg_\sigma$ representing the occupation of
$5f^1$ means when this field condenses,  it not only 
breaks the local $U (1)$ gauge symmetry, but also the global $SU(2)$ spin symmetry.
However, it breaks the $U(1)$ gauge symmetry as a slave boson, which 
have been well studied in the context of heavy fermions. The $U(1)$ phase of the local gauge symmetry is subject to the Anderson-Higgs mechanism, in which the difference
between the electromagnetic gauge field and the internal $U(1)$ gauge
field acquires a mass\cite{marston03}. It is this process that gives the $\chi $
fermions a physical charge. The combination of the global and local
symmetry breaking processes means that the hastatic order parameter 
can be thought of as a magnetic Higgs boson. 

With this slave particle factorization, we can reformulate 
\begin{equation}\label{}
\sum_{j}H_{a} (j) \rightarrow 
 \Delta E \sum_{j,\sigma = \pm }\Psi \dg_{\sigma } (j)\Psi_{\sigma } (j).
\end{equation}
The valence fluctuation term at each site takes the form 
\begin{eqnarray}\label{l}
\label{VF} H_{VF}(j) &=& 
V_6 {c}_{\Gamma_6 \pm}\dg(j) \Psi_\pm\dg(j) \chi_\pm (j) + V_7 {c}_{\Gamma^{-}_7 \mp}\dg(j)
\Psi\dg_\mp (j)\chi_\pm (j) \cr &+& \mathrm{H.c.}.
\end{eqnarray}
We now rewrite this expression 
in terms of Bloch waves by absorbing the momentum
dependent Wannier form-factors into the spin-dependent hybridization
matrix, introducing the operator
\bea\label{calV}
\hat{\mathcal{V}}(\bk,j)  = V_6 \Phi_{\Gamma_6} B\dg_j +   V_7
\Phi_{\Gamma_7^-} \hat B\dg_j\sigma_1.
\eea
where the matrix
\begin{equation}\label{thebigB}
\hat B\dg_{j} = 
\zmatrix{\hat\Psi\dg_{j\up}}{0}{0}{\hat\Psi\dg_{j\dw }}
\end{equation}
contains the hastatic boson. 
The valence fluctuation term then becomes
 \bea \label{VF3} H_{VF}(j)  =  \sum_\bk
c\dg_{\bk\sigma} \hat{\mathcal{V}}_{\sigma \alpha}(\bk,j) \chi
_\alpha(j) \mathrm{e}^{-i\bk\cdot \bR_j} + \mathrm{H.c.}
\eea

\begin{widetext}
Putting these results all together, our model for \urs is given by
\begin{eqnarray}\label{thefullhammy}
H &=&  \sum_{ \bk \sigma }\epsilon_{\bk }c\dg_{\bk \sigma }c_{\bk \sigma}
+ \sum_{\bk } 
\left[
c\dg_{\bk \sigma  }\left( V_6 \Phi_{\Gamma_6} B\dg_j
 +   V_7
\Phi_{\Gamma_7^-} \hat B\dg_j\sigma_1
 \right)
\chi_{\alpha }(j)
e^{-i \bk \cdot \bR_{j}} +{\rm H.c.}\right]
+ 
 \Delta E \sum_{j,\sigma = \pm }\Psi \dg_{\sigma} (j)\Psi _{\sigma} (j)
\cr
&+& 
\sum_{j}\lambda_{j}\left(
\sum_{\sigma }\Psi \dg_{\sigma} (j)\Psi _{\sigma}(j) +
\sum_{\alpha }\chi \dg_{\alpha} (j)\chi _{\alpha}(j) -1\right).
\end{eqnarray}
\\

\end{widetext}
The respective terms in this Hamiltonian describe the conduction
electrons, the hybridization between the excited Kramers and
ground-state non-Kramers doublets, the energy $\Delta E$ of the
excited Kramers doublets. The 
second line of the Hamiltonian describes the constraint $n_{\Psi }
(j)+n_{\chi } (j) =1 $ associated with the slave boson
representation. Finally notice that if we compare
(\ref{thefullhammy}) with the parent Anderson model (\ref{hybterm}),
the original f-annihilation operators in the $\Gamma_{6}$ and
$\Gamma_{7^{-}}$
channel have been replaced as follows:
\begin{equation}\label{opreplace}
f_{\Gamma_{6}\alpha } (j) \rightarrow 
 ( B\dg_{j}
\chi_{j})_{\alpha } , \qquad
f_{\Gamma_{7}\alpha } (j) \rightarrow (B\dg_{j} \sigma_{1}\chi_{j})_{\alpha }.
\end{equation}
So while $\chi_\pm$ is a slave particle representing the non-Kramers doublet, these operators represent \emph{composite fermions} in the two hybridization channels, with all the quantum numbers of an electron. 

\subsection{Mean Field Theory for Hastatic Order}\label{}

\subsubsection{Mean Field Hamiltonian: a spinorial  order parameter}\label{}
The central element of the mean field theory is the hastatic 
order
parameter, described by a two-component spinor.  
We consider the following configurations
\begin{equation}\label{}
{\bf \Psi } 
\equiv  \langle \hat \Psi_{\sigma}\rangle  = \pmat{e^{-i
(\bQ \cdot \bR_{j}+\phi) /2}\cr
e^{i (\bQ \cdot \bR_{j}+\phi) /2}
}.
\end{equation}
where $\bQ= (0,0,\frac{2\pi}{c})$,
corresponding to a hybridization  that is staggered between planes,
with a spinorial order parameter that points in the Basal
plane, rotated
through an angle $\phi $ around the c-axis.
Eventually, we shall choose $\phi = \pi/
4$ to provide a 45$^{0}$ rotation of the scattering t-matrix relative
to the x-axis, an orientation that provides consistency with the
measured $\chi_{xy}$ anomaly in the bulk
susceptibility\cite{Okazaki11}.

Next, in (\ref{calV}) we make the substitution
\begin{equation}\label{}
\langle \hat {\cal V} (\bk,j)\rangle  
= V_6 \Phi_{\Gamma_6} \langle \hat B\dg_j \rangle +   V_7
\Phi_{\Gamma_7^-} \langle 
\hat B\dg_j\rangle \sigma_1. 
\end{equation}
It is convenient to write $\langle B\dg_{j}\rangle $
in the form 
\begin{eqnarray}\label{sconfig1}
\langle \hat B\dg_{j}\rangle  = |\Psi| U_{j},
\end{eqnarray}
where 
\begin{equation}\label{sconfig2}
U_{j}= \left(\begin{matrix}
e^{i (\bQ \cdot \bR_{j}+\phi) /2}
& \cr
&e^{-i (\bQ \cdot \bR_{j}+ \phi) /2}
\end{matrix} \right) 
\end{equation}
is a diagonal unitary matrix. 

The gauge symmetry of the slave particle representation allows us to
redefine the f-electrons, $\tilde{\chi}_{j}= U_{j} \chi_{j}$ to absorb
the spatial dependence of $\langle \hat B_{j}\dg\rangle$ into the
redefined f-electrons(\ref{opreplace}), so that 
\begin{equation}\label{f1}
\hat B\dg_{j}\chi_{j} = |\Psi|\tilde{\chi}_{j}
\end{equation}
and 
\begin{eqnarray}\label{f2}
\hat B\dg_{j}\sigma_1 \chi_{j} = |\Psi| (U_j\sigma_{1}U\dg_{j})\tilde{\chi}_{j} = |\Psi| (\hat {\bf{n}}\cdot \vec{\sigma})
e^{i \bQ \cdot \bR_{j}}\tilde{\chi}_j, 
\end{eqnarray}
where $\hat {\bf{n}} = \frac{1}{\sqrt{2}}(\hat {\bf{x}}+
\hat {\bf{y}})$ is the unit-vector representing the orientation of the hastatic spinor.  The commensurate nature of the wavevector is important, as
here we have used the fact that $e^{i\bQ \cdot \bR_{j}}\equiv
(-1)^{z_{j}/c}$ is real.
In this gauge, the $\Gamma_6$ hybridization is uniform while the
$\Gamma_{7}^{-}$ hybridization is staggered. 
In preparation for our transition to momentum space, we write 
\begin{equation}\label{v6def}
{\cal V}_{6} (\bk ) = |\Psi|V_{6} \Phi^{6} (\bk ),
\end{equation}
and 
\begin{equation}\label{v7def}
{\cal V}_{7} (\bk ) 
=|\Psi| V_{7}\Phi^{7} (\bk ) (\hat {\bf{n}}\cdot
\vec{\sigma}),
\end{equation}
where we introduce the short-hand 
$
\Phi^{6} (\bk )\equiv \Phi_{\Gamma_{6}} (\bk )$
and $\Phi^{7} (\bk )\equiv \Phi_{\Gamma_{7}^{-}} (\bk )$. Notice that
${\cal V}_{6} (\bk +\bQ )= - {\cal V}_{6} (\bk ) $ 
and 
${\cal V}_{7} (\bk +\bQ )= - {\cal V}_{7} (\bk ) $  both change sign
when shifted by $\bQ $. 
In the slave formulation, the atomic Hamiltonian is $H_a(j) = \Delta E
\sum_\sigma \Psi\dg_{j\sigma} \Psi_{j\sigma} = \Delta E |\Psi|^2$.  

While we treat the hybridization between the conduction electrons and the f-moments very carefully,
we take a simple model of the
conduction electron hopping, treating them as s-wave electrons located
at the U sites, hopping on a bct lattice with dispersion
\begin{equation}
\epsilon_\bk = -8 t \cos \frac{k_x a}{2}\cos \frac{k_y a}{2}\cos \frac{k_z c}{2} - \mu.
\end{equation} 
We do, however, want to capture the essential characteristics of the
\urs bandstructure - namely nesting between an electron Fermi surface
about the zone center and a hole Fermi surface at $\bQ$\cite{Oppeneer10}.  In order to
favor a staggered hybridization, and to match up with ARPES
experiments suggesting a heavy f-band\cite{Santander09}, we take the hole Fermi surface
to be generated from a weakly dispersing $\chi$ band.  This f-electron
hopping will be naturally generated by hybridization fluctuations
above $T_{HO}$, effectively where $\langle \hat B\dg \hat B\rangle \neq 0$ while
$\langle \hat B \rangle = 0$.  A large $N$ expansion of this problem would capture these
fluctuation effects, but is overly complicated for this problem so we
put this dispersion in by hand, $\epsilon_{f\bk} = -8 t_f \cos \frac{k_x
a}{2}\cos \frac{k_y a}{2}\cos \frac{k_z c}{2}$.  
\begin{widetext}
So to summarize, our mean-field Hamiltonian is,
\begin{eqnarray}\label{l}
H &=& \sum_\bk \epsilon_\bk c\dg_{\bk \sigma} c_{\bk\sigma  } + \sum_\bk
(\epsilon_{f \bk}+\lambda) \chi\dg_{\bk \eta} \chi_{\bk \eta} + 
{\cal N}_{s}
\left[(\Delta E
+\lambda)|\Psi|^{2}-1
 \right]
\cr &+& \sum_{\bk } \left[c\dg_{\bk\sigma   }
\left(
{\cal V}_{6\sigma \alpha } (\bk )
\chi_{\bk\alpha }+ {\cal V}_{7\sigma \alpha } (\bk )
\chi_{\bk+\bQ\alpha  }
 \right) +{\rm H.c} \right].
\end{eqnarray}
where we have dropped the tilde's on the $\chi_{\bk }$ and ${\cal
N}_{s}$ is the number of sites in the lattice. 
We rewrite this Hamiltonian  in matrix form
\begin{eqnarray}\label{thematrix}
H  =  \sum_\bk \left( c\dg_\bk, c\dg_{\bk+\bQ}, \chi\dg_\bk,
\chi\dg_{\bk+\bQ} \right) 
\overbrace {\left(\begin{matrix}\epsilon_\bk & 0 & \mathcal{V}_6(\bk) & \mathcal{V}_7(\bk)\cr 0 & \epsilon_{\bk+\bQ} & -\mathcal{V}_7(\bk) & -\mathcal{V}_6(\bk)\cr
\mathcal{V}_6\dg(\bk) & -\mathcal{V}_7\dg(\bk) & \lambda_\bk & 0\cr
\mathcal{V}_7\dg(\bk) & -\mathcal{V}_6\dg(\bk) & 0 &
\lambda_{\bk+\bQ}\end{matrix}\right)
}^{{\cal H}_{\alpha \beta } (\bk )}
\left(\begin{array}{c}
c_\bk \\ c_{\bk+\bQ} \\ \chi_{\bk} \\ \chi_{\bk+\bQ}
\end{array} \right)
 + {\cal N}_{s} \left[(\Delta E + \lambda) |\Psi|^{2} - \lambda\right].
\end{eqnarray}
where we have suppressed spin indices, made the assumption that the Lagrange multiplier
$\lambda_j = \lambda$ is uniform, equivalent to enforcing the
constraint on average, introduced $\lambda_\bk = \lambda +\epsilon_{f\bk}$, and
used the simplification that $\bQ$ is half a reciprocal lattice
vector, making $\mathcal{V}(\bk+\bQ) = -\mathcal{V}(\bk)$, as shown
above. When we diagonalize this Hamiltonian, we obtain a
a set
of four doubly degenerate bands, $E_{\bk \eta}$.  
These eigenvalues can be obtained analytically in the special case 
where the band-structure has particle-hole symmetry, but more
generally they must be obtained numerically. 
\\

\end{widetext}

The corresponding mean field free
energy is then
\bea
F[b,\lambda] & = & -\frac{\beta^{-1}}{2}\sum_{\bk, \eta}\log\left[1 +\mathrm{e}^{-\beta E_{\bk\eta}}\right]\cr 
& & + \mathcal{N}_s \left[2(\Delta E + \lambda) |\Psi|^2 - \lambda\right],
\eea
where $\beta = (k_B T)^{-1}$.   In the work presented here, 
the mean field parameters, $|\Psi|$ and
$\lambda$ are obtained by numerically finding a stationary point that minimizes $F$ with respect to $|\Psi|$ and maximizes
it with respect to $\lambda$.  
\fight=0.9\columnwidth
\fg{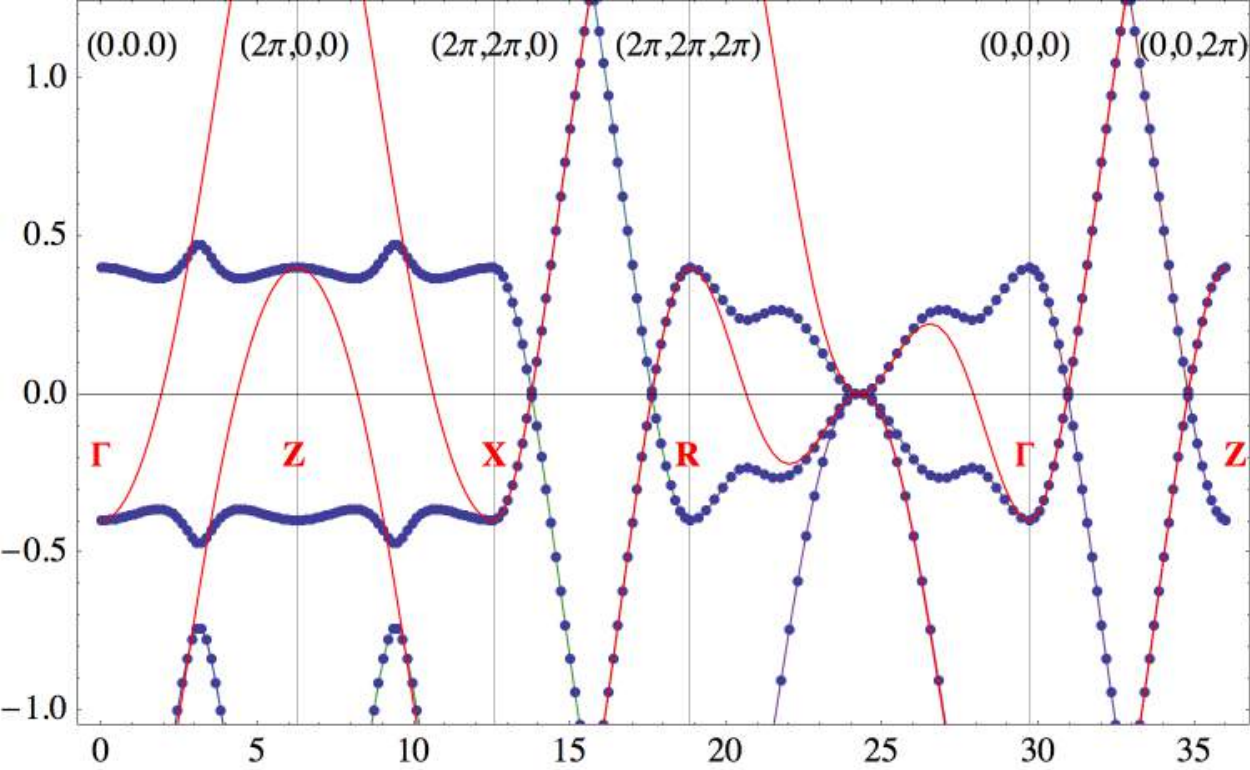}{fig5Anew}{Band structure of the
hastatic order is shown in solid blue, while the bare conduction (red)
and f (green) bands are dashed. 
The parameters used for this
calculation are given in section \ref{parameters}.
 }
\subsubsection{Mean field parameters}
\label{parameters}

In order to plot the band structure and calculate semi-realistic
experimental quantities, we have chosen the following parameters: the
internal hastatic angle, $\phi = \pi/4$ is chosen to reproduce
$\chi_{xy} \neq 0$ type tetragonal symmetry breaking; t = 12.5 meV is
taken to match the magnitude of $\chi_{xy}$ from the torque
magnetometry data\cite{Okazaki11}; $\mu/t = -.075$ gives the slight
particle-hole asymmetry essential to reproduce the flattening of
$\chi_{xy}$ at low temperatures, and has also been adjusted so that
$\mu+\lambda = 0$ at $T=0$ for consistency with the dI/dV calculations
(see later section); $t_f/t = -.025$ gives a weak f-electron
dispersion; the crystal field angle $\xi = .05$ is taken to be small,
as it is in CeRu$_2$Si$_2$\cite{Haen92} and NdRu$_2$Si$_2$; $V_6/V_7 =
1$ is arbitrary; and finally $V^2/\Delta E = 2 t$ is chosen to give
$2|\Psi|^2 \approx 15\%$ mixed valency.  {The actual degree of mixed valency in \urs is unknown, with 15\% being an upper estimate.}  The band-structure
corresponding to these parameters is shown in Fig. \ref{fig5Anew}.

\fg{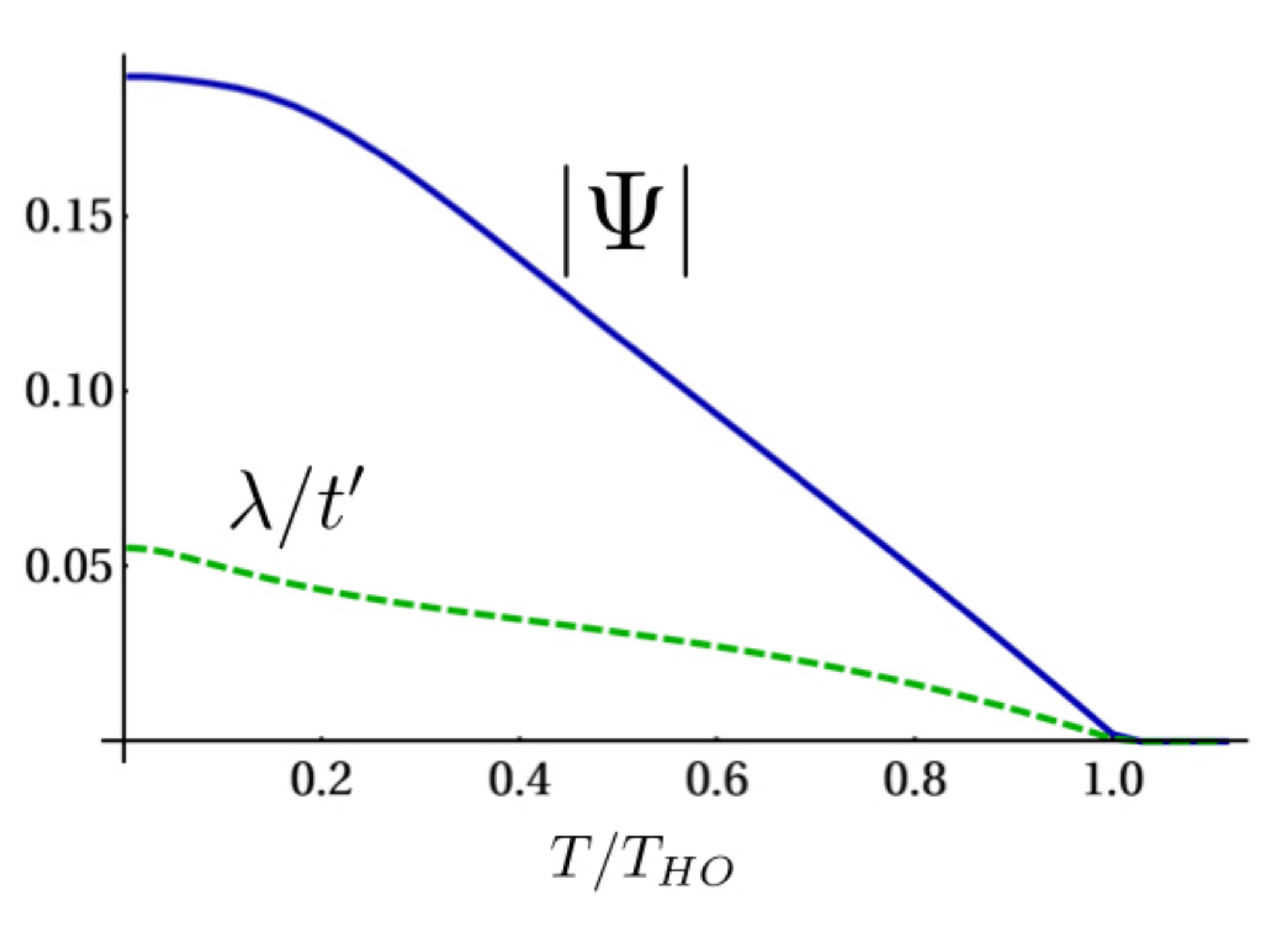}{fig5Bnew}{Mean field parameters $|\Psi|$ and
$\lambda$ as a function of temperature. The parameters used for this
calculation are given in section \ref{parameters}.}

A plot of $|\Psi|$ and $\lambda$ as a function of temperature is shown
in Figure \ref{fig5Bnew} B, for these parameters.  $|\Psi|$ controls the
amplitude of the hybridization gap opening at $T_{HO}$, while
$\lambda$ controls the location of the hybridization gap center in
energy.  We plot the band structure, both above $T_{HO}$ (dashed
lines) and at zero temperature (solid lines) in the hastatic phase,
and the integrated density of states, to show how hastatic order opens
up a gap above $E_F$.


\fg{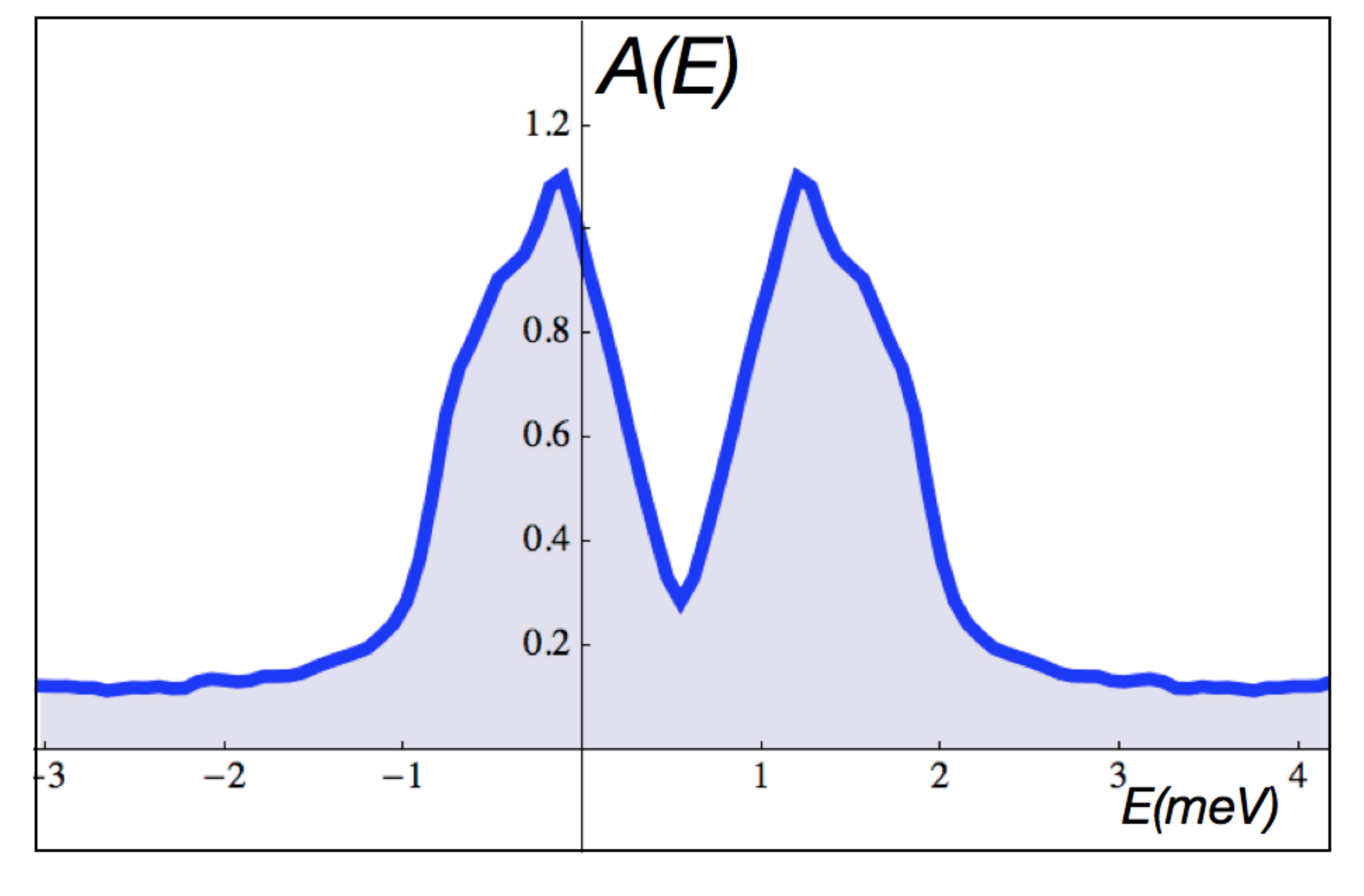}{fig5Cnew}{
Density of states in with hastatic order, for the region close to the
Fermi energy containing the hybridization gap. The parameters used for this
calculation are given in section \ref{parameters}.}
\figwidth=0.8\textwidth

The total density of states is given by 
\begin{eqnarray}\label{l}
A (\omega) &=& \sum_{\bk \eta } \delta (\omega- E_{\bk
\eta })\cr
&=& \frac{1}{\pi}{\rm Im}\int \frac{d^{3}k}{(2\pi)^{3}} \sum_{  \eta}
\frac{1}{\omega - E_{\bk \eta }-i \delta 
}
\end{eqnarray}
where the integral is over the Brillouin zone.
The results of a numerical calculation of the above density of states
are shown in Fig. \ref{fig5Cnew} C.
The calculation was carried out 
with a discrete summation over momenta,
dividing the Brillouin zone into $40^{3}$ points and
using a small value of $\delta $ to broaden the delta-function into a Lorentzian.

\begin{widetext}

\subsection{Conduction and f-electron Green's functions}\label{greensfns}

In order to calculate various moments and susceptibilities, we will require
the full conduction electron Green's function, which can be found from the Hamiltonian
by integrating out the f-electrons,
\bea
\left[\mathcal{G}^c(\bk,i\omega)\right]^{-1} = \zmatrix{i\omega_n - \epsilon_\bk}{0}{0}{i\omega_n - \epsilon_{\bk+\bQ}} - \mathcal{V}_{\bk} \zmatrix{i\omega_n - \lambda_\bk}{0}{0}{i\omega_n - \lambda_{\bk+\bQ}}^{-1}\mathcal{V}\dg_\bk,
\eea
where 
\bea
\mathcal{V}_\bk = \zmatrix{\mathcal{V}_{6\bk}}{\mathcal{V}_{7\bk}}{-\mathcal{V}_{7\bk}}{-\mathcal{V}_{6\bk}}.
\eea
Using isospin, $\vec{\tau}$ to represent $\bk, \bk+\bQ$ space, we 
split the conduction electron energy, $\epsilon_\bk$ into
$\epsilon_{0\bk} = \frac{1}{2}\left(\epsilon_{\bk} +
\epsilon_{\bk+\bQ}\right)$, $\epsilon_{1\bk} =
\frac{1}{2}\left(\epsilon_{\bk} - \epsilon_{\bk+\bQ}\right)$ into the
particle-hole symmetric and antisymmetric parts (and similarly with
$\lambda_{0\bk}, \lambda_{1\bk}$).  So now we can write the conduction
electron Green's function as:

\begin{equation}\label{86}
\left[\mathcal{G}^c(\bk,i\omega)\right]^{-1}  = (i\omega_n -
\epsilon_{0\bk}) - \epsilon_{1\bk} \tau_3 - \Sigma_{c} (\bk ,i\omega_{n}),
\end{equation}
where
\begin{equation}\label{}
\Sigma_{c} (\bk ,i\omega_n)=
\mathcal{V}\frac{i\omega_n - \lambda_{0\bk} + \lambda_{1\bk}\tau_3}{(i \omega_n - \lambda_{0\bk})^2-\lambda_{1\bk}^2} \mathcal{V}\dg
\end{equation}
is the self-energy generated by resonant Kondo scattering off the
hastatic order. The scattering terms in this quantity are determined
by two matrices, a diagonal, symmetry-preserving matrix
$\mathcal{V}_{\bk }\mathcal{{V}}\dg_{\bk }
$ 
and a symmetry-breaking
$\mathcal{V}_{\bk }\tau_{3}\mathcal{{V}}\dg_{\bk }$.  We decompose
these matrices into their channel and spin components
as follows:
\begin{eqnarray}\label{l}
\mathcal{V}_{\bk }\mathcal{{V}}\dg_{\bk }&=&V^{2}_{\bk c +}+ (\vec{\Delta}_{\bk c+}\cdot \vec{\sigma })\tau_{1} , \cr
\mathcal{V}_{\bk }\tau_3\mathcal{{V}}\dg_{\bk }&=& 
V^{2}_{\bk c -}+ \Delta_{\bk c-}\tau_{2}
,
\end{eqnarray}
where the only non-vanishing components
are:
\begin{eqnarray}\label{88}
V_{\bk c+}^2 & =& \frac{1}{4}\tr 
\left[
\mathcal{V}_\bk\mathcal{V}_\bk\dg \right] = \frac{1}{2}\tr
\left[\mathcal{V}_{6\bk} \mathcal{V}_{6\bk}\dg + \mathcal{V}_{7\bk}
\mathcal{V}_{7\bk}\dg\right]\\
 V_{\bk c-}^2 & =& 
\frac{1}{4}\tr\left[ 
\mathcal{V}_\bk\tau_3\mathcal{V}_\bk\dg\tau_3\right] = \frac{1}{2}\tr
\left[\mathcal{V}_{6\bk} \mathcal{V}_{6\bk}\dg - \mathcal{V}_{7\bk}
\mathcal{V}_{7\bk}\dg\right]\\ 
\overrightarrow{\Delta}_{\bk c+} & = & \frac{1}{4}\tr
\left[
\mathcal{V}_\bk\mathcal{V}_\bk\dg\tau_1\vec{\sigma} \right] = -\frac{1}{2} \tr
\left[\left(\mathcal{V}_{6\bk} \mathcal{V}_{7\bk}\dg + \mathcal{V}_{7\bk}
\mathcal{V}_{6\bk}\dg\right)\vec{\sigma}\right]\\
\Delta_{\bk c-} & = & \frac{1}{4}\tr\left[
\mathcal{V}_\bk\tau_3\mathcal{V}_\bk\dg\tau_2 \right] = \frac{i}{2} \tr
\left[\mathcal{V}_{6\bk} \mathcal{V}_{7\bk}\dg - \mathcal{V}_{7\bk}
\mathcal{V}_{6\bk}\dg\right].
\end{eqnarray}
We note that the non-zero form factor $\Delta_{\bk c-}$ 
which involves a product of hybridization in different channels
${\cal V}_{6\bk }$
and ${\cal V}_{7 \bk }$ has a d-wave form factor, reflecting the fact
that electron 
scattering off hastatic order
parameter breaks tetragonal lattice symmetry. 
The related form factor that describes the c-electron moments, $
\overrightarrow{\Delta}_{\bk c-}$  is a
vector in spin space which is even parity in momentum space.  Under
time-reversal, this component changes sign and therefore breaks
time-reversal symmetry.


The
Green's function can now 
be written in the form $\left[\mathcal{G}\right]^{-1} = A \tau_0 + B\tau_3 +
\vec{C}\cdot \vec{\sigma}\tau_1 + D \tau_2$.  The eigenvalues are 
found by taking $\det \left[\mathcal{G}^c(\bk,\omega)\right]^{-1} = 0$, leading to an eighth order polynomial that cannot be generically solved analytically, except in the special case of particle-hole symmetry.
The four
(doubly degenerate) eigenvalues are $E_{\bk\eta}$ and are
found numerically on a grid of $\bk$ points.  Due to the
structure of the Green's function, we can write,
\begin{eqnarray}\label{condG}
\mathcal{G}^c(i\omega_n, \bk) & = & \frac{1}{\prod_\eta (i\omega_n - E_{\bk \eta})} \left(A_{c\bk}(i\omega_n) \tau_0 - B_{c\bk}(i\omega_n)\tau_3 - \vec{C}_{c\bk}(i\omega_n)\cdot \vec{\sigma}\tau_1 - D_{c\bk}\tau_2\right),\cr
A_{c\bk}(i\omega_n) & = & (i\omega_n-\epsilon_{0\bk})\left[(i \omega_n - \lambda_{0\bk})^2-\lambda_{1\bk}^2\right] - (i\omega_n -\lambda_{0\bk})V_{\bk c +}^2\cr
B_{c\bk}(i\omega_n) & = & -\epsilon_{1\bk}\left[(i \omega_n - \lambda_{0\bk})^2-\lambda_{1\bk}^2\right]-\lambda_{1\bk}V_{\bk -}^2\cr
\vec{C}_{c\bk}(i\omega_n) & = & -(i\omega_n - \lambda_{0\bk})
\overrightarrow{\Delta}_{\bk c +} \cr
D_{c\bk}(i\omega_n) & = & -\lambda_{1\bk}\Delta_{\bk c -}.
\end{eqnarray}
Once $E_{\bk \eta}$ is
obtained numerically,
this structure makes calculations with the conduction electron
Green's function fairly straightforward, as we shall illustrate in
sections \ref{susc} and \ref{mag}.  The f-electron Green's function can be obtained by
integrating out the conduction electrons, and is quite similar \bea
\left[\mathcal{G}^f(\bk,i\omega)\right]^{-1} = (i\omega_n -
\lambda_{0\bk}) - \lambda_{1\bk} \tau_3 -
\mathcal{V}\dg\frac{i\omega_n - \epsilon_{0\bk} +
\epsilon_{1\bk}\tau_3}{(i \omega_n -
\epsilon_{0\bk})^2-\epsilon_{1\bk}^2} \mathcal{V}.  \eea 
The main
difference is that the hybridization terms will be of the form $\tr
\mathcal{V}_\bk\dg ({\bf 1}, \tau_3) \mathcal{V}_\bk \tau_a \sigma_b$,
as given below.  We note that $V_{\bk f+}^2 = V_{\bk c+}^2 = \tr
\mathcal{V}_\bk\dg \mathcal{V}_\bk$, and $\Delta_{\bk f -} =
\Delta_{\bk c -}$ are the same for the c- and f-electrons.  There is
only one new, non-zero term,
\begin{eqnarray}\label{newformfactor} \overrightarrow{\Delta}_{\bk f+} &
= & \frac{1}{4}\tr\mathcal{V}\dg_\bk \mathcal{V}_\bk\tau_1 \vec{\sigma} = -\frac{1}{2}\tr
\left[\left(\mathcal{V}\dg_{6\bk} \mathcal{V}_{7\bk} +
\mathcal{V}_{7\bk}\dg \mathcal{V}_{6\bk}\right) \vec{\sigma}\right].
\end{eqnarray} 
that breaks time-reversal symmetry and generally has a d-wave symmetry.

The f-electron Green's function is then,
\begin{eqnarray}\label{fgreens}
\mathcal{G}^f(i\omega_n, \bk) & = & \frac{1}{\prod_\eta (i\omega_n - E_{\bk \eta})} \left(A_{f\bk}(i\omega_n) \tau_0 - B_{f\bk}(i\omega_n)\tau_3 - \vec{C}_{f\bk}(i\omega_n)\cdot \vec{\sigma}\tau_1 - D_{f\bk}\tau_2\right),\cr
A_{f\bk}(i\omega_n) & = & (i\omega_n-\lambda_{0\bk})\left[(i \omega_n - \epsilon_{0\bk})^2-\epsilon_{1\bk}^2\right] - (i\omega_n -\epsilon_{0\bk})V_{\bk c +}^2\cr
B_{f\bk}(i\omega_n) & = & -\lambda_{1\bk}\left[(i \omega_n - \epsilon_{0\bk})^2-\epsilon_{1\bk}^2\right]-\epsilon_{1\bk}V_{\bk c -}^2\cr
\vec{C_{f\bk}}(i\omega_n) & = & -(i\omega_n - \epsilon_{0\bk})\overrightarrow{\Delta}_{\bk f+}\cr
D_{f\bk} & = & -\epsilon_{1\bk} \Delta_{\bk c -}.
\end{eqnarray}
We will examine the $\bk$-space structure of these terms, and the related moments, in section \ref{mag}.

\subsection{Particle-Hole Symmetric Case}

For the special case of a particle-hole symmetric dispersion, where
$(\epsilon_{\bk+\bQ}+\mu) = - (\epsilon_\bk+\mu)$ and
$\epsilon_{f\bk +\bQ }=-\epsilon_{f\bk }$, we can solve the Hamiltonian
(\ref{thematrix}) exactly provided  $\lambda +\mu =0$, so that 
$
\epsilon_{0\bk }= \lambda_{0\bk }= \lambda=-\mu$ are both dispersionless.
In fact, the simple c- and f-dispersions we have chosen
already satisfy particle-hole symmetry, so the special case
$\lambda=-\mu$ provides a limit where we can obtain analytic results. 
In this case, 
$\epsilon_{0\bk } = \lambda_{0\bk } =\lambda$, and the 
the determinant of the Green's function can be calculated from
(\ref{86}).  We wish to evaluate the determinant 
$\Det{\omega - H (\bk )}$, where $H(\bk )$ is the matrix Hamiltonian
given in (\ref{thematrix}). By integrating out the f-electrons we can
factorize the determinant into a product of the full-conduction
electron determinant and the bare f-electron determinant as follows
obtain 
\begin{eqnarray}\label{l}
\Det{\omega-H (\bk )} &=& \Det{- {\cal G}^{c} (\bk,\omega)^{-1}}
\Det{\omega - \lambda_{0\bk }- \lambda_{1\bk }\tau_{3}}\cr
&=&  \Det{- {\cal G}^{c} (\bk,\omega)^{-1}} ((\omega- \lambda_{0\bk
})^{2}- \lambda_{1\bk }^{2})^{2}
\end{eqnarray}
where the overall square in the second factor results from the
two-fold spin degeneracy 
and 
\begin{equation}\label{}
{\rm det}[-{\cal G}^{c} (\bk ,\omega)^{-1}]=\Det{
(\omega -\epsilon_{0\bk })
-\epsilon_{1\bk }\tau_{3} - {\cal V}_{\bk }
\frac{\omega -\lambda_{0\bk } + \lambda_{1\bk }\tau_{3}}
{(\omega - \lambda_{0\bk })^{2}- \lambda_{1\bk }^{2}}
{\cal V}\dg_{\bk }
}
\end{equation}
We now impose particle-hole symmetry, setting
$\lambda_{0\bk }= \lambda = \epsilon_{0\bk } = -\mu$. 
For convenience, we redefine $z = \omega - \lambda$.
Then
\begin{eqnarray}\label{99}
\Det{\omega - H (\bk )}&=& {\rm det}[-{\cal G}^{c} (\bk ,\omega)^{-1}]
(z^{2}- \lambda_{1k}^{2})^{2}\cr
&=&\Det{
z -\epsilon_{1\bk }\tau_{3} - {\cal V}_{\bk }
\frac{z + \lambda_{1\bk }\tau_{3}}
{z^{2}- \lambda_{1\bk }^{2}}
{\cal V}\dg_{\bk }
}(z^{2}- \lambda_{1k}^{2})^{2}
\cr
&=& {\cal D}^{2} (z)/
(z^{2}-\lambda_{1\bk }^{2})^{2}
\end{eqnarray}
where  we have multiplied all four rows of the determinant 
by $z^{2}-\lambda_{1\bk }^{2}$
and have defined 
\begin{equation}\label{}
{\cal D}^{2} (z) = \Det{
( z -\epsilon_{\bk }\tau_{3}) (z^{2}-\lambda_{1\bk }^{2}) 
- {\cal V}_{\bk }
({z + \lambda_{1\bk }\tau_{3}}){\cal V}\dg_{\bk }
}.
\end{equation}
Note that since this is a four dimensional determinant, ${\cal D}^{2} (z)$
is a twelfth order polynomial.
Now by employing the shorthand
$V_{+}^{2}\equiv V_{\bk  c+}^{2}  
$,
$V_{-}^{2}\equiv V_{\bk  c -}^{2}$, 
$\vec{\Delta }_{+}\equiv \vec{\Delta }_{\bk c+}$ and 
$\Delta_{-}\equiv \Delta_{\bk c-}$,
and substituting 
$\mathcal{V}_{\bk }\mathcal{{V}}\dg_{\bk }=V^{2}_{+}+
(\vec{\Delta}_{+}\cdot \vec{\sigma })\tau_{1}
$ and $\mathcal{V}_{\bk }\tau_3\mathcal{{V}}\dg_{\bk }= 
V^{2}_{-}+ \Delta_{-}\tau_{2}$
from (\ref{88}), 
we obtain 
\begin{equation}\label{}
{\cal D}^{2} (z) = \Det{z (z^{2}-\lambda_{1\bk }^{2}-V_{+}^{2})- (z^{2}-\lambda_{1\bk }^{2})\epsilon_{1\bk }\tau_{3}- z
\vec{\Delta}_{+}\cdot \vec{\sigma }\tau_{1}
- \lambda
V_{-}^{2}\tau_{3}- \Delta_{-}\lambda_{1\bk }\tau_{2}}.
\end{equation}
If we normalize $\vec{\Delta}_{+}= \Delta_{+}\hat n$, then 
the triplet of matrices $(\gamma_{1},\gamma_{2},\gamma_{3})\equiv 
((\hat {n}\cdot\vec\sigma)\tau_{1},\tau_{2},\tau_{3})$ forms a triplet
of anticommuting 
Dirac matrices, ($\{\gamma_{i},\gamma_{j} \}=2 \delta_{ij}$), 
which satisfy the charge conjugation symmetry 
$U\gamma_{i}U\dg = -\gamma_{i}^{T}$, ($i\in [1,3]$), where $U=
\sigma_{2}\tau_{2}$. Since the determinant is unchanged under this
transformation, it is unchanged under a reversal
$\gamma_{i}\rightarrow -\gamma_{i}$. If we take the product of the
$\gamma$-reversed determinant with itself, 
the resulting ``squared'' determinant is then diagonal, giving
\begin{equation}\label{}
{\cal D}^{4} (z)= {\rm Det}\biggl[
\biggl(
z^{2} (z^{2}- \lambda_{1\bk }^{2}- V_{+}^{2})^{2}-
 (z^{2}-\lambda_{1\bk }^{2})^{2}\epsilon_{1\bk }^{2}-
z^{2}\Delta_{+}^{2}- \lambda_{1\bk }^{2}V_{-}^{4}-
\Delta_{-}^{2}\lambda_{1\bk }^{2}\biggr) \mathds{1}_{4}\biggr].
\end{equation}
And since the argument of the determinant is a diagonal
four-dimensional matrix, 
\begin{equation}\label{}
{\cal D} (z)= 
z^{2} (z^{2}- \lambda_{1\bk }^{2}- V_{+}^{2})^{2}-
 (z^{2}-\lambda_{1\bk }^{2})^{2}\epsilon_{1\bk }^{2}-
z^{2}\Delta_{+}^{2}- \lambda_{1\bk }^{2}V_{-}^{4}-
\Delta_{-}^{2}\lambda_{1\bk }^{2},
\end{equation}
Now there can be no poles in ${\rm
det}[\omega-H (\bk )]$ at $z=\lambda_{1\bk }$, so 
${\cal D} (z)$ must have zeros at
 $z= \pm\lambda_{1\bk}$ to cancel the denominator in
$\Det{\omega- H (\bk )}={\cal D}^{2} (z)/
(z^{2}-\lambda_{1\bk }^{2})^{2}$. 
We can
factorize the determinant as follows:
\begin{equation}\label{}
{\cal D} (z)=  (z^{2}-\lambda_{1\bk }^{2}) \biggl(z^{2}
(z^{2}-\lambda_{1\bk }^{2}-2V_{+}^{2})
- \epsilon_{1\bk }^{2} (z^{2}-\lambda_{1\bk }^{2})- 2 \epsilon_{1\bk }\lambda_{1\bk} V_{-}^{2}
\biggr)+ z^{2} ( V_{+}^{4}-\Delta_{+}^{2})- \lambda_{1\bk }^{2} (V_{-}^{4}+ 
\Delta_{-}^{2}).
\end{equation}
Now since $D (z=\lambda_{1\bk })=0$ is a zero, 
it follows
that 
\begin{equation}\label{}
V_{+}^{4}-V_{-}^{4}= \Delta_{+}^{2}+\Delta_{-}^{2},
\end{equation}
which also follows somewhat nonintuitively from the form-factor definitions, allowing us to rewrite 
\begin{eqnarray}\label{l}
{\cal D} (z)&=&  (z^{2}-\lambda_{1\bk }^{2}) \biggl(z^{2}
(z^{2}-\lambda_{1\bk }^{2}-2V_{+}^{2} - \epsilon_{1\bk }^{2})
+ (\epsilon_{1\bk }\lambda_{1\bk }- V_{-}^{2})^{2} + \Delta_{-}^{2}
\biggl)\cr
&=& (z^{2}-\lambda_{1\bk }^{2}) \biggl(z^{4}- 2 z^{2}
\left(V_{+}^2 +
\frac{1}{2}(\epsilon_{1\bk}^2+\lambda_{1\bk}^2)\right)
+ (\epsilon_{1\bk }\lambda_{1\bk }- V_{-}^{2})^{2} + \Delta_{-}^{2}
\biggl).
\end{eqnarray}
Therefore, by (\ref{99})
\begin{equation}\label{}
\Det{\omega- H (\bk )} = 
\left[
(\omega-\lambda)^4 - 2 \alpha_\bk
(\omega-\lambda)^2+\gamma_\bk^2 
\right]^{2}
\end{equation}
where 
\begin{eqnarray}\label{l}
\alpha_\bk & = & V_{\bk c+}^2 +
\frac{1}{2}\left(\epsilon_{1\bk}^2+\lambda_{1\bk}^2\right)
\cr
\gamma_\bk^2 & = & \left(\epsilon_{1\bk}\lambda_{1\bk}-V_{\bk
c-}^2\right)^2 + \Delta_{\bk c-}^2.
\end{eqnarray}
The square in the determinant reflects a two-fold Kramers degeneracy
associated with the invariance of the physics under a combined
translation and time reversal. 
The energy eigenstates are then determined by the condition
\begin{equation}\label{}
(\omega-\lambda)^4 - 2 \alpha_\bk
(\omega-\lambda)^2+\gamma_\bk^2 
=0,
\end{equation}
The four bands are then given
by,
\begin{equation}
\label{PH}
E_{\bk \eta} = \lambda \pm \sqrt{\alpha_{\bk} \pm \sqrt{\alpha_\bk^2 - \gamma_\bk^2}},
\end{equation}
where $\eta = 1 - 4$ labels the $++$, $+-$, $-+$ and $--$ bands.

\end{widetext}

\subsection{Hybridization gaps}

As hastatic order is, at its heart, a spinorial hybridization, the
nature of the resulting hybridization gaps is essential to
understanding the order.  In fact, there are generically two types of
gaps: those that break one or more symmetries: translation, spin
rotation, crystal or time-reversal symmetries, and those that break no
symmetries.  In our mean-field picture, the intra-channel gaps,
$V_{\bk c \pm}^2$ break no symmetries and are proportional to the
amplitude of the hastatic spinor, $\langle \Psi\dg \Psi\rangle$, while
the inter-channel gap, $\Delta_{\bk c-}$ breaks all of the above
symmetries and is proportional to $\langle \Psi\dg \vec{\sigma}
\Psi\rangle$.  The role of these form-factors as hybridization gaps is
especially clear in the particle-hole symmetric case, (\ref{PH}),
where their relative roles may be distinguished.  
While in the
mean-field theory, all hybridization gaps will develop at $T_{HO}$, we
believe that hastatic order will melt via phase fluctuations,
destroying the coherence of the symmetry breaking gap, but keeping the
symmetry preserving gaps.  The existence of \emph{two} types of
hybridization gaps that turn on at different temperatures can
reconcile the number of experiments that find hybridization gaps
turning on either at $T_{HO}$\cite{Schmidt10,Aynajian10} or around the
coherence temperature $T^* \sim 50-70$K\cite{Park11,Haraldsen11}.  
{In addition, these hybridization gaps will connect different parts of the Fermi surface as the intra-channel gaps carry ${\bf Q} = 0$, while the inter-channel gap connects the folded bands.}

{The symmetry-breaking hybridization gap $\Delta_{\bk c-}$,
shown in Fig.  \ref{hybridgap} (a)
has an approximate four-fold ``d-wave'' symmetry about a nematic axis $\hat n$
lying in the basal plane (Fig.  \ref{hybridgap} (b,c)).  While only the square of the inter-channel gap appears in equation (\ref{PH}), it plays the same role as the superconducting gap in composite pairing\cite{nphysus}, and the nodal structure has corresponds to a changing phase of the hybridization between c- and f-electrons around the Brillouin zone.
The modulus of the gap, $\vert\Delta_{\bk c-}\vert$ carries a nematicity, whereby the moments of the gap function squared averaged over the Fermi surface,}
\begin{equation}\label{}
\left\langle \Delta_{\bk c-}^2\hat k_{\alpha }\hat k_{\beta } \right
\rangle_{FS}\propto  \hat n_{\alpha }\hat n_{\beta }
\end{equation}
define a secondary nematic director $\hat n = (n_{x},n_{y})$
of magnitude
\begin{equation}\label{}
\hat n_{x,y} \propto  \Psi \dg {\sigma }_{x,y}\Psi 
\end{equation}
proportional
 to the {\sl square} of the hastatic order
parameter.
The orientation of the nematicity 
is set by $\phi$, here chosen to be $\phi = \pi/4$.  The tetragonal symmetry breaking
can be tuned by changing the crystal field parameter, $\xi$, but
cannot be eliminated.  
However, as the tetragonal symmetry breaking is
d-wave in nature, it can not couple linearly to strain and
therefore will not lead to a first order structural
transition.

\fight=8.5cm
\fg{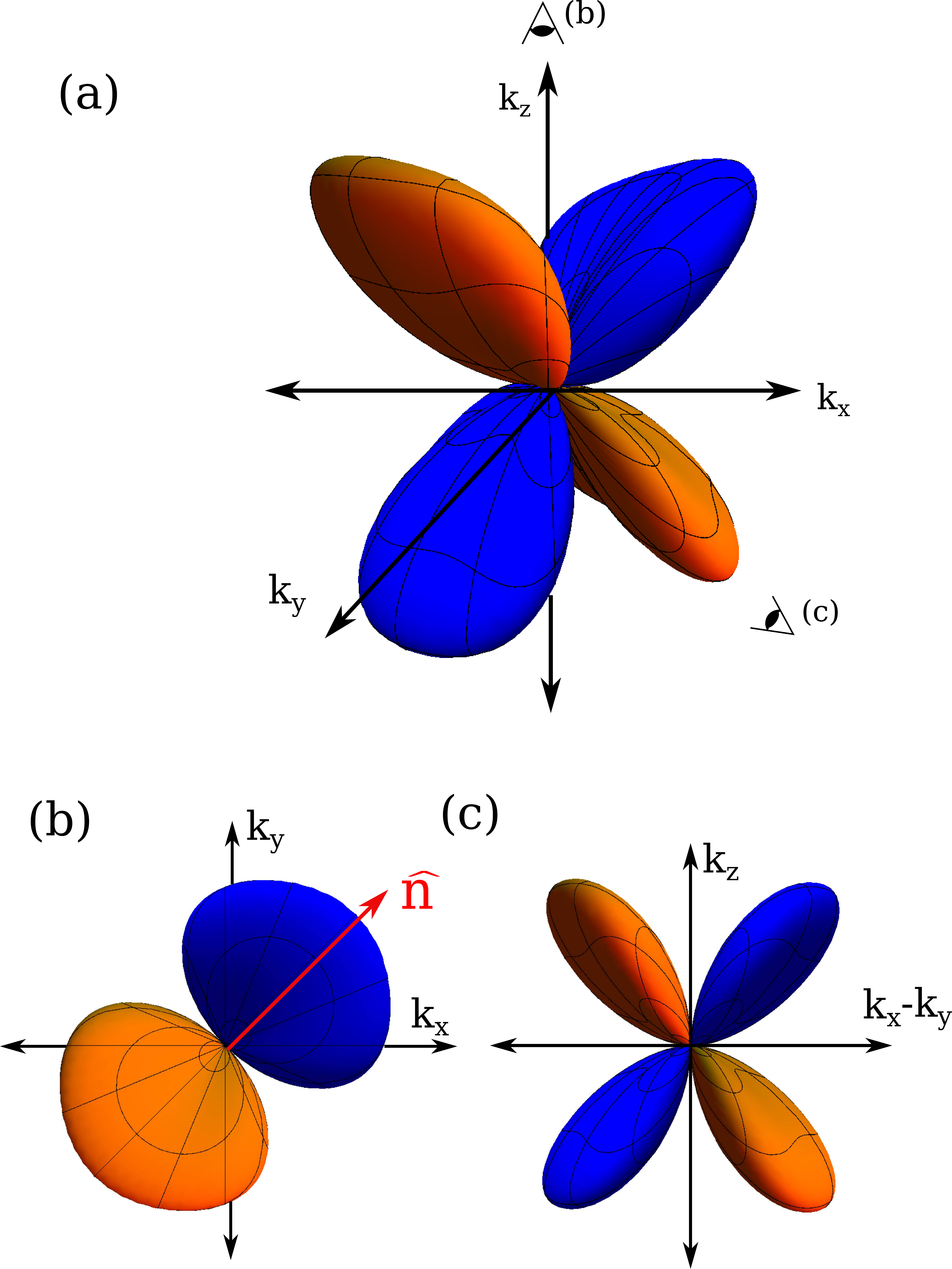}{hybridgap}{(a)
Three dimensional plot showing the 
symmetry breaking component of the hybridization gap, $\Delta_{\bk c-}$
for the chosen crystal field parameters, $\xi = .05$ and for $\phi
=\pi/4$.  Orange and blue represent positive and negative values of
the gap.  The gap lobes are oriented in the plane along the
nematic director $\hat n$ shown in the figure. (b) Top view showing
the nematic character of the gap function, aligned along the nematic director $\hat  n$ (c) Side view along
the axis of the nematic director $\hat n$ showing its nodal 
d-wave structure. 
}

%

\section{Comparison to Experiment: Postdictions}

\subsection{g-factor Anisotropy}

The Zeeman energy is determined by the Hamiltonian 
\begin{equation}\label{}
- \vec{B}\cdot\vec{M} = -\sum_{\bk  \in \frac{1}{2}BZ} \psi \dg_{\bk }{\vec{\cal M}}
\psi_{\bk }\cdot \vec{B}
\end{equation}
where $\psi_{\bk }= (c_{\bk }, c_{\bk +\bQ }, \chi _{\bk },\chi_{\bk
+\bQ } )^{T}$ and 
\begin{equation}\label{}
\vec{{\cal M}}= \frac{1}{2}\left(\begin{matrix} 
2 \mu_B \vec{\sigma}
& 0  & 0 & 0 \cr
0 & 2 \mu_B \vec{\sigma}
 & 0  & 0 \cr
0 & 0 & {g_{f}} \mu_B {\sigma^{z}} & 0 \cr
0 & 0 & 0 & {g_{f}}\mu_B {\sigma^{z}}
\end{matrix}
 \right), 
\end{equation}
where 
$g_{f}$ is the effective g-factor of the Ising Kramers doublet.
In a field, the doubly-degenerate energies, $\vert  \bk \eta \sigma \rangle $
($\sigma = \pm 1$) are split apart so that 
$\Delta E_{\bk \eta }= |E_{\bk \eta \up}-E_{\bk \eta \dw }|=  g_{\bk \eta
} (\theta ) B$, 
so the g-factor is given by  $g_{\bk \eta } (\theta ) = \left \vert \frac{d\Delta E_{\bk \eta }}{dB}\right\vert_{B\rightarrow 0}
$.
Now we are interested in the Fermi surface average of the g-factor,
given by 
\[
g (\theta ) = \frac{
\sum_{\bk \eta } g_{\bk \eta } (\theta )\delta (E_{\bk\eta })
}{\sum_{\bk \eta } \delta (E_{\bk\eta })
}
\]
These quantities were 
calculated numerically, on a $40^{3}$ grid, 
using $g_{f}=2.9$ for 
the effective g-factor of the local non-Kramers doublet. The resulting 
g-factor in the $z-$ direction is reduced to $g_{eff} (\theta =0)=
2.6$ because of the admixture with conduction electrons. 
The 
delta-functions were treated as narrow Lorentzians $\delta (E)= \frac{1}{\pi }Im
(E-i\eta )^{-1}$, where $\eta $ is a small positive number. 
The g-factors at each point in momentum space were computed
by introducing a small field $\delta B$
into the Hamiltonian, with the approximation $g_{\bk \eta } (\theta ) = |E_{\bk \up} -E_{\bk \dw} |/\delta B$.

\subsection{Anisotropic Linear Susceptibility}\label{susc}

\begin{widetext}
The uniform basal plane conduction electron magnetic susceptibility acquires a tetragonal symmetry breaking component in the hastatic phase, given by
\bea
\chi^{xy} = -(g \mu_B)^2 T \sum_{i\omega_n} \sum_{\bk}\tr\left[\sigma^{x}\mathcal{G}^c(\bk,\bk+\bQ,i\omega_n)\sigma^{y}\mathcal{G}^c(\bk+\bQ,\bk,i\omega_n)\right].
\eea
Expanding this in terms of the conduction electron Green's function,
we obtain
\bea
\chi^{xy} & = & -(g \mu_B)^2 T \sum_{i\omega_n}
\sum_{\bk}\tr\left[\sigma_x \mathcal{G}^c(\bk,i\omega_n)\sigma_y
\mathcal{G}^c(\bk,i\omega_n)\right]
\cr
& = & -(g \mu_B)^2\sum_{\bk \eta }\left[
 \frac{2 (E_{\bk \eta}-\lambda_{0\bk})f(E_{\bk\eta})+(E_{\bk \eta}-\lambda_{0\bk})^2f'(E_{\bk\eta})}{\prod_{\eta' \neq \eta} (E_{\bk\eta}-E_{\bk\eta'})^2}\right.\cr
&& \left.-\sum_{\eta'\neq \eta } \frac{2 (E_{\bk\eta}-\lambda_{0\bk})^2
f(E_{\bk\eta})}{
{(E_{\bk\eta}-E_{\bk\eta'})}\prod_{\eta'' \neq
\eta}(E_{\bk\eta}-E_{\bk\eta''})^2}\right]\Delta^{x}_{\bk +}\Delta^{y}_{\bk
+}
\eea
\end{widetext}
Note that above integral may be positive or negative.  The functions $f$ and $f'$ are the Fermi function, $f(x) = \left(\mathrm{e}^{-x/T}+1\right)^{-1}$ and it's derivative, $f'(x) = df(x)/dx$, respectively. {The exact nature of the tetragonal symmetry breaking is determined by the angle of the hastatic spinor, $\phi$; when $\phi = \pi/4$, $\chi_{xx} = \chi_{yy}$, but $\chi_{xy} \neq 0$, but changing $\phi$ can rotate the direction of the tetragonal symmetry breaking.}


In our current mean-field approach, all Kondo 
behavior develops at the hidden order transition, which would lead to an entropy of $R \log 2$ at $T_{HO}$.  Incorporating 
Gaussian fluctuations should suppress the hidden order phase 
transition, $T_{HO}$, while allowing many of
the signatures of heavy fermion physics, including the heavy mass and a partial quenching of the spin entropy to develop at a higher crossover
scale, $T_K$.  The two-channel Kondo impurity has a zero-point entropy
of $\frac{1}{2} R \log 2$\cite{Destri84,Tsvelik85,EmeryKivelson92}, which should be incorporated into the hastatic phase.  
There is considerable
uncertainty in the entropy associated with the development of hidden
order, $S(T_{HO})$, due to difficulties subtracting the phonon and
other non-electronic contributions, leading to estimates ranging from
$.15 R\log 2$\cite{Jaime02} to $.3 R \log 2$\cite{Palstra85}.  If we
take a conservative estimate of $S(T_{HO}) = .2 R\log 2$, and the
normal state $\gamma = 180 {\rm mJ/mol K^2}$\cite{Palstra85}, $S(T_K)
= .2 R \log 2 + \int_{T_{HO}}^{T_K} \gamma dT = \frac{1}{2} R \log 2$
yields $T_K = 27$K, much lower than the coherence temperature seen in
the resistivity.


\section{Comparison to Experiment: Predictions}\label{}

\subsection{Resonant Nematicity in Scanning Probes}

\fg{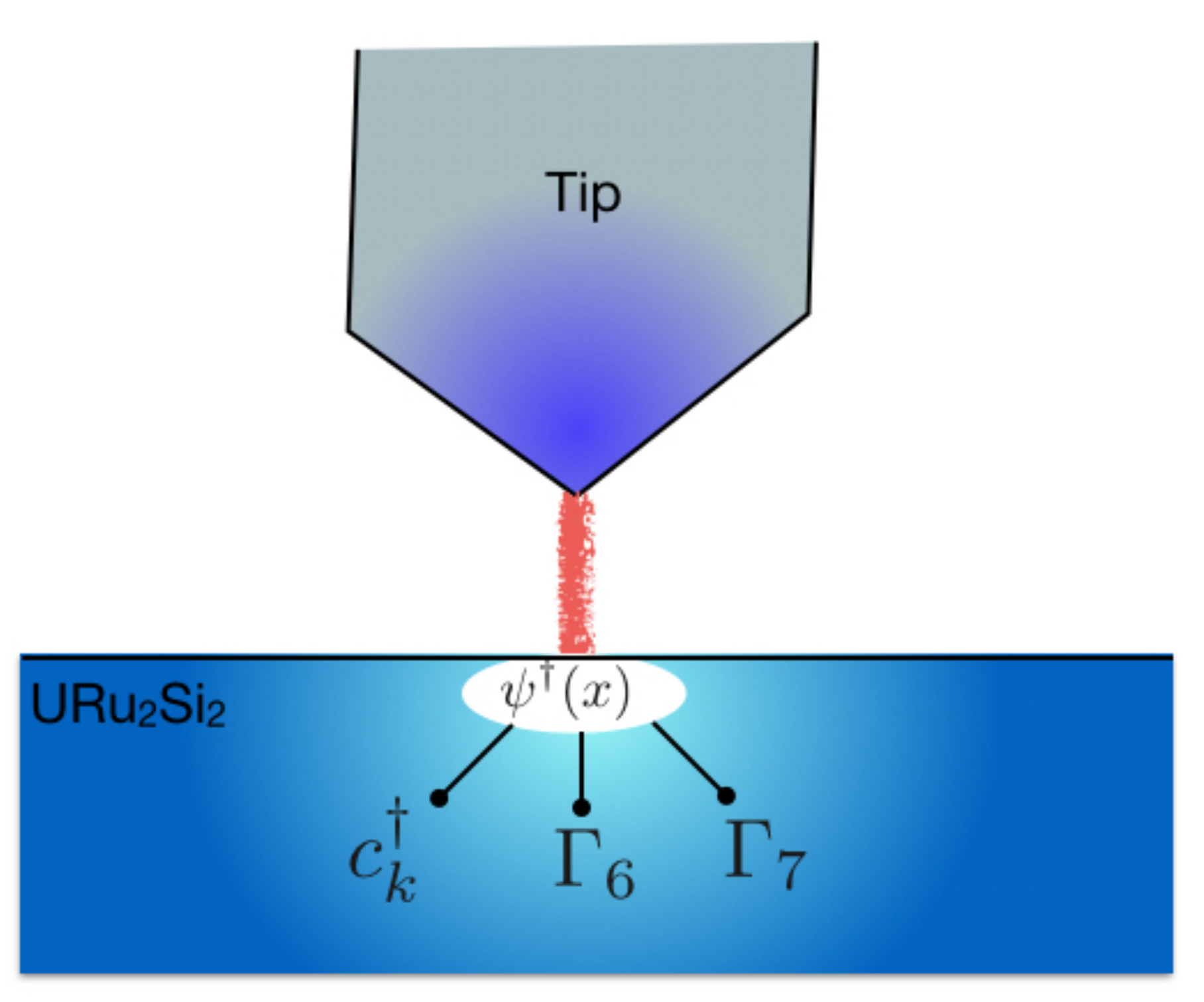}{cotunnelfig}{Schematic illustrating the
tunneling and co-tunneling into three channels: a conduction, a
$\Gamma_{6}$ and a $\Gamma_{6}$ f-electron channel. }

\figwidth=0.8\columnwidth
\fg{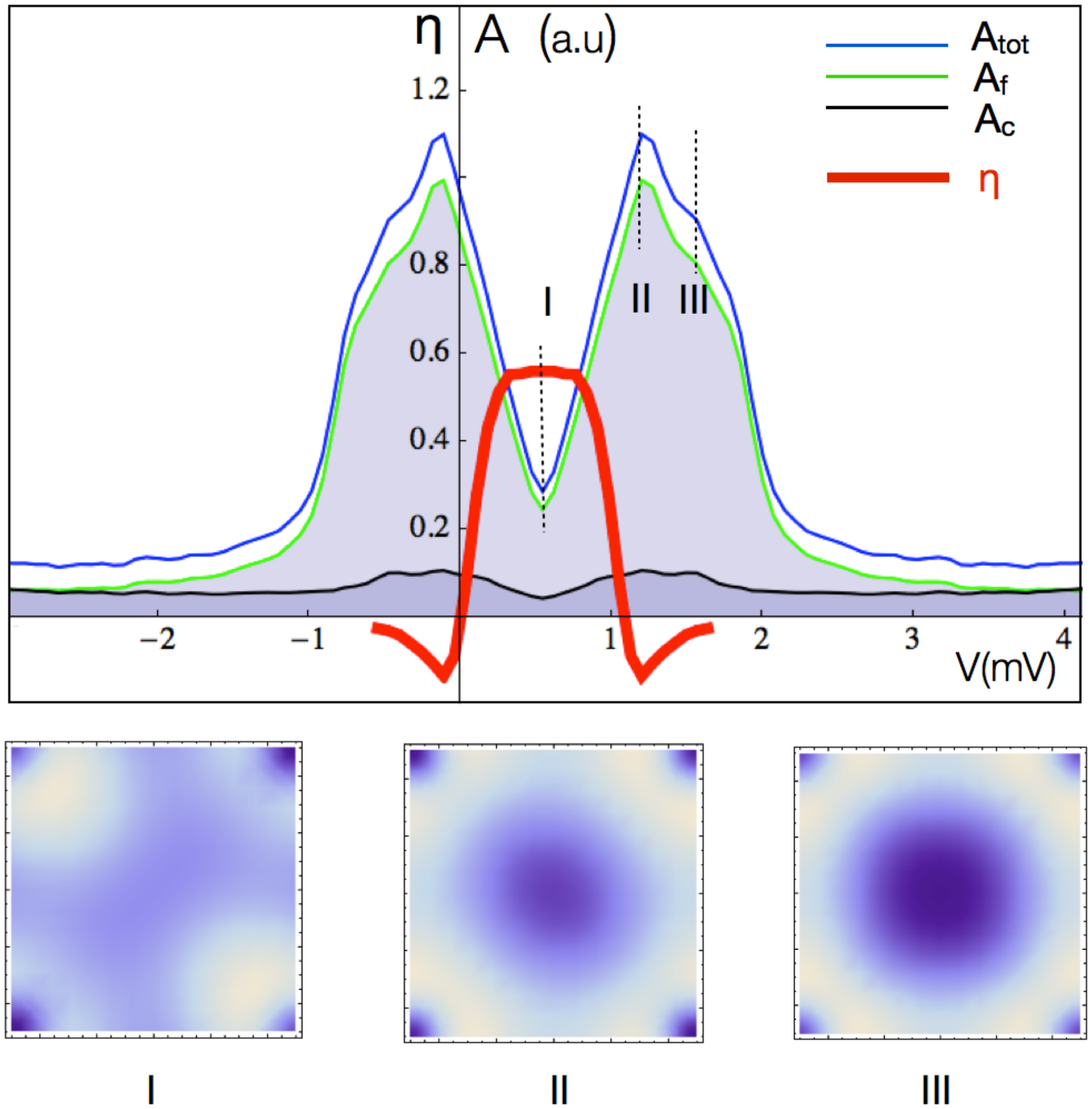}{tunnelplot}{Upper panel: showing the total 
density of states, decomposed 
according to f, conduction and total spectral weight. These curves
were calculated for a particle-hole symmetric dispersion, setting
$\lambda=-\mu$ as described in III C2 and IIIE.
The red curve shows
energy dependent nematicity.  Small panels below show the density of
states at three different energies, showing the energy dependence of
the nematicity. Reprinted from \cite{hastatic}.}

To calculate the tunneling density of states, we assume that the
differential conductance is proportional to the local Green's function
on the surface of the material
\begin{equation}\label{l}
\frac{dI}{dV} (\bx ) \propto  A (\bx, eV)
\end{equation}
where
\begin{eqnarray}\label{l}
A (\bx ,\omega ) &=& \frac{1}{\pi} {\rm Im }G_{\sigma \sigma } (\bx, \omega- i\delta )\cr
&=&\sum_{\sigma } \int_{-\infty }^{\infty }dt  \langle \{
\psi_{\sigma } (\bx ,t), \psi_{\sigma } \dg(\bx,0)\}\rangle  e^{i \omega t}
\end{eqnarray}
is the imaginary part of the local electronic Green's function.

To calculate this quantity, we  decompose the local electron
field in terms of the low energy fermion modes of the system.
Typically, in a Kondo system, there are two channels - a conduction
channel, and a f-electron channel into which the electron may tunnel
\cite{maltseva09}. 
However, in \urs the presence of a non-Kramer's doublet now involves two
f-tunneling channels - the $\Gamma_{6}$ and $\Gamma_{7^{-}}$
channel, and tunneling through these three channels can mutually
interfere (see Fig. \ref{cotunnelfig}).   We therefore decompose the electron field $\psi_{\sigma}(\bx)$
in terms of a conduction and two f-electron channels, writing, 
\begin{widetext}
\begin{eqnarray}\label{l}
\psi _{\sigma } (\bx ) = \sum_{j}
\left( \phi_c (|\bx - \bR_{j}|), \phi^{6}_{\sigma \alpha } (\bx -
\bR_{j}) , \phi^{7}_{\sigma \alpha }  (\bx - \bR_{j})  \right) \cdot
\pmat{
c_{j\sigma } \cr
f_{j\Gamma_{6}\alpha}\cr
f_{j\Gamma_{7}\alpha} \cr}
\end{eqnarray}
where $\phi_{c} (|\bx -\bR_{j}|)$ is the wavefunction of the conduction electron
centered at site j, while
\begin{eqnarray}\label{l}
\phi^{6}_{\sigma \alpha } (|\bx  - \bR_{j}|) &=& \phi^{6} (|\bx  - \bR_{j}|){\cal
Y}^{6}_{\sigma \alpha } (\bx  - \bR_{j})\cr
\phi^{7}_{\sigma \alpha } (|\bx  - \bR_{j}|) &=& \phi^{7} (|\bx  - \bR_{j}|){\cal
Y}^{7}_{\sigma \alpha } (\bx  - \bR_{j})
\end{eqnarray}
are the wave functions of the $\Gamma_{6} $ and $\Gamma_{7^{-}}$ 
f-orbitals  centered at site $j$.

Projected into the low energy subspace, following eqs (\ref{opreplace}) we have 
$f_{j\Gamma_{6}\alpha } \rightarrow (\langle B\dg_{j}\rangle
\chi_{j})_{\alpha } $ and 
$f_{j\Gamma_{7}\alpha } \rightarrow (\langle B\dg_{j}\rangle \sigma_{1}\chi_{j})_{\alpha }$.
Writing $B\dg _{j} = b U_{j}$, and $\tilde{\chi }_{j}= U_{j}\chi_{j}$
the expression for the electron field operator becomes 
\begin{eqnarray}\label{l}
\psi _{\sigma } (\bx ) = \sum_{j}
\left( \phi_c (|\bx - \bR_{j}|), \phi^{6}_{\sigma \alpha } (\bx -
\bR_{j}) , \phi^{7}_{\sigma \alpha }  (\bx - \bR_{j})  \right) \cdot
\pmat{
c_{j\sigma } \cr
b \tilde{ \chi }_{j\alpha }\cr
 b (\hat {\bf n}\cdot \vec{\sigma
})e^{-i \bQ  \cdot \bR_{j}}\tilde\chi_{j\alpha}
\cr}
\end{eqnarray}
Next, rewriting the field operators in momentum space, 
\begin{eqnarray}\label{l}
c_{j\sigma } &=& \sum_{\bk  \in \frac{1}{2}BZ}e^{i\bk \cdot \bR_{j}}
(c_{\bk \sigma }+ e^{i\bQ \cdot \bR_{j}}c_{\bk +\bQ \sigma })\cr
\chi_{j\alpha }&=& \sum_{\bk  \in \frac{1}{2}BZ}e^{i\bk \cdot \bR_{j}}
(\chi _{\bk \sigma }+ e^{i\bQ \cdot \bR_{j}}\chi _{\bk +\bQ \alpha })
\end{eqnarray}

we can decompose
the electron field operator as the dot product of two four component
vectors
\begin{eqnarray}\label{l}
\psi _{\sigma } (\bx )  
= \sum_{j\bk \in \frac{1}{2}BZ}
e^{-i\bk \cdot \bR_{j}}
\Lambda_{j}(\bx - \bR_{j})
\cdot \left(
\begin{matrix} c_{\bk \alpha }\cr c_{\bk+\bQ\alpha}\cr
\chi_{\bk \alpha }\cr
\chi_{\bk+\bQ \alpha }
\end{matrix} \right)
\end{eqnarray}
where 
\begin{eqnarray}\label{l}
\Lambda_{j} (\bx ) =
\left( \right.
\phi_c (|\bx |)\delta_{\sigma \alpha },
e^{i \bQ \cdot \bR_{j}}
\phi_c (|\bx |)\delta_{\sigma \alpha },
\left. b\phi^{6}_{\sigma
\alpha } (\bx ) + e^{-i \bQ \cdot \bR_{j}}
b\phi^{7}_{\sigma \alpha }  (\bx ) (\hat  {\bf n}\cdot \vec{\sigma })
,
 e^{i \bQ \cdot \bR_{j}}b\phi^{6}_{\sigma
\alpha } (\bx ) +
b\phi^{7}_{\sigma \alpha }  (\bx ) (\hat  {\bf n}\cdot \vec{\sigma })
 \right)
\end{eqnarray}
\end{widetext}
We choose a layer where $e^{-i(\bQ\cdot\bR_{j})}=+1$, then on this layer
the local Green's function is given by 
\begin{eqnarray}\label{l}
G (\bx ,\omega)=
\sum_{j,l}
\tilde{ \Lambda} (\bx
 - \bR_{j})\cdot {\cal G}_{jl} (\omega)\cdot \tilde{\Lambda}\dg (\bx -\bR_{l}) 
\end{eqnarray}
where 
\[
{\cal G}_{jl} (\omega)= \sum_{\bk  \in \frac{1}{2 }BZ}{\rm Tr}[(1+ \tau_{1}) {\cal G} (\bk ,\omega)] e^{-i \bk \cdot(\bR_j-
\bR_{l})}
\]
is a trace only over the momentum degrees of freedom, so ${\cal G}_{jl}$ is a
four by four matrix for each pair of lattice points $j$ and $l$, where
\begin{equation}\label{}
\tilde{ \Lambda} (\bx) = 
\left( 
\phi_c (|\bx |)\delta_{\sigma \alpha },
b\phi^{6}_{\sigma
\alpha } (\bx ) 
+ b\phi^{7}_{\sigma \alpha }  (\bx ) 
(\hat  {\bf n}\cdot \vec{\sigma })
 \right).
\end{equation}

The final spectral function is then 
\begin{equation}\label{}
A (\bx ,\omega ) = \frac{1}{\pi} {\rm Im \ Tr}\left[
\sum_{j,l} \tilde{ \Lambda} (\bx - \bR_{j})\cdot {\cal G}_{jl} (\omega-i\delta )\cdot \tilde{\Lambda}\dg (\bx -\bR_{l}) 
 \right]
\end{equation}
To evaluate this quantity,  the summations were limited to the four
nearest neighbor sites at the corner of a plaquette.  The positions $\bx
$ were taken to lie in the plane of the U atoms.
The wavefunctions 
$\phi^{6} (\vert \bx \vert )= e^{-\vert  \bx  \vert /a}$, $\phi^{7}
(\vert \bx \vert )= e^{-\vert  \bx  \vert /a}$ and 
$\phi_{c} (\vert \bx \vert )= e^{-\vert  \bx  \vert /a}$ were each
taken to be simple exponentials of characteristic range equal to the
$U-U$ spacing $a$.

The nematicity of the tunneling conductance was then calculated
numerically from the spatial integral
\begin{equation}
\eta (eV) = \frac{
\int A (\bx ,eV)
\ {\rm sgn} (xy)dxdy
}
{\left(
\int dx dy A (\bx ,eV)^{2}
- 
\left[ \int dx dy A (\bx ,eV)\right]^{2}
\right)^{1/2}
}.
\end{equation}
{This nematicity is shown in Fig. \ref{tunnelplot}.  Note that the nematicity near the Fermi energy is nearly zero, consistent with the absence of quadrupolar moments, but is largest at the hybridization gap energy.}


\subsection{Anisotropy of the Nonlinear Susceptibility Anomaly}

\subsubsection{Landau Theory}\label{}
The origin of the large c-axis nonlinear susceptibility anomaly in \urs\cite{ramirez94} has been a long-standing mystery.  It has been understood phenomenologically 
within a Landau theory as a consequence of a large $\Psi^2 B_z^2$ coupling of unknown origin\cite{ramirez94,newrefschi3}, which can now be understood
within the hastatic proposal.  While the conduction electrons couple isotropically to an applied field, the non-Kramers doublet linearly couples only to the z-component of the magnetic field, $B_{z}= B\cos \theta$, which splits the doublet as it begins to suppress the Kondo effect.  When we include the effect of the magnetic field in the Landau theory, we obtain
\begin{equation}\label{}
f[\Psi] = \bigl [\alpha (T_{c}-T)- \eta_z B_{z}^{2} -\eta_\perp B_\perp^2\bigr ]\Psi^{2} + \beta
\vert \Psi |^{4}  +\gamma (\Psi \dg \sigma_{z}\Psi )^{2},
\end{equation}
where the coefficients of the $\Psi^2 B_z^2$ and the $\Psi^2 B_\perp^2$ terms, $\eta_z$ and $\eta_\perp$, will be estimated using a simplified microscopic approach
discussed shortly in Section \ref{microscopic}.
Minimizing this functional with respect to $\Psi$, we obtain
\begin{equation}\label{}
f = - \frac{1}{4 \beta }\left[\alpha (T_{c}-T)- \eta_z (B\cos \theta)^{2} - \eta_\perp (B \sin \theta)^2 \right]^{2}.
\end{equation}
Following the arguments of \cite{newrefschi3}, 
we can calculate the jump in the specific heat 
$\Delta C_{v}$ and the linear
and nonlinear susceptibility anomalies $\frac{d\Delta\chi_1}{dT}$ and $\Delta\chi_{3}$
respectively, to find
\begin{eqnarray}\label{l}
\frac{\Delta C_{V}}{T_{HO}} &=&
\frac{\alpha^{2}}{2 \beta }\\
\frac{d\chi_{1}}{dT} &=& - \frac{\alpha }{2 \beta }\left(\eta_z \cos^2
\theta + \eta_\perp \sin^2 \theta\right) 
\cr
&\approx& - \frac{\alpha \eta_z}{2 \beta } \cos^2 \theta \\
\Delta \chi_{3} &= & 
 \frac{6}{\beta } \left(\eta_z \cos^2 \theta + \eta_\perp \sin^2 \theta\right)^2 \approx \frac{6\eta_z^2}{\beta }\cos^4\theta
\end{eqnarray}
where $\frac{d\Delta\chi_1}{dT}$ and $\Delta\chi_{3}$ are the anomalous
components of the linear and nonlinear susceptibilities that develops at $T_{HO}$.  These results
show that $\Delta\chi_{3}$ will exhibit a giant Ising
anisotropy; we note that these results are compatible with the observed Fermi surface magnetization results
that indicate that $g(\theta) \sim \cos\theta$ so that $\chi_1 \sim \cos^2 \theta$.  The thermodynamic relation
\begin{equation}\label{}
\frac{\Delta  C}{T} \chi_{3}= 12 \left(\frac{ d \chi_{1}}{dT} \right)^{2}
\end{equation} is maintained for all angles $\theta$; the important point here is that the
anisotropy in $\Delta \chi_3$ is significantly larger than that
in $\Delta \chi_1$ and numerical estimates will be discussed once we have introduced the microscopic approach to
hastatic order.

\subsubsection{$\eta_z$ and $\eta_\perp$ from 
Microscopics}\label{microscopic}

To complete this simple Landau theory, we will calculate
$\vec{\eta}$ in a simplified model: we will neglect the momentum
dependence of both the f-level and the hybridization and take the
hastatic order to be uniform.  None of these assumptions qualitatively
changes the results.  The $|\Psi|^2$ coefficient is calculated from
the microscopic theory (see the next section) by expanding the action,
$S = -\tr \log\left[ 1 - \mathcal{F}_0 (V\Psi) \mathcal{G}_0 (V
\Psi\dg)\right]$ in $\Psi$, where $\mathcal{F}_0 = (i\omega_n -
\lambda - g_f \mu_f B_z \sigma_3)^{-1}$ and $\mathcal{G}_0 =
(i\omega_n - \epsilon_\bk-g/2 \vec{B}\cdot \vec{\sigma})^{-1}$ are the
bare $\chi$ and conduction electron Green's functions (remember,
$\chi$ are the fermions representing the non-Kramers doublet).  $V$
represents the hybridization matrix elements, which are
momentum-independent here, and proportional to the unit matrix.  Note
that while the conduction electrons are isotropic, the $\chi$'s are
perfectly Ising.
The coefficient of $|\Psi|^2$ is then,
\upit=-0.25in
\bxwidth = 0.95in
\bea
\raiser{\frm{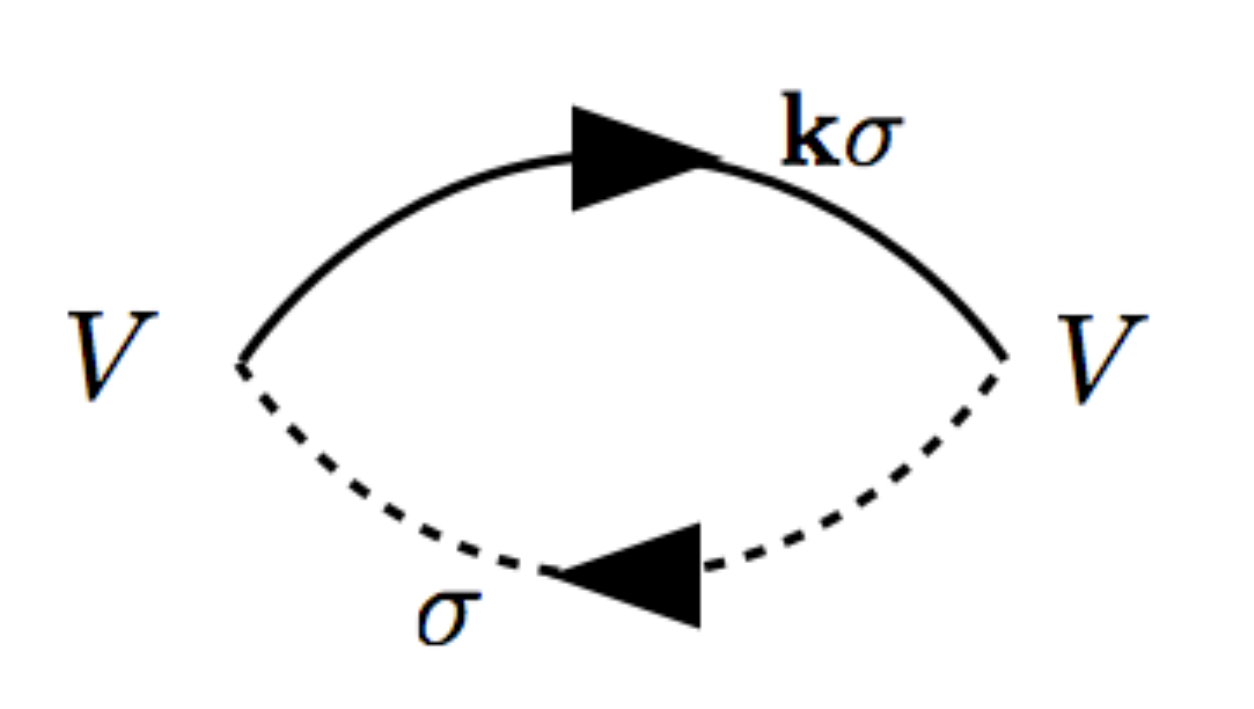}}=
V^2 T \sum_{i \omega_n} \sum_{\bk\sigma} \frac{1}{i\omega_n - \epsilon_{\bk \sigma}}\frac{1}{i\omega_n - \lambda_\sigma},
\eea
where $\epsilon_{\bk \sigma} = \epsilon_\bk - g/2\vec{\sigma}\cdot
\vec{B}$ and $\lambda_\sigma = \lambda - g_f \mu_f \sigma B_z$ are the
dispersions in field. Performing the Matsubara sum, we obtain
\begin{widetext}
\begin{eqnarray}\label{l}
V^2 \sum_{\sigma}\int_{-\infty}^{\infty}d\epsilon
\mathcal{D}(\epsilon) \frac{\tanh \frac{\epsilon_{\bk \sigma}}{2T} -
\tanh \frac{\lambda_\sigma}{2T}}{2 (\lambda_\sigma - \epsilon_{\bk
\sigma})}
= \rho V^2 \sum_{\sigma}\int_{-D}^{D}d\epsilon \frac{\tanh \frac{\epsilon-g/2\vec{\sigma}\cdot \vec{B}}{2T} - \tanh \frac{\lambda_\sigma}{2T}}{2 (\lambda_\sigma - \epsilon + \frac{g}{2} \vec{\sigma}\cdot \vec{B})},
\end{eqnarray}
where we approximated the conduction electron density of states as a
constant, $\rho$ within the bandwidth, $2D$.
Let us first calculate $\eta_\perp$, quantizing the field along the x-direction and taking g = 2,
\begin{eqnarray}\label{l}
\eta_\perp 
=
\raiser{\frm{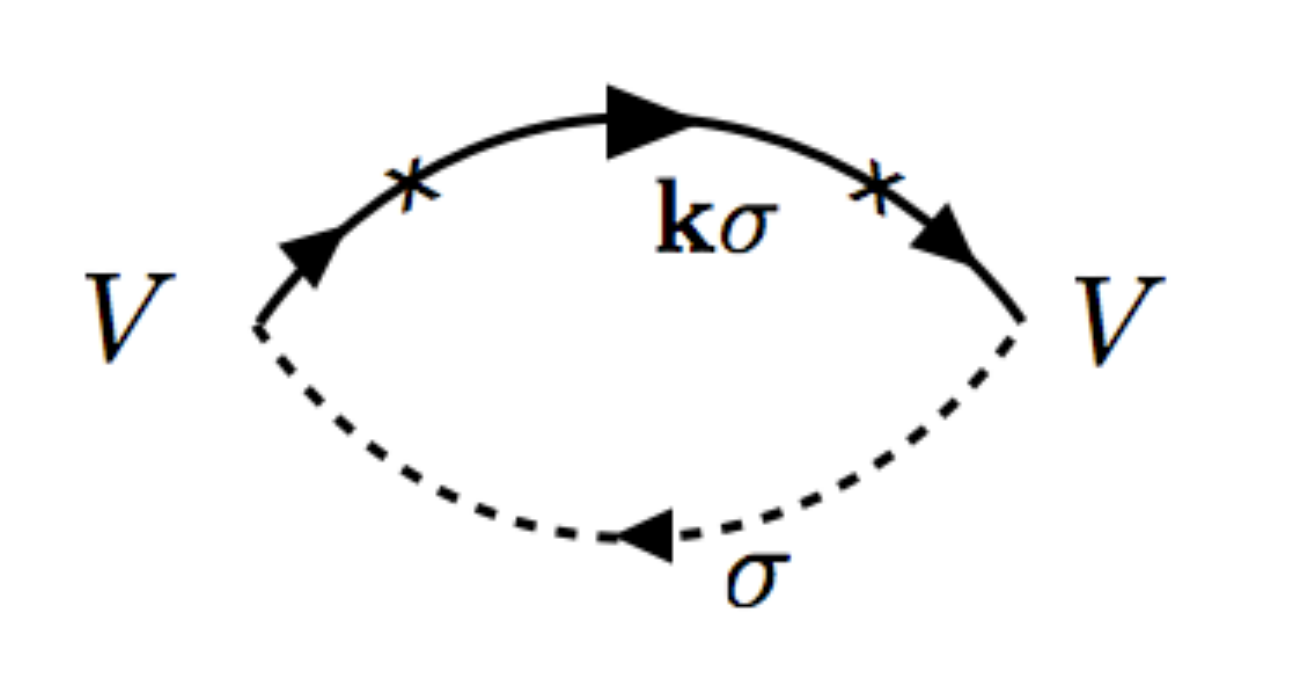}}
= -\rho V^2 \sum_\sigma \int_{-D}^{D}d\epsilon \left.\frac{\partial^2}{\partial B_\perp^2}\frac{\tanh \frac{\epsilon-\sigma B_\perp}{2T} - \tanh \frac{\lambda}{2T}}{2 (\lambda- \epsilon + \sigma B_\perp)}\right|_{B_\perp = 0}.
\end{eqnarray}
As the integrand is a function of $\epsilon - \sigma B$, the integral is straightforward.  And as $D \gg \lambda, T$, the dominant term will be:
\bea
\eta_\perp = -\rho V^2 \left.\left( \frac{{\rm sech}^2\frac{\epsilon}{2T}}{4 T(\epsilon-\lambda)}+\frac{\tanh \frac{\epsilon}{2T}-\tanh\frac{\lambda}{2T}}{(\epsilon-\lambda)^2}\right)\right|_{-D}^{D} = \frac{\rho V^2}{D^2}.
\eea
$\eta_z$ will have three contributing terms: one purely from the
conduction electrons that is $\eta_\perp$, one arising from cross
terms between the conduction and f-electrons, and finally one solely
from the f-electrons that dominates the other two.  We shall focus on
this last term,
\bea
\eta_z =
\raiser{\frm{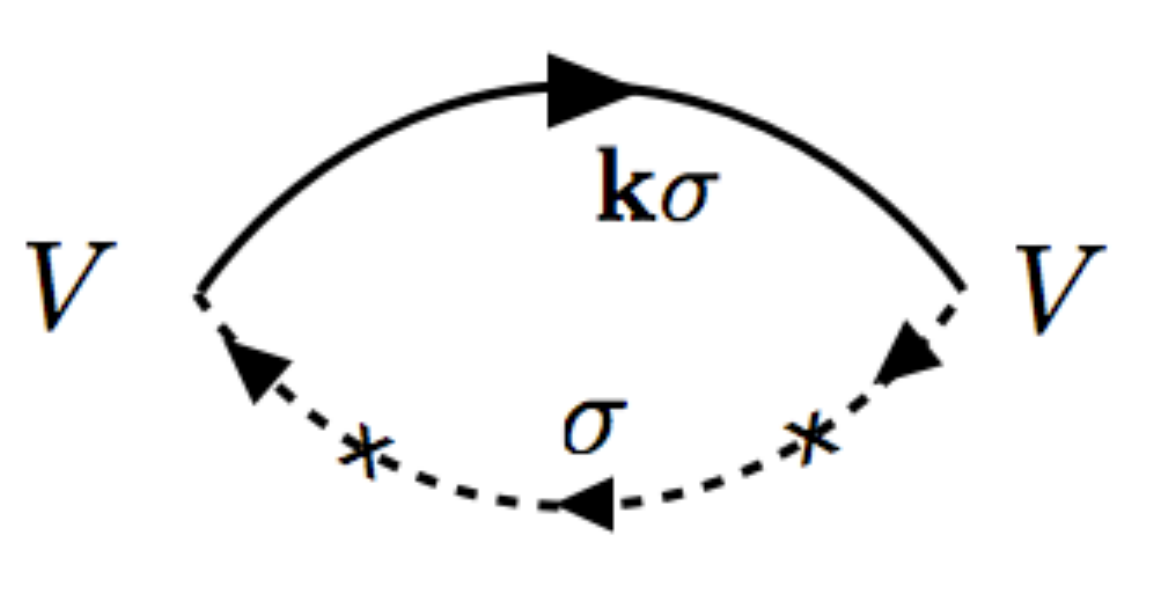}}=
-\rho V^2 \sum_\sigma \int_{-D}^{D} d\epsilon \left.\frac{\partial^2}{\partial B_z^2}\frac{\tanh \frac{\epsilon}{2T} - \tanh \frac{\lambda-g_f \mu_f \sigma B_z}{2T}}{2 (\lambda- \epsilon-g_f \mu_f \sigma B_z)}\right|_{B_z = 0}.
\eea
This integral cannot be done analytically at finite temperature, so we take $T\rarrow 0$.
\bea
\eta_z \approx -\rho V^2\left.\frac{\partial^2}{\partial B_z^2} \sum_\sigma \int_{-D}^0 \frac{1}{2(\lambda_\sigma - \epsilon)} \right|_{B_z = 0} = \frac{\rho V^2}{\lambda^2} - \frac{\rho V^2}{(D+\lambda)^2} = \frac{\rho V^2}{T_{HO}^2},
\eea
as $\lambda = T_{HO}$ at zero temperature.  So $\eta_\perp/\eta_z =
\frac{T_{HO}^2}{D^2}$.  
\end{widetext}
Using a conservative value of
$T_{H0}/D\sim 1/30$, we predict an anisotropy of 
about $900$ in $d\chi_{1}/dT$ and nearly $10^6$ in
$\Delta \chi_{3}$.  However, in a realistic model, there will be f-electron contributions to $\eta_\perp$ involving fluctuations to excited crystal field states that may reduce the anisotropy somewhat.
The important point here is that the anisotropies will
be orders of magnitude larger than the single ion anisotropy in
$\chi_{1}$ (approximately $3$), 
and furthermore, that they will develop exclusively at the hidden order
transition.  

\subsection{Basal-Plane Moment}\label{mag}




Another key aspect of the hastatic picture is the presence of broken
time-reversal symmetry in \emph{both} the
HO and AFM phases, manifested
by a staggered moment of wavevector $\bQ  = (0, 0, \pi)$
\begin{equation}\label{}
\vec{m} (\bQ ) =\vec{m}_{c} (\bQ ) + \vec{m}_{f} (\bQ ).
\end{equation}
$\vec{m}$ contains two parts: a conduction electron component
\begin{eqnarray}\label{l}
\vec{m}_{c} (\bQ )&=&  \frac{g_{c}\mu_{B}}{2}\sum_{\bk }\langle c\dg_{\bk+\bQ  \alpha
}\vec{\sigma }_{\alpha \beta }c_{\bk \beta } + {\rm H.c}\rangle \cr
&=& 
-\frac{g \mu_B}{4}
 T\sum_{i\omega_n} \sum_\bk \tr \left[
\vec
{\sigma} \mathcal{G}^c(\bk,i\omega_n) \tau_{1}\right].
\end{eqnarray}
which involves the off-diagonal component 
of the conduction electron Green's function (see section
\ref{greensfns}) and an f-electron component, 
\begin{equation}\label{}
\vec{m}_f (\bQ ) = m_{I} (\bQ )\hat z + \vec{m}_{f^{3}} (\bQ ).
\end{equation}
Here
\begin{equation}\label{}
m_{I} (\bQ ) = 
-\frac{g_{f^{2}} \mu_B }{4}
T\sum_{i\omega_n} \sum_\bk \tr
\left[
 {\sigma}_{z}
\mathcal{G}^f(\bk,i\omega_n) \tau_{1} \right]
\end{equation}
is the Ising $5f^{2}$ contribution,
and 
\begin{equation}\label{}
\vec{m}_{f^{3}} (\bQ )
 =  \frac{g_{f^{3}}
\mu_{B}}{2}\langle \Hast\dg \vec{\sigma} \Hast \rangle, 
\end{equation}
is the contribution  derived from 
valence fluctuations into the $5f^{3}$ Kramer's doublet, where $\Hast
$  is the staggered component of the hastatic order parameter. 
In the antiferromagnet, the hastatic order parameter at site $j$ is
given by 
\begin{eqnarray}\label{l}
\Hast_{j} = 
\exp \left[- i (\bQ  \cdot \bR_{j})
\frac{\sigma_{y}}{2}\right] \Hast
\end{eqnarray}
where
\begin{equation}\label{}
\Hast  = \pmat {\psi_0\cr 0}.
\end{equation}
points out of the plane.  By contrast, in the hidden order phase  
\begin{equation}\label{}
\Hast_{j} = \exp \left[- i (\bQ  \cdot \bR_{j})
\frac{\sigma_{z}}{2}\right] \Hast
\end{equation}
where \begin{equation}\label{}
\Hast = \frac{\psi_0}{\sqrt{2}}\pmat {e^{i\phi /2}
\cr 
e^{-i \phi /2}}, 
\end{equation}
lies in the basal plane, and $\phi $ determines the angle of moment from the $x$-axis in the
plane; now
the 
magnetic moment lies 
entirely in the basal plane, determined by
\begin{equation}\label{l}
\vec{m} (\bQ )=\vec{m}_{c} (\bQ )+ 
\frac{g_{f^{3}}
\mu_{B}}{2}\langle \Hast\dg \vec{\sigma} \Hast \rangle .
\end{equation}
According to the Clogston-Anderson compensation
theorem\cite{Anderson61}, 
the magnetic polarization of the conduction electrons
$\vec{ m}_{c}\sim O (T_{K}/D)$ is small and set by the 
the same ratio $T_{K}/D$ that determines the g-factor anisotropy.  The
magnitude of the second $f^{3}$
term is set by the overall magnitude of $\Psi
$, which in turn is determined by the overall amount of mixed valent
admixture of $5f^{3}$ configuration into the ground-state. 

Writing out the conduction electron polarization $\vec{m}_{c}$in detail using 
Eq (\ref{condG})  we have
\begin{eqnarray}\label{condmom}
\vec{m}_c(\bQ)  &= & -(g \mu_B) T\sum_{\bk, \omega_{n}}\tr
\left[\vec{\sigma} \mathcal{G}^c(\bk,i\omega_n) \tau_1  \right]\cr
& &
=  -(g \mu_B)\sum_{\bk \eta }\frac{(E_{\bk\eta} -
\lambda_{0\bk})}{\prod_{\eta'\neq \eta} (E_{\bk\eta}- E_{\bk
\eta'})}f(E_{\bk\eta})\overrightarrow{\Delta}_{\bk c +}.\cr
&&
\end{eqnarray}

\fg{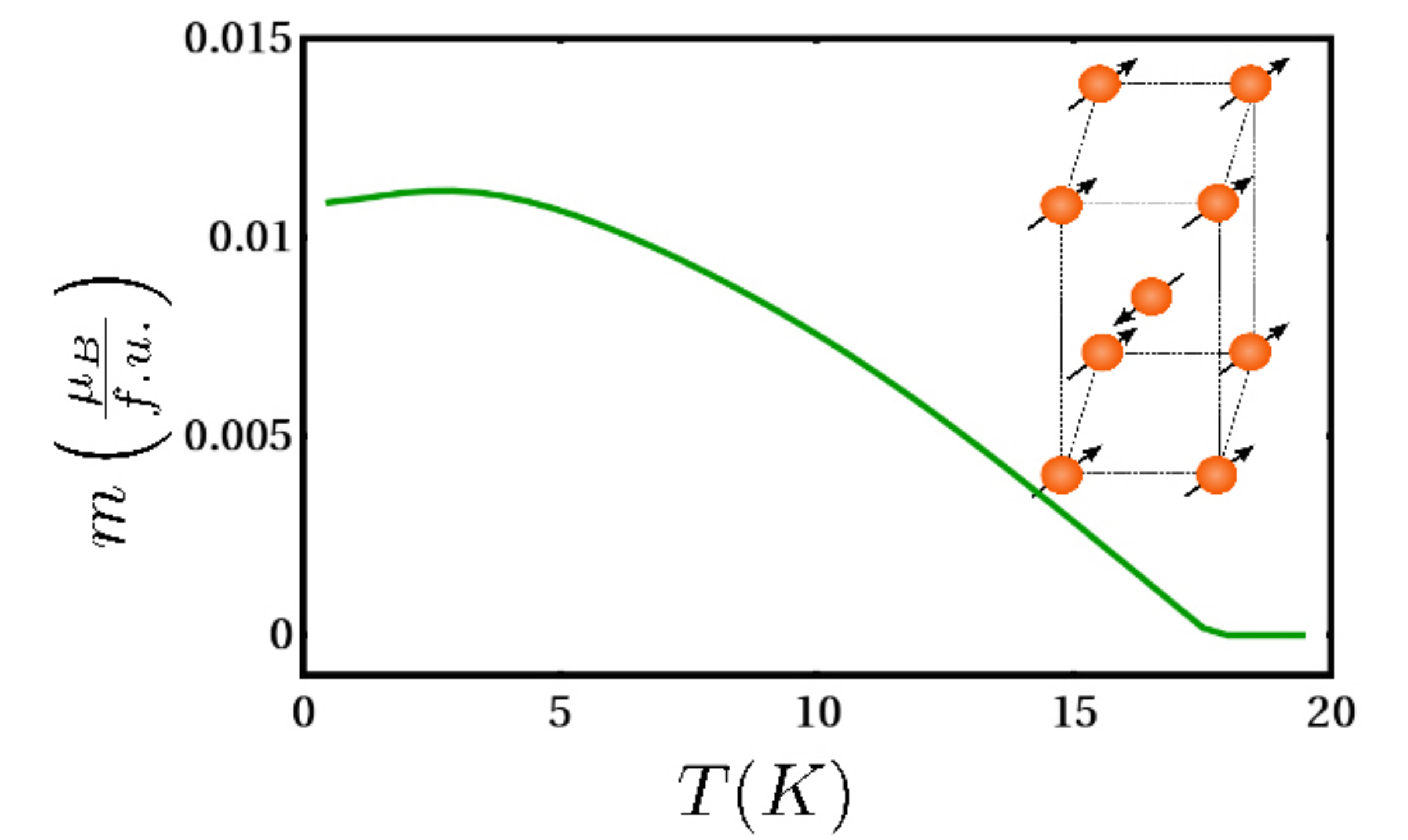}{mag_chi}{Predicted temperature dependence of the
basal plane moment. Parameters used for this calculation are given in section. }
Fig. (\ref{mag_chi}) shows the temperature dependence of the 
magnetic moment calculated 
for a case where $D/T_{K} \approx 30$, 
 for which $m_\perp(0)$ =0.015$\mu_{B}$, which is an upper bound
for the predicted conduction electron moment.
Neutron scattering measurements on \urs have 
placed bounds on the c-axis magnetization of the f-electrons 
using a momentum transfer $Q$ in the basal plane.
Detection of an $m_\perp(0)$ carried by conduction electrons, 
with a small scattering form-factor requires
high-resolution measurements with a c-axis momentum transfer.
Recent high resolution neutron measurements 
to detect this small transverse moment
have not detected a signal, and quote a bound on the f-component of the moment
$|\vec{ m}_{f}| < 0.001\mu_{B}$.  We discuss the implications of this
result in section \ref{moretests}. 
We note that there have been reports from 
$\mu$SR and NMR measurements\cite{Amitsuka03,Bernal04} 
of very small intrinsic basal plane fields in \urs  comparable with
this bound.

We can also examine the quadrupolar moment associated with the HO
state. This is set by the expectation of the transverse components of
the non-Kramers doublet, 
\begin{eqnarray}\label{l}
Q_{x,y}& \propto & \langle \chi \dg_{\alpha } (\sigma_{x,y})_{\alpha \beta }\chi_{\beta }
\rangle \cr
&=&-
T\sum_{i\omega_n} \sum_\bk \tr
\left[
 {\sigma}_{x,y}
\mathcal{G}^f(\bk,i\omega_n) \tau_{1} \right].
\end{eqnarray}
If we expand this using the f-Green's function from  eq (\ref{fgreens}), we find
\begin{eqnarray}\label{fquaddy}
Q_{xy}\propto   -\sum_{\bk \eta }\frac{(E_{\bk\eta} -
\epsilon_{0\bk})}{\prod_{\eta'\neq \eta} (E_{\bk\eta}- E_{\bk
\eta'})}f(E_{\bk\eta})\overrightarrow{\Delta}_{\bk f +}.
\cr
&&.
\end{eqnarray}
The f-electron quadrupole moment (\ref{fquaddy}) has 
an identical form to the
conduction electron moment, (\ref{condmom}), with $\epsilon \leftrightarrow
\lambda$ everywhere, and the relevant form-factor is
(\ref{newformfactor})
\begin{eqnarray}\label{l}
 \overrightarrow{\Delta}_{\bk f+} &
= & -\tr\mathcal{V}\dg_\bk \mathcal{V}_\bk\tau_1 \vec{\sigma} = -2\tr
\left[\left(\mathcal{V}\dg_{6\bk} \mathcal{V}_{7\bk} +
\mathcal{V}_{7\bk}\dg \mathcal{V}_{6\bk}\right) \vec{\sigma}\right]\cr
&&.
\end{eqnarray}
which has a d-wave form-factor.
This means that the summation over momentum vanishes, so that 
the staggered quadrupolar moments $Q_{x,y}$
must vanish. 
(Indeed, there would be
no associated lattice distortion, even for
a uniform hastatic order. )
As a d-wave quadrupole is an $L=4$-tupole, or ``hexadecapole'',
this means that like Haule and Kotliar\cite{Haule09}, the hastatic
order  has staggered $(J_x J_y+J_y J_x)(J_x^2-J_y^2)$ hexadecapolar moments.
However, unlike Haule and Kotliar, where the hexadecapolar moments are
the primary order parameter (and thus of order one), here the
hexadecapolar moments are a secondary effect of the composite hastatic
order, and like the conduction electron moments, will be of order
$T_K/D$.  
Given how difficult it is to observe large
hexadecapolar moments, the hexadecapolar moments associated with
hastatic order will almost certainly be unobservably small.  By
contrast, in the antiferromagnetic phase, the f-electrons develop a
large c-axis magnetic moment.  

{We will return to the predicted basal plane moment in the HO
phase of \urs in Section VI B. when we discuss recent experimental
constraints.}

\section{Discussion and Open Questions}

\fg{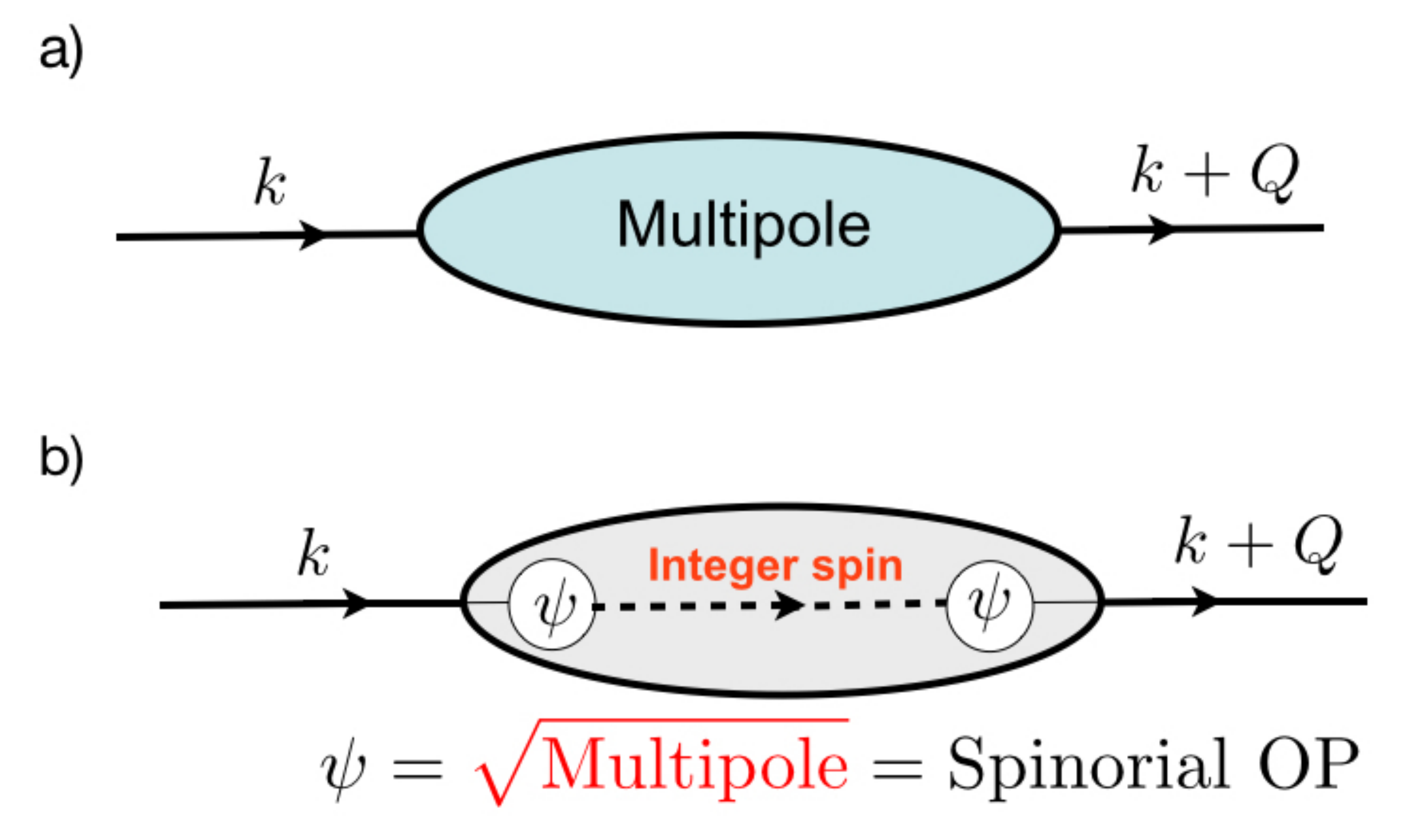}{Fsqrt}{Schematic contrasting the  multipolar and
spinorial theories of Hidden order. (a) in a multipolar scenario, the
heavy electrons Bragg diffract off a staggered spin or charge
multipole (b) in the hastatic scenario, the development of a spinor
hybridization opens up resonant scattering with a an integer spin
state of the ion. The multipole is generated as a consequence of two
spinorial scattering events. In this way, the Hastatic spinor order
parameter can be loosely regarded as the square root of a multipole. }


In summary, the key idea of the hastatic proposal for hidden order in \urs
is that observation of heavy Ising  quasiparticles implies
the development of resonant scattering between half-integer spin
electrons and integer spin local moments.  It is perhaps useful to
contrast the various staggered multipolar scenarios for the hidden order with
the hastatic one proposed here. In the former, mobile f-electrons Bragg 
diffract off a multipolar density wave (see Fig \ref{Fsqrt} (a)), 
whereas in the latter, the multipole contains an internal structure,
associated with the resonant scattering into an integer spin f-state. 
(see Fig \ref{Fsqrt} (b)).   Hastatic order can thus be 
loosely regarded as the ``square root'' of a multipole order
parameter,
\begin{equation}\label{}
\Psi \sim  \sqrt{\hbox{multipole OP}}.
\end{equation}
In fact, as we have seen the square of the hastatic order parameter
breaks tetragonal symmetry, and is thus nematic (see Fig. 9.), with a
director $\vec{n}= (n_{x},n_{y})$ of magnitude determined by the 
square of the hastatic order parameter, 
\begin{equation}\label{}
(n_{x}+ i n_{y}) \propto \psi^*_{\up}\psi_{\dw}.
\end{equation}
It can also be viewed to result from a symmetry-breaking Kondo effect 
between non-Kramers and Kramers doublets.  Hastatic order should be present 
in any f-electron material whose unfilled f-shell contains a geometrically
stabilized non-Kramers doublet, and we expect its realization in other 
$5f$ uranium and $4f$ praseodymium compounds.  Praseodymium compounds are particularly promising tests for hastatic order, as the presence and nature of any non-Kramers doublets can be determined via inelastic neutron scattering.  Any non-Kramers doublet Pr compound must order either magnetically or quadrupolarly or form hastatic order - there is no non-symmetry breaking option, as in Kramers materials.

\subsection{Broader Implications of Hastatic Order}\label{}

At a microscopic level hastatic order demands a
new kind of particle condensation, one that gives rise
to a  Landau order parameter that transforms under
half-integer spin  or {\sl double-group } 
representations.\cite{Chandra14}
Conventionally Landau theory in electronic systems 
is based on the formation and
condensation of two-body bound-states.
For example the
development of a magnetic order parameter $\vec{M} (x)$ is given
by the contraction
\begin{equation}\label{}
\contracty{\psi \dg_{\alpha } (x)\psi_{\beta } (x)}= \vec{\sigma}_{\alpha \beta }
\cdot \vec{M } (x)
\end{equation}
and s-wave superconductivity is based on the formation of
spinless
bosons 
\begin{equation}\label{}
\contracty{\psi _{\up} (1)\psi_{\dw} (2)}= -F(1 - 2),
\end{equation}
where $F (1-2 )= -\langle T\psi _{\up} (1)\psi_{\dw} (2)\rangle $ is
the anomalous Gor'kov Greens function
that breaks the gauge system of the underlying system ( see
Fig. \ref{fig8} a).
The take-home message from conventional two-body
condensation is that when the two-body bound-state wavefunction
carries a quantum number (e.g. charge or spin), a symmetry is broken.
However under this scheme, all order parameters are bosons that carry
integer spin.
\fg{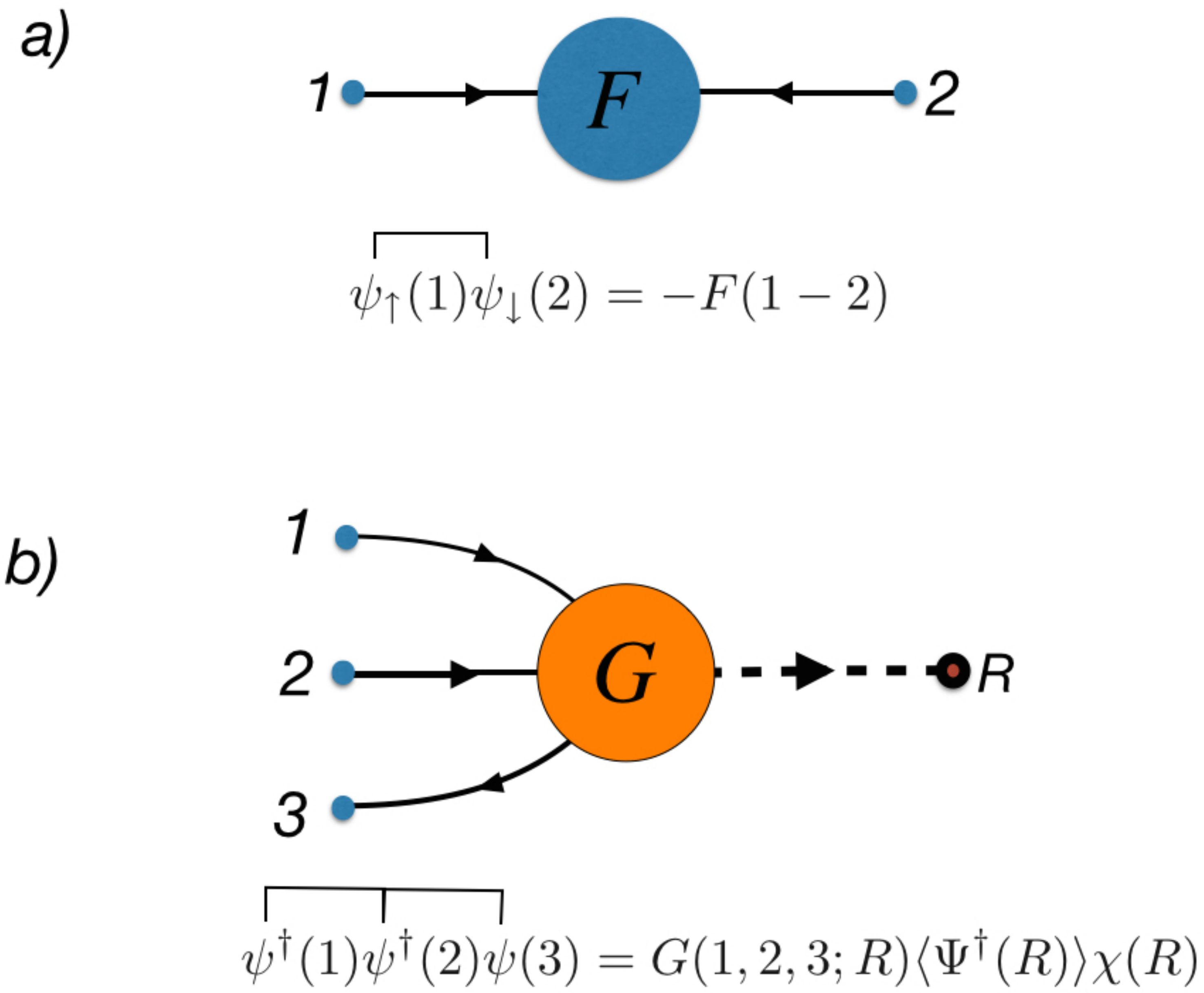}{fig8}{Schematic Feynman diagrams indicating (a) two-body  (b) and three-body electronic bound-states where in the latter case spin indices have been suppressed for pedagogical simplicity.}

Hastatic order carries half-integer spin and cannot develop via this
mechanism. We are then led to the question of whether it is possible
for Landau order parameters to transform under half-integer
representations of the spin rotation group?  At first sight this
impossible for all order parameters are necessarily bosonic and
bosons carry integer spin.  However the connection between spin and
statistics is strictly a relativistic idea that depends on the full
Poincar\'e invariance of the vacuum.  This invariance is lost in
non-relativistic condensed matter systems,{where high energy degrees of freedom are integrated out}, suggesting the possibility
of order parameters with half-integer spin that transform under
double-group representations of the rotation group.  Spinor order
parameters involving ``internal'' quantum numbers are well known in
the context of two-component Bose-Einstein condensates.  The Higgs
field of electroweak theory is also a two-component spinor.  However
in neither case does the spinor transform under the physical rotation
group.  Moreover it is not immediately obvious how such bound-states
emerge within fermionic systems.


Hastatic order is a generalization of Landau's order parameter concept
to three-body bound-states.  This is natural in heavy fermion systems
since the conventional Kondo effect is the formation of a three-body
bound state between a spin flip and a conduction electron.  However here
the three-body wavefunction carries no quantum number and thus
is not an order parameter; this is why conventional Kondo behavior
is associated with a crossover and not a true phase transition.  

In the mean-field formulation of hastatic order,\cite{hastatic} a spin-1/2 order parameter develops as a consequence
of a factorization of a Hubbard operator that connect the Kramers and
non-Kramers states; it is a tensor operator that 
corresponds to the three-body combination 
\begin{equation}\label{}
X_{\alpha \sigma } (R)\equiv
\vert f^{2}\alpha \rangle   
\langle f^{1}\sigma \vert
= \Lambda_{\alpha \sigma }^{abc} (R;1,2,3) \psi\dg _{a } (1)\psi\dg _{b} (2)\psi_{c} (3),
\end{equation}
where we have used 
the short-hand notation $1 \equiv R_{1}$ etc. and 
\begin{equation}\label{}
\Lambda_{\alpha \sigma }^{abc} (R;1,2,3)= 
\langle R_{1},a;R_{2},b\vert \hat X_{\alpha \sigma } (R)\vert R_{3},c\rangle 
\end{equation}
defines the overlap between the Hubbard operators and the bare
electron states. 
In a
simple model, this three body wavefunction is local, $\Lambda_{\alpha \sigma
}^{abc} (R;1,2,3)= \Lambda_{\alpha \sigma }^{abc}\delta
(R-1)\delta (R-2)\delta (R-3)$.
The factorization of the Hubbard operator into 
a spin-1
fermion and a spin-1/2 boson 
\begin{widetext}
\begin{equation}\label{}
X_{\alpha \sigma } (R)\rightarrow 
\chi\dg _\alpha (R)
\left\langle 
 \Psi_{\sigma } (R)\right\rangle 
,
\end{equation}
then represents a ``fractionalization'' of the three body operator. 
Written in terms of the microscopic electron fields, this
becomes 
\begin{eqnarray}\label{l}
\Lambda_{\alpha \sigma }^{abc} (R;1,2,3)\contract{\psi\dg _{a} (1)\psi\dg_{b} (2)\psi _{c} (3)}&=&
\chi\dg  _{\alpha } (R)
\left\langle \phantom{\sum}\hskip -5mm \Psi_{\sigma }(R) \right\rangle
.\cr&&
\end{eqnarray}
This expression can be inverted to give the three body contraction
\begin{eqnarray}\label{l}
\contract{\psi\dg _{a} (1)\psi\dg _{b} (2)\psi _{c} (3)}&=&
\sum_{R}G_{abc}^{\alpha \sigma}(1,2,3;R) 
\chi\dg  _{\alpha } (R)
\left\langle \phantom{\sum}\hskip -5mm \Psi_{\sigma }(R) \right\rangle
,\cr&&
\end{eqnarray}
where 
$G_{abc}^{\sigma\alpha}(1,2,3;R) =[ \Lambda_{ \sigma\alpha  }^{abc}
(R;1,2,3)]^{*}$ (see Fig. \ref{fig8} b).\\

\end{widetext}

The asymmetric decomposition 
of a three-body fermion state into a binary combination of
boson and fermion is a  fractionalization process; if
the boson carries a quantum number, when it condenses we have the
phenomenon of 
{\sl ``order parameter fractionalization''}.
Fractionalization is well-established for
excitations of low dimensional systems, such the one-dimensional
Heisenberg spin chain and the fractional quantum 
Hall effect\cite{Jackiw76,Su79,Laughlin99,Castelnuovo12}, but 
{\sl order parameter fractionalization} is a new concept.
Unlike pair or exciton condensation, the hastatic order parameter transforms
under a double-group representation of the underlying symmetry group,
and thus represents a fundamentally new class of broken symmetries. 
We are currently investigating order parameter fractionalization 
beyond the realm of \urs.  The proposed three-body bound-state has a nonlocal
order parameter, and it may be possible to identify a dual theory with
a local order parameter that breaks a global symmetry.

\subsection{Experimental Constraints and More Tests }\label{moretests}

Let us now return to the situation in \urs. As we discussed
earlier, hastatic order leads to a predication of a basal-plane
moment of order $\frac{T_K}{D}$, where $T_K$ and $D$ are the
Kondo temperature and the band-width respectively.  The transverse
moment in our mean-field treatment has contributions from both
conduction and f electrons, and the ratio $\frac{T_K}{D}$ is very
sensitive to the degree of mixed valence of the $U$ ion.  Our original
calculation assumed $20\%$ 5$f^3$, leading to a predicted basal-plane
moment of $0.01 \mu_B$. Recent high-resolution neutron 
experiments\cite{Das13,Metoki13,Ross14}
with momentum transfer along the c-axis
designed to detect this predicted transverse moment
have placed a bound $\mu_{\perp }< 0.0011\mu_{B}$ on the ordered
transverse moment of the uranium ions, constraining it to be
at best an order of magnitude smaller than what we predicted. 

Clearly we need to reconsider our calculation of the transverse moment
and understand why it is so small if not absent and we are currently
exploring a number of possibilities:

\begin{itemize}
\item {\sl Fluctuations}.
Amplitude fluctuations of the hastatic order parameter
are needed to describe the incoherent Fermi
liquid observed to develop at temperatures well above
$T_{HO}$ in optical, tunneling and thermodynamic 
measurements,\cite{Schmidt10,Aynajian10,Park11,Haraldsen11}
and they will reduce the transverse moment. We note that various
probes, including X-rays, $\mu$-spin resonance and
NMR\cite{Caciuffo14,Amitsuka03,Bernal04,Takagi12} 
have consistently  detected basal plane fields of order
$0.5G$, consistent with the presence of a tiny in-plane moment. 

\item {\sl Uranium Valence}. The predicted transverse moment
is very sensitive to the $5$f valence, decreasing with increasing 
proximity to pure $5f^2$.  More specifically it is proportional 
to the \emph{change} in valence between $T_{HO}$ and the measurement 
temperature and thus is significantly smaller than the high-temperature 
mixed valency. It would be very helpful to have low temperature 
probes of the 5f-valence. 

\item {\sl Domains}. X-ray,\cite{Caciuffo14} 
muon,\cite{Amitsuka03} torque
magnetometry\cite{Okazaki11}, cyclotron resonance\cite{Matsuda11} and NMR measurements\cite{Bernal04,Takagi12} 
that have indicated either
a static moment or broken tetragonal symmetry were performed on
small samples. By contrast, the neutron measurements that show no
measurable moment use large samples
\cite{Das13,Metoki13,Ross14}. The apparent inconsistency
between these two sets of measurements 
may be due to domain formation of hidden order.
Such domain structure could result from 
random pinning\cite{Imry75} of the transverse moment 
by defects of random strain fields. 
The situation in \urs is somewhat analogous
to that in Sr$_{2}$RuO$_{4}$, where there is evidence for broken
time-reversal symmetry breaking with a measured Kerr effect
and $\mu$SR to support chiral p-wave superconductivity, but
no surface currents have yet been observed.\cite{Kallin12}  Domains are
an issue in this system too.

\item {\sl x-y order and spin superflow}. 
The current mean-field theory
has the transverse hastatic vector $\Psi \dg \vec{\sigma }\Psi $
pointing in one of four possible directions at each site,
corresponding to a four-state clock model.  The
tunneling barrier between these configurations is very small. When we
expand the effective action as a function of $\phi $, the leading
order anisotropy will have the form 
\begin{equation}\label{}
\Delta E (\phi  ) =  E_{4} \cos 4 \phi .
\end{equation}
where $E_{4 \phi }$ determines the magnitude of the tunneling
barrier. 
Now the anisotropic terms have the form $e^{\pm i 4 \phi }$,  and since
the $\phi  $ dependence in $\Hast$ enters as $e^{\pm i \frac{\phi
}{2}}$, the leading dependence of this term on $\Hast $ has the form 
\begin{equation}\label{}
E_{4} \sim  T_{K }|\Hast |^{8} 
\end{equation}
Now since $|\Hast|^{2}\sim  \frac{T_{K}}{D}$, this implies that the
tunneling barrier has magnitude 
\begin{equation}\label{}
E_{4  } \sim  T_{K }\left(\frac{ T_{K}}{D} \right)^{4}
\end{equation}
In our theory we have estimated $T_{K}/D\sim 0.01$, so that the
tunneling barrier is of order $10^{-8}$ times smaller than the Kondo
temperature.  
In practice, the XY-like basal plane hastatic moments will be extremely weakly pinned, 
with large domain walls between $Z_4$ domains, with widths $\sim \frac{D}{T_K} \sim 100$ lattice spacings.
To our knowledge, such nearly perfect XY order, which can lead to spin
superflow\cite{chandracolemanlarkinsuperflow,fomin,borovik} is completely unknown in
magnetism: its only counterpart occurring in neutral superfluids.  
This opens the interesting
possibility that the presence of persistent spin currents in the
hastatic phase $j_{\hbox{spin}}\propto \nabla \phi $ give rise to a
destruction of the staggered moment associated with hastatic order.
By contrast, near the surfaces, where the tetragonal symmetry is
broken, the $Z_2$ pinning is expected to be much greater. This might
account for why large moments and broken tetragonal symmetry only
appear in tiny crystals. 
\end{itemize}  


The central tenet of
the hastatic proposal is that Ising quasiparticles are associated
with the development of hidden order and there remain several tests 
of this aspect that can be made. In particular: 

\begin{enumerate}

\item {\sl Giant Anisotropy in $\Delta \chi_3\propto \cos^4 \theta $}.
In this measurement the temperature-dependence of the Ising anisotropy
of the conduction fluid can be probed to confirm that it is associated
with the development of hidden order.
      
\item {\sl dHvA on all the heavy Fermi surface pockets}. 
Based on the upper-critical field results, we expect that the heavy 
quasiparticles in the $\alpha $ $\beta $ and $\gamma$
orbits will exhibit the multiple spin zeros of Ising
quasiparticles but to date only the $\alpha$ orbits have been
measured as a function of field orientation. 

\item {\sl Spin zeros in the AFM phase? (Finite pressure)}
If the antiferromagnetic phase is also hastatic, then we expect 
the spin zeros to persist at finite pressures. 

\end{enumerate}



\subsection{Future Challenges} 

The observation of Ising quasiparticles in the hidden order state
\cite{Brison95,Ohkuni99,Altarawneh11,Altarawneh12}
represents a major challenge to our understanding of \urs; to our
knowledge this is the only example of such anisotropic mobile
electrons, and as we have emphasized, completely unexpected for
f-electrons  in a tetragonal environment. 
As we have emphasized throughout this paper, 
Ising quasiparticles  are the central motivation for the 
hastatic proposal, and a key
question is whether this phenomenon can be described by other HO
theories? In particular:

\begin{itemize}  

\item {\sl Can band theory account for the $g(\theta)$ observed in \urs?} 
Recent advances in the understanding of orbital 
magnetization\cite{Xiao05,Thonhouser05,Xiao06} suggest it may be possible
to compute the g-factor associated with conventional Bloch waves; in
a strongly spin-orbit coupled system, the orbital contributions to the total
energy in a magnetic field are significant.  
It would be particularly interesting to compare the $g(\theta)$ 
computed in a density functional treatment of \urs with
that observed experimentally.  

\item {\sl Can other  $5f^2$ theories account for the multiple spin
zeroes and the upper bound $\Delta  < 1K$ 
on the spin degeneracy of the heavy fermion bands? }
In particular, is it possible to account for the observed spin 
zeros without invoking a non-Kramers $5f^{2}$ doublet? 
\end{itemize}

We have benefitted from inspiring discussions with our colleagues who
include C. Batista, C. Broholm, K. Haule, N. Harrison, G. Kotliar, P. A. Lee,
G. Lonzarich, J. Mydosh, K. Ross and J. Schmalian. PC and PC are
grateful to the hospitality of the Institute for the Theory of
Condensed Matter, Karlsrule Instiute for Technology, the Centro
Brasileiro de Pesquisas Fisicas (CBPF) and Trinity College, Cambridge
where parts of this paper were written. PC and PC gratefully
acknowledge the support of the Conselho Nacional de Desenvolvimento
Científico e Tecnológico-CNPq Brasil (CBPF), supported by CAPES and FAPERJ
grants CAPES - AUX–PE-EAE-705/2013 and 
FAPERJ - E-26/110.030/2013  during their stay.
The three of us also acknowledge the hospitality of the Aspen Center
for Physics, supported by National Science Foundation Grant
No. PHYS-1066293 where we worked together on this project.  This work
was supported by the National Science Foundation grants
NSF-DMR-1334428 (P. Chandra) and DMR-1309929 (P. Coleman), and by the
Simons Foundation (R. Flint).


\begin{thebibliography}{99}

\bibitem{Palstra85} T.T.M. Palstra et al., ``Superconducting and Magnetic Transitions in the Heavy-Fermion System $URu_2Si_2$,'' {\sl Phys. Rev. Lett.} 
{\bf 55} 2727-2730 (1985).

\bibitem{Schlabitz86}W. Schlabitz et al.,
``Superconductivity and Magnetic Order in a Strongly Interacting
Fermi System: \urs'', Z. Phys. B.,
62, 171-177 (1986).

\bibitem{broholm91}C. Broholm et al., ``Magnetic excitations in the heavy-fermion superconductor \urs,'' {\sl Phys. Rev B.} {\bf 43}, 12809 (1991).


Si$_2$single crystals $(0 \leq x \leq 1)$ J. Appl. Phys. 70, 5791–5793 (1991).




\bibitem{walter93} M.B. Walter et al., ``Nature of the order parameter in the Heavy-Fermion system \urs,'' {\sl Phys. Rev. Lett.} 71, 2630 (1993).

\bibitem{takagi07} S. Takagi et al., ``No Evidence for ``Small-Moment Antiferromagnetism'' under Ambient Pressure in \urs: Single-Crystal $^{29}$Si NMR Study,'' J. Phys. Soc. Jpn. {\sl 76}, 033708 (2007).


\bibitem{Amitsuka99}H. Amitsuka, M. Sato, N. Metoki, M. Yokoyama,
K. Kuwahara, T. Sakakibara, H. Morimoto, S. Kawarazaki, Y. Miyako, and
J. A. Mydosh, Phys. Rev. Lett. 83, 5114 (1999).

\bibitem{Amitsuka07} H. Amitsuka et al., ``Pressure-Temperature Phase Diagram of the Heavy-Electron Superconductor $URu_2Si_2$,'' {\sl J. Magn. Magn. Mater.} {\bf 310}, 214-220 (2007).

\bibitem{Butch2010}Nicholas P. Butch, Jason R. Jeffries, Songxue
Chi, Juscelino Batista Leão, Jeffrey W. Lynn, and M. Brian Maple,
``Antiferromagnetic critical pressure in URu2Si2 under hydrostatic
conditions'', Phys. Rev. B 82, 060408 (R), (2010).

\bibitem{Amitsuka94} H. Amitsuka and T. Sakakibara,''Single Uranium-Site Properties of the Dilute Heavy Electron System $U_xTh_{1-x}Ru_2Si_2$ $(x\le 0.07)$,'' {\sl J. Phys. Soc. Japan} {\bf 63} 736-747 (1994).

\bibitem{Haule09} K. Haule and G. Kotliar, ``Arrested Kondo Effect and Hidden Order in $URu_2Si_2$,'' {\sl Nature Phys.} {\bf 5}, 796-799 (2009).

\bibitem{Santini94} P. Santini and G. Amoretti, ``Crystal Field Model of the Magnetic Properties of \urs'' {\sl Phys. Rev. Lett}, {\bf 73}, 1027-1030 (1994).



\bibitem{Varma} C. M. Varma and L. Zhu,  ``Helicity Order: Hidden Order Parameter
in \urs '', Phys. Rev. Lett. 96, 036405-036408 (2006).

\bibitem{pepin} C. P\'{e}pin, M. R. Norman, S. Burdin, and A. Ferraz,
''Modulated Spin Liquid: A New Paradigm for \urs'', Phys. Rev. Lett. 106, 106601-106604 (2011).

\bibitem{Morr10}Ting Yuan, Jeremy Figgins and  Dirk K. Morr, 
``Hidden order transition in \urs: Evidence for the emergence of a coherent Anderson lattice from scanning tunneling spectroscopy'', Phys. Rev. B 86, 035129-035134 (2012).

\bibitem{Dubi10} Y. Dubi and A.V. Balatsky, ``Hybridization Wave as the `Hidden Order' in $URu_2Si_2$,'', {\it Phys. Rev. Lett.} {\bf 106}, 086401-086404 (2011).

\bibitem{Fujimoto11} S. Fujimoto, ``Spin Nematic State as a Candidate
of the Hidden Order Phase of $URu_2Si2$, {\sl Phys. Rev. Lett.} {\bf 106}, 196407-196410 (2011).

\bibitem{Ikeda11} H. Ikeda et al, ``Emergent Rank-5 `Nematic' Order in $URu_Si_2$'', Nature Physics 8, 528–533 (2012).

\bibitem{Mydosh11}J. A. Mydosh and P. M. Oppeneer, ``Colloquium:
Hidden Order, Superconductivity and Magnetism -- The Unsolved Case of
\urs'', Rev. Mod. Phys. 83, 1301–1322 (2011).

\bibitem{hastatic}Premala Chandra, Piers Coleman and Rebecca Flint, 
{\sl ``Hastatic order: a theory for the hidden order in
URu$_{2}$Si$_{2}$''}, Nature, 493, 621-626 (2013).

\bibitem{Flint14} R. Flint, P. Chandra and P. Coleman, 
{\sl ``Hidden and Hastatic orders in \urs"}, J. Phys. Soc. Jpn. 83, 061003 (2014). 

\bibitem{Chandra14} P. Chandra, P. Coleman and R. Flint, 
``Ising Quasiparticles and Hidden Order in \urs''
Phil. Mag 94:32-33, 3803 - 3819 (2014).


\bibitem{Ohkuni99} H. Ohkuni, Y. Inada, Y. Tokiwa, K. Sakurai, R. Settai
, T. Honma, Y. Haga, E. Yamamoto, Y. Obarnuki
, H. Yamagami, S. Takahashi \& T. Yanagisawa, ``Fermi surface properties and de
Haas-van Alphen oscillation in both the normal and superconducting
mixed states of \urs'',
{\sl Phil. Mag. B} {\bf 79}, 1045 (1999).



\bibitem{Altarawneh11}M. M. Altarawneh, N. Harrison, S. E. Sebastian,
L. Balicas, P. H. Tobash, J. D. Thompson, F. Ronning, and E. D. Bauer
{\sl Phys. Rev. Lett.} {\bf 106}, 146403-146416 (2011).

\bibitem{Hc2}J. P.  Brison et al, ``Anisotropy of the upper critical
field in URu2Si2 and FFLO state in antiferromagnetic
superconductors'', Physica C {\bf 250}, 128-138 (1995).


\bibitem{Altarawneh2}M. M. Altarawneh et al. , 
``Superconducting pairs with extreme uniaxial anisotropy in \urs''
Phys. Rev. Lett. {\bf 108}, 066407-066410 (2012). 

\bibitem{palstra86}T. T. M. Palstra, A. A. Menvosky and J. A. Mydosh,
``Anisotropic electrical resistivity of the magnetic heavy-fermion
superconductor \urs", {\sl Phys.Rev. B}, {\bf 33}, 6528 (1986).

\bibitem{Jo07}Y.J. Jo et al.,''Field-Induced Fermi Surface Reconstruction and Adiabatic Continuity between Antiferromagnetism
and the Hidden-Order State in $URu_2Si_2$,'' {\sl Phys. Rev. Lett.} {\bf 98}, 166404-165407 (2007).

\bibitem{Villaume08} A. Villaume et al., ``Signature of Hidden Order in Heavy Fermion Superconductor $URu_2Si_2$:  Resonance at the wave vector $Q_0 = (1,0,0)$,''{\sl Phys. Rev. B} {\bf 78} 
5114-5117 (2008).

\bibitem{Hassinger10} E. Hassinger et al, ``Similarity of the Fermi Surface in the Hidden Order State and in the Antiferromagnetic State of $URu_2Si_2$,''
{\sl Phys. Rev. Lett.} {\bf 105}, 216409-216412 (2010).

\bibitem{Haule10} K. Haule and G. Kotliar, ``Complex Landau-Ginzburg Theory of the Hidden Order in $URu_2Si_2$,'' {\sl Eur. Lett.} {\bf 89} 57006:p1-57006:p6 (2010).


\bibitem{Ramirez92} A.P. Ramirez et al., ``Nonlinear Susceptibility as a Probe of Tensor Spin Order in $URu_2Si_2$,'', {\sl Phys. Rev. Lett.} {\bf 68}, 2680-2683 (1992).

\bibitem{Ohkawa99} F.J. Ohkawa and H. Shimizu, ``Quadrupole and Dipole Orders in $URu_2Si_2$,'' {\sl J. Phys:  Cond. Mat.} {\bf 11}, L519-L524 (1999).

\bibitem{flint12}R. Flint, P. Chandra and P. Coleman, 
{\sl ``Basal-Plane Nonlinear Susceptibility: A Direct Probe of the Single-Ion Physics in URu$_{2}$Si$_{2}$''},
Phys. Rev. B 86, 155155-155160(2012).

\bibitem{Goremychkin00} E. A. Goremychkin, R. Osborn, B. D. Rainford, and A. P. Murani, ``Evidence for Anisotropic Kondo Behavior in Ce$_{0.8}$La$_{0.2}$Al$_3$,''  {\sl Phys. Rev. Lett.} {\bf 84}, 2211 (2000).

\bibitem{sikkema96} A.E. Sikkema, W.J.L. Buyers, I. Affleck and J.Gan, ``Ising-Kondo lattice with transverse field: A possible f-moment Hamiltonian for \urs,'' {\sl Phys. Rev. B} {\bf 54}, 9322 (1996).''

\bibitem{CoxZawad}D. L. Cox and A. Zawadowski, ``Exotic Kondo Effects
in Metals'', Taylor \& Francis, London. (2002).

\bibitem{Timusk11} U. Nagel et al., ``Optical spectroscopy shows that
the normal state of \urs is an anomalous Fermi liquid'', Proc. National Academy Science, {\bf 109}, 1916-1965 (2012).  


\bibitem{Schmidt10}
A. R. Schmidt
et al., 
``Imaging the Fano lattice to ‘hidden order’ transition in \urs'',  {\sl Nature} {\bf 465}, 570-576 (2010).

\bibitem{Aynajian10}P. Aynajian et al.,
``Visualizing the Formation of the Kondo Lattice and the Hidden Order in \urs'', {\sl PNAS} {\bf 107}, 10383-10388 (2010).


\bibitem{Park11} W.K. Park et al, ``Fano Resonance and Hybridization Gap in Kondo Lattice $URu_2Si_2$'', 
{\sl Phys. Rev. Lett.} {\bf 108}, 246403-24646 (2012).

\bibitem{Coleman83} P. Coleman, ``A New approach to the Mixed Valence Problem'', Phys. Rev. B 29, 3035-3044 (1984)

\bibitem{Cox96} D.L. Cox and M. Jarrell, ``The Two-Channel Kondo route to non-Fermi liquids,''
{\sl J. Phys. Cond. Mat.}{\bf 8} 9825-9853 (1996).

\bibitem{CATK} P. Coleman, A. M. Tsvelik, N. Andrei \& H. Y. Kee, ``Co-operative Kondo effect in the two-channel Kondo lattice'' {\it Phys. Rev. B }{\bf 60}, 3608-3628 (1999).

\bibitem{Hoshino11}S. Hoshino, J. Otsuki and Y. Kuramoto, ``Diagonal Composite Order in a Two-Channel Kondo Lattice'' Phys. Rev. Lett. 107, 247202-247205 (2011).

\bibitem{Bolech02}C. Bolech and N. Andrei, 
{{\sl ``Solution of the Two-Channel Anderson Impurity Model:
Implications for the Heavy Fermion UBe$_{13}$'' }},
{Phys.  Rev. Lett.}, {\bf 88}, 237206-237209 (2002).


\bibitem{marston03}P. Coleman, J. B.  Marston and A. J. Schofield,
{\sl ``Transport anomalies in a simplified model for a heavy-electron
quantum critical point},  Phys. Rev. B {\bf 72},  245111-245116 (2003).

\bibitem{Broholm91} C. Broholm et al., ``Magnetic excitations in the heavy-fermion superconductor $URu_2Si_2$,'' {\sl Phys. Rev B.} {\bf 43}, 12809-12822 (1991).

\bibitem{Wiebe07} C.R. Wiebe et al, ``Gapped Itinerant Spin Excitations Account for Missing Entropy in the
Hidden Order State of $URu_2Si_2$,'' {\sl Nature Physics} {\bf 3},
96-99 (2007).


\bibitem{Niklowitz11}P. G. Niklowitz, S. Dunsiger, C. Pfleiderer,
P. Link, A. Schneidewind, E. Faulhaber, M. Vojta, Y.-K. Huang, and
J. A. Mydosh, {\sl ``Role of commensurate and incommensurate
low-energy excitations in the paramagnetic to hidden-order transition
of \urs''}, arXiv:1110.5599 (2011).







\bibitem{butch}N. Kanchanavavatee et al, ``Twofold enhancement of the hidden-order/large-moment antiferromagnetic phase boundary in the URu$_{2-x}$Fe$_x$Si2 system'', Phys. Rev. B 84, 245122 (2011).

\bibitem{Oppeneer10} P. M. Oppeneer, J. Rusz, S. Elgazzar, M.-T. Suzuki, T. Durakiewicz, and J. A. Mydosh Phys. Rev. B {\bf 82}, 205103 (2010).

\bibitem{slater}J. C. Slater and G. F. Koster, 
``Simplified LCAO Method for the Periodic Potential Problem'',
Phys. Rev. 94, 1498-1524, (1954).


\bibitem{Okazaki11} R. Okazaki et al., ``Rotational Symmetry Breaking in the 
Hidden Order Phase of $URu_2Si_2$'', {\sl Science} {\bf 331} 439-442 (2011).

\bibitem{Santander09} A. F. Santander-Syro et al., ``Fermi-surface instability at the 'hidden order' transition
of $URu_2Si_2$'' {\sl Nature Physics} {\bf 5}, 637 - 641 (2009). 

\bibitem{Haen92}P. Haen , F. Lapierre, P. Lejay, J. Voiron, J. Magnetism and Magnetic Materials {\bf 116}, 108-110 (1992).

\bibitem{Haraldsen11}
J.T. Haraldsen et al., {\sl Phys. Rev. B} {\bf 84}, 214410 (2011).

\bibitem{nphysus}R. Flint, M. Dzero \& P. Coleman, Nat. Phys. {\bf 4}, 643 (2008).

\bibitem{Destri84}N. Andrei and C. Destri, ``Solution of the
Multichannel Kondo Problem'', Phys. Rev. Lett., 52, 364 (1984).

\bibitem{Tsvelik85}A. M. Tsvelik and P. B. Wiegmann, ``Exact solution
of the multichannel Kondo problem, scaling, and integrability'',  J.Stat. Phys. 38, 125 (1985).

\bibitem{EmeryKivelson92}V. J. Emery and S. Kivelson, ``Mapping of
the two-channel Kondo problem to a resonant-level model'', Phys. Rev. B 46,  10812 (1992).

\bibitem{Jaime02} M. Jaime, K. H. Kim, G. Jorge, S. McCall, and J. A. Mydosh, ``High Magnetic Field Studies of the Hidden Order Transition in \ursp,'' {\sl Phys. Rev. Lett.} {\bf 89}, 287201 (2002).

\bibitem{maltseva09}
M. Maltseva, M. Dzero and P. Coleman, 
``Electron Cotunneling into a Kondo Lattice'', 
Physical Review Letters 103, 206402 (2009).


\bibitem{ramirez94}A. P. Ramirez, P. Coleman, P. Chandra, E. Br\:{u}ck, A. A. Menovsky,
Z. Fisk, and E. Bucher, ``Nonlinear susceptibility as a probe of
tensor spin order in \urs '', {\sl Phys. Rev. Lett.}{\bf  68}, 2680 (1992).

\bibitem{newrefschi3}P. Chandra et al., {\sl Physica B}, {\bf 199 \& 200} 426 (1994).

\bibitem{Anderson61}P.W. Anderson, ``Localized magnetic states in
metals'',  Phys. Rev. 124,41-53 (1961). 

\bibitem{Amitsuka03} H. Amitsuka et al., ``Inhomogeneous magnetism in URu2Si2 studied by muon spin relaxation under high pressure," Physica B 326, 418-421 (2003).

\bibitem{Bernal04} O.O. Bernal et al., ``Ambient Pressure $^{99}$Ru NMR in $URu_2Si_2$:  Internal Field Anisotropy,''
J. Mag. Magn. Mat. 272, {E59-60} (2004).


\bibitem{Jackiw76} R. Jackiw and C. Rebbi, {\sl Phys. Rev. D} {\bf 13} 3398 
(1976).

\bibitem{Su79}  W.P. Su, J.R. Schrieffer and A.J. Heeger, {\sl Phys. Rev. Lett.} {\bf 42} 1698 (1979).

\bibitem{Laughlin99} R.B. Laughlin, {\sl Rev. Mod. Phys.} {\bf 71}, 863 (1999).

\bibitem{Castelnuovo12}  C Castelnovo, R. Moessner and S.L. Sondhi, {\sl Ann. Rev. Cond. Mat.} {\bf 3} 35 (2012).


\bibitem{Das13}
P. Das et al., {\sl New J. Phys.} {\bf 15}, 053031 (2013).

\bibitem{Metoki13}
N. Metoki et al., {\sl J. Phys. Soc. Jpn.} {\bf 82}, 055004 (2013).

\bibitem{Ross14}
K.A. Ross, L. Harriger, Z. Yamani, W. J. L. Buyers, J. D. Garrett,
A. A. Menovsky, J. A. Mydosh and  C. L. Broholm,
 arXiv:1402.2689 (2014).

\bibitem{Caciuffo14} 
R. Caciuffo, Private Communication.


\bibitem{Takagi12}
Shigeru Takagi, Shu Ishihara, Makato Yokoyama and Hiroshi Amitsuka, 
``Symmetry of the Hidden Order in URu2Si2 from Nuclear Magnetic Resonance Studies'', {\sl J. Phys. Soc. Jpn.} {\bf 81}, 114710 (2012).


\bibitem{Matsuda11} S. Tonegawa et al, ``Cyclotron Resonance in the Hidden-Order Phase of \urs'' PRL 109, 036501-036504 (2012)





\bibitem{Imry75} Y. Imry and S.-K. Ma, {\sl Phys. Rev. Lett.} {\bf 35}, 
1399 (1975).

\bibitem{Kallin12} 
C. Kallin, {\sl Rep. Prog. Phys.} {\bf 75}, 042501 (2012).


\bibitem{chandracolemanlarkinsuperflow}P. Chandra ,  P. Coleman \&
A. I. Larkin, , {\sl ``Quantum fluids approach to Frustrated Heisenberg Models''},  Journal of Condensed Matter Physics,  {\bf 2} 7933,  (1990).

\bibitem{fomin}I. A. Fomin, {\sl Long-lived induction signal and
spatially nonuniform spin precession in $^{3}$He-B},  Pisma Zh. Eksp. Teor. Fiz {\bf 40}, 260
(1984)[JETP Lett. {\bf 40}, 1037 (1984).

\bibitem{borovik}A. S. Borovik-Romanov, Yu. M. Bunkov, V. V. Dmitriev,
Yu. M. Mukharskii and K. Flachbart, {\sl ``Experimental study of
separation of magnetization precession in $^{3}$He-$B$ into two
magnetic domains}, Zh. Eksp Teor Fiz {\bf 87}, 2025-2038 (1985)[
Sov. Phys JETP, {\bf 61}, 1199 (1985)].

\bibitem{Brison95} 
J.P. Brison , N. Keller , A. Verni\` ere , P.  Lejay , L. Schmidt,
A. Buzdin, J. Flouquet , S.R. Julian  and  G.G. Lonzarich, 
{\sl Physica C} {\bf 250}, 128-138 (1995).

\bibitem{Altarawneh12}M. M. Altarawneh, N. Harrison, G. Li,
L. Balicas, P. H. Tobash, F. Ronning, and E. D. Bauer,
{\sl Phys. Rev. Lett.} {\bf 108}, 066407-066410 (2012). 

\bibitem{Xiao05} D. Xiao, J. Shi, and Q. Niu, 
{\sl Phys. Rev. Lett.} {\bf 95}, 137204 (2005).

\bibitem{Thonhouser05} 
T. Thonhauser, D. Ceresoli, D. Vanderbilt, and R. Resta,
{\sl Phys. Rev. Lett.} {\bf 95}, 137205 (2005).

\bibitem{Xiao06}
D. Xiao, Y. Yao, Z. Fang and Q. Niu, 
{\sl Phys. Rev. Lett.} {\bf 97}, 026603 (2006).




\end{thebibliography}
\end{document}